\documentclass{article}
\usepackage{arxiv}
\usepackage{savesym}
\usepackage[T1]{fontenc}
\restoresymbol{TXF}{iint}
\usepackage{amsfonts}
\usepackage{amsmath}
\usepackage{amssymb}
\usepackage{tikz}
\usepackage{pgfplots}
\usepackage{subfigure}
\usepackage{mathtools}
\usepackage{balance}
\usepackage{xspace}
\usepackage{url}
\usepackage{textcomp}
\usepackage{multirow}
\usepackage{pbox}
 \usepackage{float}
\usepackage{algorithm}
\usepackage{algorithmicx}
\usepackage{algpseudocode}
\usepackage[british]{babel}
\usepackage{enumitem}
\usepackage[normalem]{ulem}
\usepackage{lmodern}
\usetikzlibrary{patterns}
\usetikzlibrary{positioning}
\pgfplotsset{compat=1.17}

\colorlet{boxcolor}{teal!10!white}

\newcommand{\greenp}[2][green,fill=green]{\tikz[baseline=-0.5ex]\draw[#1,radius=#2] (0,0) circle ;}
\newcommand{\yellowp}[2][yellow,fill=yellow]{\tikz[baseline=-0.5ex]\draw[#1,radius=#2] (0,0) circle ;}
\newcommand{\bluep}[2][cyan,fill=cyan]{\tikz[baseline=-0.5ex]\draw[#1,radius=#2] (0,0) circle ;}
\newcommand{\orangep}[2][orange,fill=orange]{\tikz[baseline=-0.5ex]\draw[#1,radius=#2] (0,0) circle ;}

\newcommand{\sys}{Qanaat\xspace}
\newcommand{\REQ}{\textsf{REQUEST}\xspace}
\newcommand{\req}{{\sf \small request}\xspace}
\newcommand{\ONE}{\textsf{PROPOSE}\xspace}
\newcommand{\one}{{\sf \small propose}\xspace}
\newcommand{\TWO}{\textsf{ACCEPT}\xspace}
\newcommand{\two}{{\sf \small accept}\xspace}
\newcommand{\THREE}{\textsf{COMMIT}\xspace}
\newcommand{\three}{{\sf \small commit}\xspace}
\newcommand{\PRE}{\textsf{PREPARE}\xspace}
\newcommand{\pre}{{\sf \small prepare}\xspace}
\newcommand{\PRED}{\textsf{PREPARED}\xspace}
\newcommand{\pred}{{\sf \small prepared}\xspace}
\newcommand{\abort}{{\sf \small abort}\xspace}

\newcommand{\reply}{{\sf \small reply}\xspace}
\newcommand{\cmtq}{{\sf \small commit-query}\xspace}
\newcommand{\predq}{{\sf \small prepared-query}\xspace}

\newtheorem{metalemma}{Lemma}[section]

\newtheorem{Lemma}[metalemma]{Lemma}
\newtheorem{Proposition}[metalemma]{Proposition}

\newenvironment{prop}{\begin{Proposition}\em}{\end{Proposition}}
\newenvironment{lmm}{\begin{Lemma}\em}{\end{Lemma}}

\newenvironment{prf}{\noindent{\bf Proof:}\rm}

\newif\ifextend
\extendtrue

\title{\sys: A Scalable Multi-Enterprise Permissioned Blockchain System with Confidentiality Guarantees}

\author{Mohammad Javad Amiri$^1$~~~Boon Thau Loo$^1$~~~Divyakant Agrawal$^2$~~~Amr El Abbadi$^2$\\
$^1$University of Pennsylvania, $^2$University of California Santa Barbara\\
$^1$\{mjamiri, boonloo\}@seas.upenn.edu~~$^2$\{agrawal, amr\}@cs.ucsb.edu
\vspace{2em}
}

\begin{document}

\maketitle

\begin{abstract}

Today's large-scale data management systems need to address
distributed applications' {\em confidentiality} and {\em scalability} requirements
among a set of collaborative enterprises.
This paper presents \textit{\sys}, a scalable multi-enterprise permissioned blockchain system
that guarantees the confidentiality of enterprises in collaboration workflows.
\sys presents {\em data collections} that enable
any subset of enterprises involved in a collaboration workflow
to keep their collaboration private from other enterprises.
A transaction ordering scheme is also presented to enforce only the necessary and sufficient constraints
on transaction order to guarantee data consistency.
Furthermore, \sys supports data consistency across collaboration workflows
where an enterprise can participate in different collaboration workflows with different sets of enterprises.
Finally, \sys presents a suite of consensus protocols
to support intra-shard and cross-shard transactions
within or across enterprises.
\end{abstract}
\section{Introduction}
\label{sec:intro}

Emerging multi-enterprise applications, e.g.,
supply chain management \cite{korpela2017digital}, multi-platform crowdworking \cite{amiri2021separ}, and healthcare \cite{azaria2016medrec},
require extensive use of {\em collaboration workflows} where multiple mutually distrustful 
distributed enterprises collaboratively process a mix of public and private transactions.
Despite years of research in distributed transaction processing, data management systems today have not yet achieved a good balance among the {\em confidentiality} and {\em scalability} requirements of these applications.

One of the main requirements of multi-enterprise applications is confidentiality with respect to
{\em data sharing} and {\em data leakage}.
First, while public collaboration among all enterprises is visible to everyone,
specific subsets of the data may need to be shared only with specific subsets of involved enterprises.
For example, in a product trading,
the {\sf distributor} may want to keep collaboration with a {\sf farmer} and a {\sf shipper}
involving terms of the trades {\em confidential} from the {\sf wholesaler} and the {\sf retailer},
so as not to expose the premium they are charging \cite{fabric2018collections}.
Second, the enterprises must prevent malicious nodes from leaking confidential data, e.g., requests, replies, and stored data.

Multi-enterprise applications also need to scalably
process a large number of transactions within or across enterprises.
Although sharding has been used in single-enterprise applications to address scalability,
it is challenging to use sharding in multi-enterprise applications where confidentiality of data is paramount.

In multi-enterprise applications, furthermore, data {\em provenance} is needed to
provide a detailed picture of how the data was collected, where it was stored, and how it was used.
The information should be stored in a {\em transparent} and {\em immutable} manner
which is {\em verifiable} by all participants, e.g., transparent end-to-end tracking of goods to detect possible fraud in a supply chain collaboration workflow.

The decentralized nature of blockchain and its unique features such as
provenance, immutability, and tamper-resistant, make it appealing to a wide range of applications, e.g.,
supply chain management \cite{chen2020blockchain,tian2017supply}, crowdsourcing \cite{han2019fluid,amiri2021separ},
contact tracing \cite{peng2021p2b} and federated learning \cite{peng2021vfchain}.

In recent years, various permissioned blockchain systems have been proposed to
address the confidentiality and/or scalability of multi-enterprise applications.
Hyperledger Fabric~\cite{androulaki2018hyperledger}
and its variants \cite{gorenflo2019fastfabric,gorenflo2020xox,sharma2019blurring,ruan2020transactional}
supports multi-enterprise applications and
provides scalability using channels~\cite{androulaki2018channels} managed by subsets of enterprises.
However, all transactions of a channel are sequentially ordered, resulting in reduced performance.
Fabric also uses private data collections~\cite{fabric2018collections}
to manage confidential collaboration among a subset of enterprises.
However, appending a hash of all private transactions to
a global ledger replicated on every enterprise
increases computational overhead and hence reduces throughput.
Furthermore, Fabric does not address confidential data leakage
by malicious nodes.

Caper~\cite{amiri2019caper} supports collaborative enterprises
and provides confidentiality by maintaining a partial view of the global ledger on each enterprise.
Caper, however, does not support:
(1) confidential collaboration among subsets of enterprises (i.e., it supports internal or public transactions),
(2) data consistency across collaboration workflows that an enterprise is involved in,
(3) data confidentiality in the presence of malicious nodes, and
(4) multi-shard enterprises.

Scalability has also been studied in the context of applications that are used within a single enterprise, e.g.,
AHL~\cite{dang2018towards}, SharPer~\cite{amiri2021sharper} and
ResilientDB~\cite{gupta2020resilientdb}.
However, since these systems are restricted to single-enterprise applications, they
do not address the confidentiality requirement of multi-enterprise applications.

To address the above scalability and confidentiality challenges of multi-enterprise applications,
we present {\em \sys}\footnote{\sys is a scalable
underground network consisting of private channels to transport water
from an aquifer to the surface (i.e., an underground aqueduct).},
a permissioned blockchain system that
supports collaboration workflows across enterprises.
In \sys, each enterprise partitions its data into multiple {\em shards} to improve scalability.
Each shard's transactions are then processed by a disjoint {\em cluster} of nodes.

To support confidential collaboration (i.e., data sharing),
in addition to public transactions among all enterprises and internal transactions within each enterprise,
any subset of enterprises might process transactions {\em confidentially} from other enterprises.
As a result, \sys uses a novel hierarchical data model consisting of a set of {\em data collections}
where each transaction is executed on a data collection.
The root data collection consists of the public records from executing public transactions
replicated on all enterprises.
Each local data collection includes the private records of a single enterprise.
Intermediate data collections, which are optional and are needed in case of {\em confidential}
collaborations among subsets of enterprises,
maintain private records shared among subsets of collaborating enterprises.
\sys replicates such shared data collections on the involved enterprises to facilitate the use of shared data by enterprises in their internal transactions.
For example, a {\sf supplier} in a supply chain collaboration workflow
requires the order data stored in the shared data collection with a {\sf consumer}
to perform its internal transactions that produce the products.

To support this collaborative yet confidential data model, \sys proposes a transaction ordering scheme
to preserve the data consistency of transactions.
In \sys, an enterprise might be involved in multiple data collections.
Hence, the traditional transaction ordering schemes used in fault-tolerant protocols where each cluster
orders transactions on a single data store do not work.
In fact, while ordering transactions, \sys needs to capture the state of all data collections
that might affect the execution of a transaction.

Finally, to prevent confidential data leakage despite the Byzantine failure of nodes,
\sys utilizes the privacy firewall technique \cite{yin2003separating,duan2016practical} and
(1) separates ordering nodes that agree on the order of transactions
from execution nodes that execute transactions and maintain the ledger,
and (2) uses a {\em privacy firewall} between execution nodes and ordering nodes.
The privacy firewall restricts communication from execution nodes in order to filter out incorrect messages, possibly including confidential data.

The main contributions of this paper are:
\begin{itemize}[nosep]
    \item An infrastructure consisting of ordering nodes, execution nodes, and a privacy firewall and
    a data model consisting of data collections to
    preserve confidentiality, followed by
    a transaction ordering scheme that enforces only the {\em necessary} and {\em sufficient} constraints
    to guarantee data consistency,
    \item A suite of consensus protocols to process transactions
    that accesses single or multiple data shards within or across enterprises in a {\em scalable} manner, and
    \item \sys, a scalable multi-enterprise permissioned blockchain system that guarantees confidentiality.    
\end{itemize}

The rest of this paper is organized as follows.
Section~\ref{sec:motiv} motivates \sys
by discussing the vaccine supply chain.
The \sys model is introduced in Section~\ref{sec:model}.
Section~\ref{sec:protocol} presents transaction processing in \sys.
Section~\ref{sec:eval} evaluates the performance of \sys.
Section~\ref{sec:related} discusses related work, and
Section~\ref{sec:conc} concludes the paper.
\section{Motivation}
\label{sec:motiv}

\sys is designed to support multi-enterprise applications such as supply chain management \cite{korpela2017digital},
multi-platform crowdworking \cite{amiri2021separ}, and healthcare \cite{azaria2016medrec} in a scalable and confidentiality-preserving manner.
To motivate \sys, we consider an example application based on the COVID-19 vaccine supply chain, given that it is a demanding application.
The COVID-19 outbreak has demonstrated diverse challenges that supply chains face \cite{alam2021challenges}.
While the vaccine supply chain seems to work fine on the surface, it suffers from several crucial challenges.
First, current vaccine supply chains are subjected to different types of vulnerabilities, e.g., 
attacks on $44$ companies involved in the COVID-19 vaccine distribution in $14$ countries \cite{covid21},
rapidly growth of the black market for COVID-19 vaccines \cite{counterfeit2021stone},
issuing fake vaccine cards \cite{fake2020reilly}, and
loss of $400$ millions of vaccine doses \cite{emergent2022bruggeman}.
Moreover, while vaccines are well distributed in most developed countries, getting them to developing countries and accounting for them remains a challenge, e.g., fake vaccines \cite{aborode2022fake,choudhary2021fake}
and falsified CoviShield vaccines \cite{medical2021WHO}.
Currently, the COVID-19 vaccine supply chains have been over-stretched given the global supply chain challenges and the possibility of abuse and fraud stated above.
At the same time, other supply chains are now under unprecedented pressure \cite{Shanghai2022Schiffling,global2022liberty}.
Even in a developed country, supply chains can be subjected to attacks that can potentially disrupt the distribution \cite{covid21}.
Vaccines are vulnerable to counterfeiting \cite{VaccineCounterfeit,amankwah2022covid}, tampering and theft \cite{VaccineTheft}
especially since products and components often pass through multiple locations and even countries
and it has become difficult to trace the source of vaccines,
creating space for substandard and falsified vaccines \cite{world2017study}.

Multiple enterprises collaborate in a vaccine supply chain to get vaccines from manufacturers to citizens.
Enterprises involved in the vaccine ecosystem need to
collaborate and share data based on agreed-upon service level agreements.
Vaccines, on one hand, need to be tracked and monitored throughout the process to ensure
that they are stable and not tampered with, and on the other hand, this data should be accessible
to the public to increase their trust in the vaccines.
While distributed databases can be used to address these challenges, they lack mechanisms
to first prevent any malicious entity from altering data and second enable enterprises to verify the data and the state of collaboration.
\sys uses blockchains to provide verifiability, provenance, transparency, and
assurances to end-users of the vaccines’ safety and efficacy.
In particular, blockchains enable pharmaceutical manufacturers to easily track on-time shipment and delivery of vaccines,
provide efficient delivery tracking for transportation companies, and
help hospitals manage their stocks and mitigate supply and demand constraints.

\begin{figure}[t] \center
\includegraphics[width=0.6\linewidth]{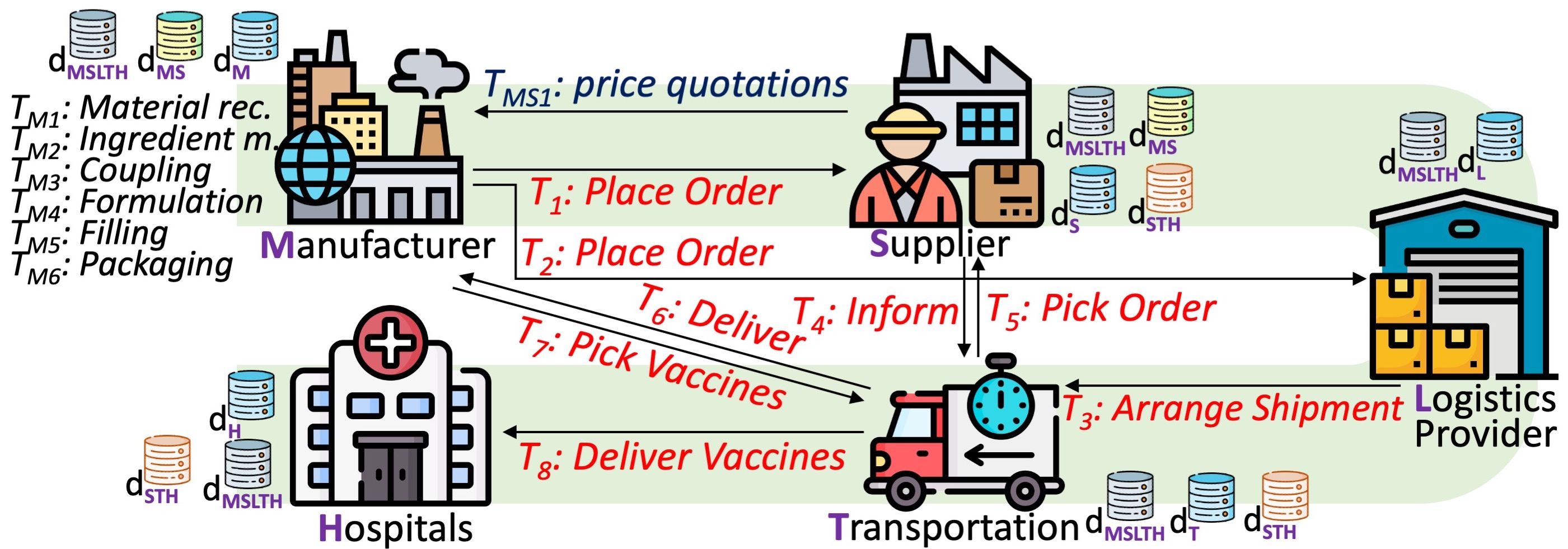}
\caption{A vaccine supply chain collaboration workflow}
\label{fig:chain}
\end{figure}

Figure~\ref{fig:chain} shows a simplified vaccine supply chain collaboration workflow consisting of
the \textsf{pharmaceutical manufacturer (M)}, a \textsf{supplier (S)}, a \textsf{logistics provider (L)},
a \textsf{transportation company (T)}, and \textsf{hospitals (H)}.
Using \sys, the public transactions of the collaboration workflow ($T_1$ to $T_8$) are executed on
{\em data collection} $d_{MSLTH}$ maintained by all enterprises.
In particular, the \textsf{pharmaceutical manufacturer} {\tt \small places orders} ($T_1$ and $T_2$)
of materials via the \textsf{supplier} and \textsf{logistics provider}.
The \textsf{logistics provider} {\tt \small arranges shipment} ($T_3$) by the \textsf{transportation company}, 
Once the materials are ready, the \textsf{supplier} {\tt \small informs} ($T_4$) the \textsf{transportation company} to {\tt \small pick order} ($T_5$)
and {\tt \small deliver order} ($T_6$) to the \textsf{manufacturer}. 
Once the vaccines become available, the \textsf{transportation company} {\tt \small pick vaccines} ($T_7$)
and {\tt \small deliver} ($T_8$) them to \textsf{hospitals}.
These transactions are public and executed on data collection $d_{MSLTH}$.

In addition, the collaboration workflow includes the internal transactions of each enterprise.
Internal transactions are executed on the enterprise's {\em private} data collection and might reflect patented and
copyrighted formulas and information.
For example, producing vaccines in a \textsf{pharmaceutical manufacturer} consists of {\tt \small
material reception} ($T_{M1}$) (some vaccines requiring ${\sim} 160$ consumables),
{\tt \small ingredient manufacturing} ($T_{M2}$), {\tt \small coupling} ($T_{M3}$), \texttt{\small formulation} ($T_{M4}$),
{\tt \small filling} ($T_{M5}$), and {\tt \small packaging} ($T_{M6}$) \cite{sanofi2019,vaccine2016}.
The \textsf{pharmaceutical manufacturer} executes these transactions on its private data collection $d_{M}$.
The \textsf{supplier}, \textsf{logistics provider}, \textsf{transportation company}, and \textsf{hospitals} also
execute their internal transactions on private
data collections $d_{S}$, $d_{L}$, $d_{T}$, and $d_{H}$ respectively.

A blockchain-enabled multi-enterprise application, e.g., vaccine supply chain, needs to support the following requirements:

\noindent
{\bf R1. Confidential collaborations across enterprises.}
Any subset of enterprises involved in a collaboration workflow
might want to keep their collaboration private from other enterprises.
For example, the \textsf{supplier} might want to make private transactions with the \textsf{pharmaceutical manufacturer}
to keep some terms of trade confidential from other enterprises, e.g.,
\texttt{\small price quotation} ($T_{MS1}$) transaction.
Such private transactions need to be executed on a data collection shared between {\em only} the involved enterprises, e.g.,
$d_{MS}$ between the \textsf{pharmaceutical manufacturer} and the \textsf{supplier}.

\noindent
{\bf R2. Consistency across collaboration workflows.}
An enterprise might be involved in multiple collaboration workflows with different sets of enterprises.
For example, a \textsf{transportation company} that
distributes both Pfizer and Moderna vaccines
needs to assign trucks based on the total number of vaccine packages and deal with interleaving transactions.
In case of data dependency among transactions that span collaboration workflows,
data integrity and consistency must be maintained.

\noindent
{\bf R3. Confidential data leakage prevention.}
An enterprise needs to ensure that even if its infrastructure includes malicious nodes
(i.e., an attacker manages to compromise some nodes),
the malicious nodes cannot leak any confidential data, e.g., requests, replies, processed data and stored data.

\noindent
{\bf R4. Scaling multi-shard enterprises.}
Every day, millions of people get vaccinated, thousands of shipments take place and many other
transactions are executed.
All these activities need to be immediately processed by the system and specifically by all involved enterprises.
A \textsf{transportation company} that distributes vaccines across the world might
maintain its data in multiple data shards.
This highlights the need for a system that can scale as demand increases and
efficiently process distributed transactions that access multiple data shards across multiple enterprises.

Existing blockchain solutions, however, are not able to meet all requirements of multi-enterprise collaboration workflows.
While Caper \cite{amiri2019caper} enables enterprises to keep their local data confidential,
it does not address any of the {\bf R1} to {\bf R4} requirements.
Fabric \cite{androulaki2018hyperledger} addresses confidential collaboration using private data collection
(although with a high overhead) and scalability using channels.
However, Fabric does not support requirements {\bf R2} and {\bf R3}.
Several variants of Fabric, such as Fast Fabric \cite{gorenflo2019fastfabric}, 
Fabric++ \cite{sharma2019blurring},
FabricSharp \cite{ruan2020transactional}, and
XOX Fabric  \cite{gorenflo2020xox},
try to address the performance shortcomings of Fabric, especially when dealing with contentious workloads.
However, these permissioned blockchain systems, similar to Fabric,
suffer from the overhead of confidential collaboration and also
do not support requirements {\bf R2} and {\bf R3}.

\section{\sys Model}
\label{sec:model}

\sys is a permissioned blockchain system designed to support multi-enterprise applications.
This section introduces \sys model and demonstrates how it supports different requirements of multi-enterprise applications.

\subsection{The Model Assumptions}\label{sec:assumption}

\sys consists of a set of collaborative {\em enterprises}.
Each enterprise owns a set of {\em nodes} (i.e., servers) that are grouped into different {\em clusters}.
Each enterprise further partitions its data into multiple {\em data shards}.
Each cluster of nodes maintains a {\em shard} of the enterprise data and
processes transactions on that data shard.

\sys assumes the partially synchronous communication model.
In the partially synchrony model, 
there exists an unknown global stabilization time (GST), after which
all messages between correct nodes are received within some bound $\Delta$.
\sys inherits the standard assumptions of previous permissioned blockchain systems, including
a strong computationally-bounded adversary that can coordinate malicious nodes and
delay communication to compromise service
an unreliable network that connects nodes and might drop, corrupt, or delay messages, and
the existence of pairwise authenticated communication channels,
standard digital signatures and public-key infrastructure (PKI).
A message $m$ signed by node $i$ is denoted as $\langle m \rangle_{\sigma_i}$.
\sys further uses threshold signatures where each node $i$ holds a distinct private key
that is used to create a signature share $\sigma\langle m \rangle_i$ for message $m$.
A node can generate a valid threshold signature $\sigma\langle m \rangle$ for $m$
given $n-f$ signature shares from distinct nodes.
A collision-resistant hash function $D(.)$ is also used to map
a message $m$ to a constant-sized digest $D(m)$.

We next present the data model (Section~\ref{sec:data}) and the blockchain ledger (Section~\ref{sec:ledger})
of \sys, which together address 
collaboration confidentiality ({\bf R1}) and data consistency ({\bf R2}) requirements and
then demonstrate the \sys infrastructure (Section~\ref{sec:infra}) that addresses the confidential data leakage prevention requirement ({\bf R3}).
For simplicity of presentation, we first consider single-shard
enterprises and then show how the model and the ledger are extended to support multi-shard enterprises (Section~\ref{sec:multicluster}),
which addresses the scalability requirement ({\bf R4}).

\subsection{Data Model}\label{sec:data}

\sys constructs a hierarchical data model consisting of a set of {\em data collections} for each collaboration workflow.
Each data collection can be seen as a separate datastore where the execution of transactions updates its data.
Each data collection further has its own (business) logic to execute transactions.
{\em Public} transactions of a collaboration workflow, e.g.,
{\tt \small place order} or {\tt \small arrange shipment} in the vaccine supply chain,
are executed on the {\em root} data collection.
All enterprises maintain the root data collection.
This is needed because enterprises use the data maintained in the root data collection
in other transactions.
For example,
the {\tt \small order} data stored in the root data collection is used by
the {\sf supplier} in its private transactions to provide raw materials.

{\em Internal private} transactions of each enterprise, on the other hand,
are executed on its local data collection.
Internal transactions are performed within an enterprise following the logic of the enterprise, e.g.,
{\tt \small develop formulation}, and {\tt \small package vaccines} take place within the {\sf pharmaceutical manufacturer}.

Any subset of enterprises might also execute private transactions
among themselves and confidentially from other enterprises.
Such transactions are executed on a data collection shared
among and maintained by {\em only} the involved enterprises.
For example,
the {\sf supplier} executes private transactions
with the {\sf pharmaceutical company} on the material demand on
a data collection that is maintained only by the {\sf supplier} and {\sf pharmaceutical company}.

It should be noted that a data collection is not a physically separated datastore,
rather a logical partition of the data records of an enterprise that might be shared with a set of enterprises.
Hence, creating a data collection causes no overhead, e.g., configuration cost.

\begin{figure}[t] \center
\includegraphics[width=0.7\linewidth]{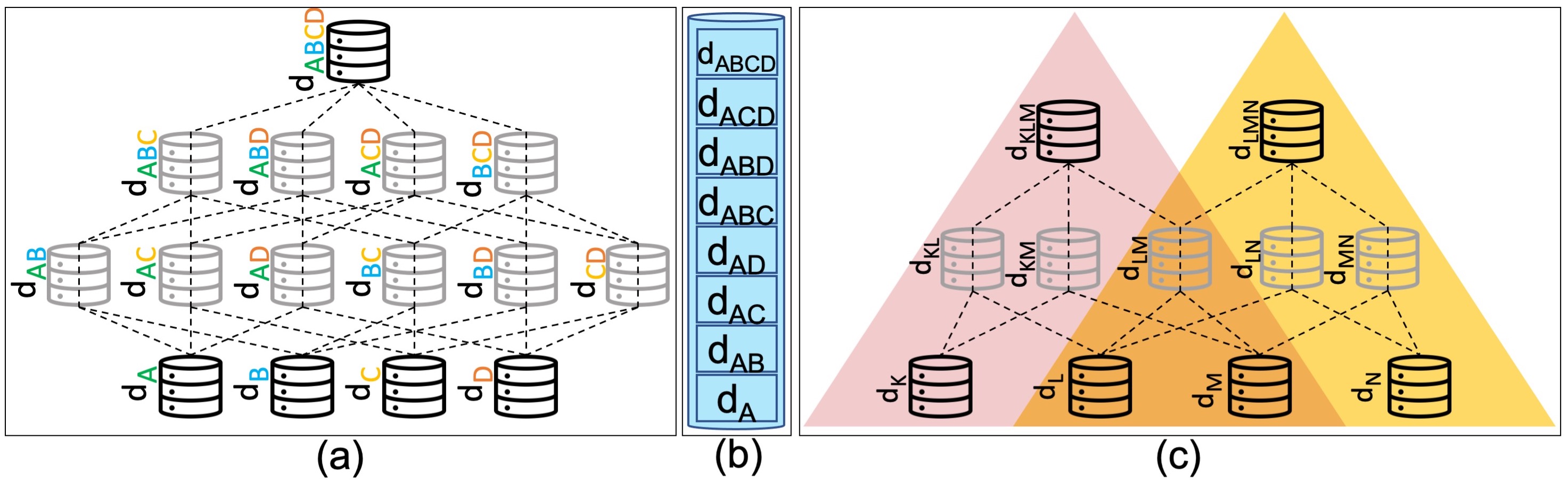}
\caption{(a) A data model for a $4$-enterprise collaboration workflow.
Intermediate data collections are optional and needed only if
there is a collaboration among a specific subset of enterprises,
(b) data maintained by enterprise $A$,
(c) data models for $3$-enterprise workflows.}
\label{fig:data}
\end{figure}

Figure~\ref{fig:data}(a) presents a data model
for a collaboration workflow with enterprises $A$, $B$, $C$, and $D$.
At the top level, a public data collection $d_{ABCD}$ is stored on all four enterprises.
At the bottom level,
there are four private data collections $d_A$, $d_B$, $d_C$, and $d_D$ stored on the corresponding enterprises.
The figure also includes all possible private data collections with different subsets of enterprises at the intermediate levels, e.g.,
$d_{AB}$ or $d_{ACD}$.
When a subset of enterprises, e.g., $A$ and $B$, creates a data collection, e.g., $d_{AB}$, to execute private transactions,
the shared data collection maintains the execution results of transactions that are executed on the data collection.
Such data records are different from the records maintained in the data collection of each individual enterprise, e.g., $d_A$ and $d_B$.
In contrast to the root and the local data collections that are needed in every collaboration workflow,
intermediate data collections are {\em optional} and needed only if
there is a collaboration among a specific subset of enterprises.

\sys defines operational primitives necessarily to capture
data consistency and data dependency across data collections.

\noindent{\bf Write.}
The execution results of transactions executed on a data collection are written on the records of the same data collection.
This is needed to guarantee collaboration confidentiality as
(intermediate) data collections are created for private collaboration, e.g., private lobbying,
among a subset of enterprises while preventing other enterprises from accessing the data.
In particular, if enterprises $A$ and $B$ make a confidential collaboration,
the resulting data should not be written on a data collection that is either shared with some other enterprise $C$ or
not shared with one of the involved enterprises, e.g., $d_A$.
Note that $d_{AB}$ is maintained by both enterprises $A$ and $B$ and they have access to its record.
This captures the confidential collaborations in real-world cross-enterprise applications.

\noindent{\bf Read.}
When a transaction is being executed on a data collection $d_X$, it might read records of a data collection $d_Y$
if enterprises sharing $d_X$ are a subset of the enterprises sharing $d_Y$.
For example, the internal transactions of the {\sf supplier} can read
the records of all data collections that the {\sf supplier} is involved in with other enterprises.
This is needed because those records might affect the internal transactions of the {\sf supplier},
e.g., the number of vaccines that the {\sf supplier} supplies depends on the orders it receives from other enterprises.

More formally, given a data collection $d_X$ where $X$ is the set of enterprises sharing $d_X$.
For each data collection $d_Y$ where $X \subseteq Y$, we define $d_X$ as {\em order-dependent} of $d_Y$.
We say transactions executed on $d_X$ can read the records of $d_Y$ if data collection $d_X$ is order-dependent on $d_Y$.
The dashed lines between data collections in Figure~\ref{fig:data}(a) present order-dependency among data collections;
the transactions executed on the lower-level data collections can read the records of the higher-level data collections, e.g.,
$d_{AB}$ can read $d_{ABC}$, $d_{ABD}$ and $d_{ABCD}$.

In some applications, transactions that are executed on data collection $d_X$ might also need to
verify the records of another data collection $d_Y$ in a {\em privacy-preserving} manner
(i.e., without reading the exact records) revealing any information about the content of the records
if enterprises sharing $d_X$ are a {\em superset} of the enterprises sharing $d_Y$, i.e., $Y \subset X$.

In particular, for intangible assets, e.g., cryptocurrencies,
if enterprise $A$ initiates a transaction in data collection $d_{AB}$ that consumes some coins,
enterprise $B$ needs to verify the existence of the coins in data collection $d_{A}$.
In supply chain workflows, however, since transactions are usually placed as a result of physical actions,
verifying the records of order-dependent data collections is unnecessary.
\sys can be extended to support privacy-preserving verifiability using advanced cryptographic primitives
like secure multiparty computation \cite{evans2017pragmatic,archer2018keys} and
zero-knowledge proofs \cite{kosba2018xjsnark,maller2019sonic,gabizon2019plonk}.

Every enterprise
maintains all data collections that the enterprise is involved in, i.e., the root, a local,
and perhaps several intermediate data collections.
\sys replicates shared data collections on all involved enterprises to facilitate the use of shared data
by the transactions on order-dependent data collections.
Note that such shared data collections can be maintained by a third party, e.g., a cloud provider, trusted by all enterprises.
However, this is a centralized solution that contradicts the decentralized nature of blockchains.
Moreover, the data maintained in a shared data collection
might be used (i.e., read) in transactions on all order-dependent data collections.
Hence, enterprises need to query the cloud for every simple read operation to access the data.

As shown in Figure~\ref{fig:data}(b),
enterprise $A$ maintains local data collection $d_A$,
root data collection $d_{ABCD}$, and all intermediate
data collections (if exist), e.g., $d_{AB}$ and $d_{ACD}$.

A data model represents data processed by each collaboration workflow among a set of enterprises.
However, an enterprise (or a group of enterprises) might be involved
in multiple collaboration workflows (instances of \sys) with different sets of enterprises.
The transactions of the same enterprise in different collaboration workflows might be data-dependent.
Creating an independent data collection for each enterprise
in each collaboration workflow might result in {\em data consistency} issues.
For example, a {\sf supplier} that provides raw materials for both Pfizer and Moderna vaccines in two different
supply chain collaboration workflows needs to know the total number of the requested materials in order to provision for them correctly.
As a result, \sys creates a single data collection for each enterprise
and if an enterprise is involved in multiple possibly data-dependent collaboration workflows, 
the transactions of the enterprise in different collaboration workflows are executed on the same data collection.
An enterprise might also collaborate with enterprises outside \sys and require to access its data.
Such data operations can be performed on the enterprise's local data collection without violating data consistency
enabling the enterprise to use such data in its collaboration workflows within \sys.

Figure~\ref{fig:data}(c) presents two data models for two collaboration workflows
where enterprises $K$, $L$, and $M$ are involved in the first and
enterprises $L$, $M$, and $N$ are involved in the second collaboration workflow.
Since $L$ and $M$ are involved in both workflows,
the local data collections $d_L$ and $d_M$ and the intermediate data collection $d_{LM}$
are shared in both collaboration workflows.

\subsection{Blockchain Ledger}\label{sec:ledger}

The blockchain ledger is an append-only data structure to maintain transaction records.
The blockchain ledger provides immutability and verifiability
that prevents any malicious enterprise from altering data and
enables enterprises to verify the state of the data.
When several enterprises execute transactions across different data collections,
maintaining consistency and preserving confidentiality in a scalable manner is challenging.
Before describing the \sys ledger,
we discuss three possible solutions for ordering transactions in cross-enterprise collaborations.

\noindent
\textbf{1. A single, global ledger.}
One possibility is to construct a linear blockchain ledger for each collaboration workflow where
all transactions on all data collections are totally ordered and appended
to a single global ledger.
To preserve the confidentiality of private transactions,
the cryptographic hash of such transactions (instead of the actual transaction)
can be maintained in the blockchain ledger.
This technique, which has been used in Hyperledger Fabric \cite{androulaki2018hyperledger},
suffers from two main shortcomings.
First, forcing {\em all} transactions into a single, sequential order
unnecessarily degrades performance, especially in the large-scale collaborations targeted by \sys.
To see this, note that while total ordering is needed for data-dependent transactions, e.g.,
transactions executed on the same data collection, to preserve data consistency,
total ordering among transactions of independent data collections,
e.g., $d_A$ and $d_B$, is clearly not needed.
Second, since this solution requires a single, global blockchain ledger to be maintained by every enterprise, 
each enterprise needs to maintain (the hash of) transactions that the enterprise was not involved in,
e.g., internal transactions of other enterprises, which increases bandwidth and storage costs. 

\noindent
\textbf{2. One ledger for each data collection.}
A second possibility is to maintain a separate transaction ordering (i.e., linear blockchain ledger) for each data collection. 
The blockchain ledger of each enterprise then consists of multiple parallel linear ledgers
(one per each data collection that the enterprise is involved in).
This solution, however, does not consider dependencies across order-dependent data collections where
the execution of a transaction might require reading records from other data collections.
Such dependencies need to be captured during the {\em ordering phase} to ensure that all replicas read the same state in the execution phase.
Note that the read-set and write-set of transactions might not be known before execution.
As a result, writing the read values into the block is not possible in the ordering phase.

For example, consider an internal transaction of the {\sf supplier} that
reads the {\sf order} record of the root data collection and based on that, computes and stores the results.
In the meantime (during the period from the initiation to the execution of the transaction), the value of the {\sf order} record might change.
As a result, if the state of the root data collection has not been captured,
different nodes (i.e., replicas) of the {\sf supplier},
might read different values (old or new) of the {\sf order} in the execution phase resulting in inconsistency.

\noindent
\textbf{3. One ledger for each enterprise.}
A third solution is to maintain a total order among all transactions in which an enterprise is involved.
This solution, in contrast to the second solution, guarantees data consistency. 
However, similar to the first solution, this solution prevents order-independent transactions from being appended to the ledger in parallel.
For example, transactions on data collections $d_{AB}$ have no data dependency
with transactions on $d_{AC}$ and can append to the ledger of enterprise $A$ in parallel.

\sys addresses the shortcomings of the existing solutions
in guaranteeing data consistency and preserving collaboration confidentiality while
creating a ledger in an efficient and scalable manner.
In \sys, the blockchain ledger guarantees two main properties:
\begin{itemize}
\item {\bf Local consistency.} 
A total order on the transactions of each data collection is enforced
due to data dependency among transactions executed on a data collection, and

\item {\bf Global consistency.} The order of the transactions of data collection $d_X$
with respect to the transactions of all data collections that $d_X$ is order-dependent on,
is determined by capturing the state of such data collections.
Moreover, transactions are ordered across the enterprises consistently.

\end{itemize}

\begin{figure}[t] \center
\includegraphics[width=0.7\linewidth]{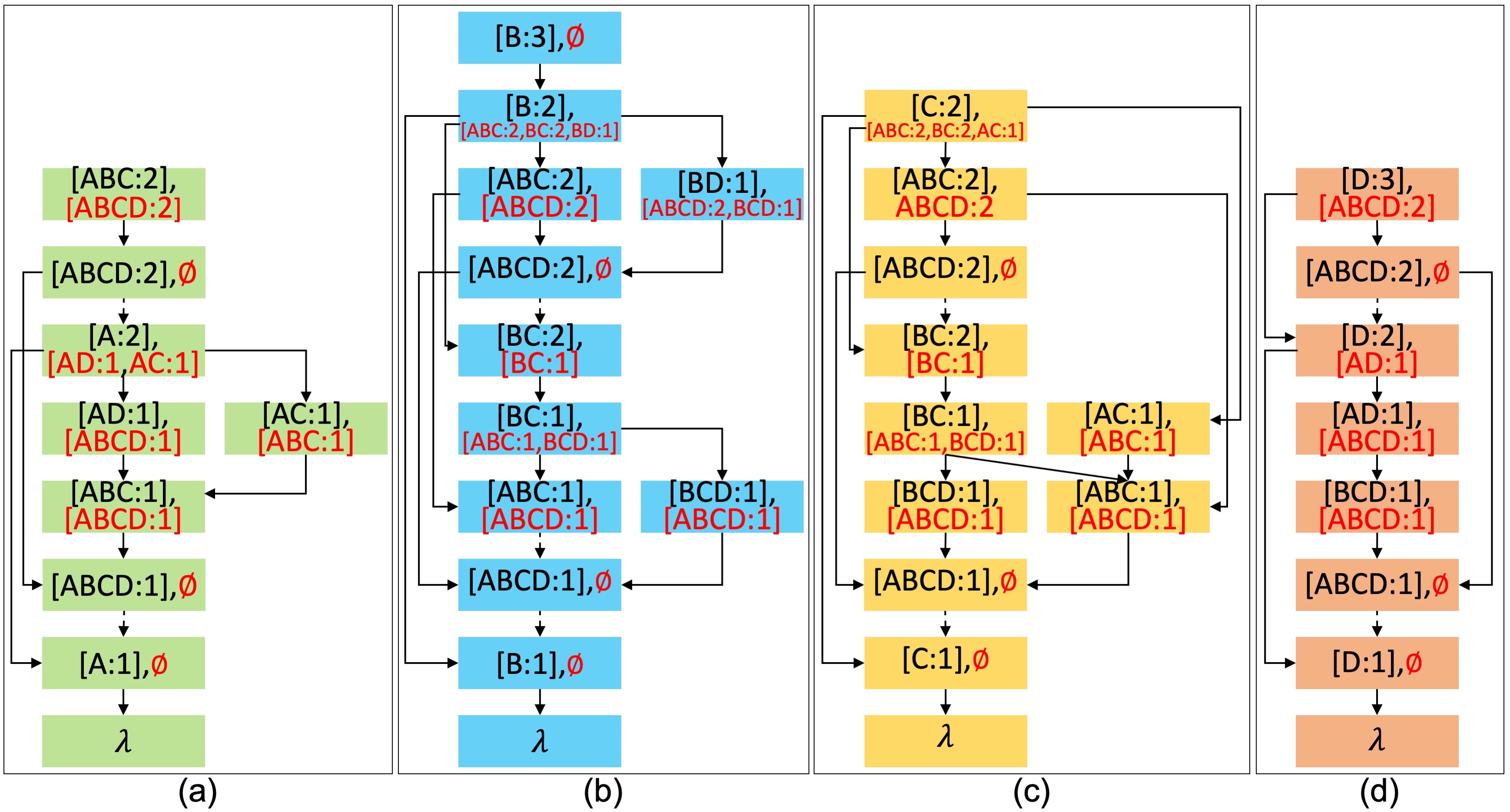}
\caption{(a)-(d) The blockchain ledger for enterprises $A$, $B$, $C$, and $D$ in a collaboration workflow}
\label{fig:ledger}
\end{figure}

\sys constructs a single DAG-structured ledger for each enterprise consisting of
transaction records of the data collections that are maintained by the enterprise.
The first rule guarantees data consistency within each single data collection and the second rule
ensures global data consistency across all order-dependent data collections maintained by an enterprise.
Given the above rules,
each transaction $t$ initiated on data collection $d_X$ has
an identifier, \texttt{ID} = $\langle \alpha, \gamma \rangle$,
composed of a {\em local} part $\alpha$ and possibly a {\em global} part $\gamma$
where the \texttt{ID} is assigned during the ordering phase.
The local part $\alpha$ is $[X{:}n]$ where $X$ is a {\em label} representing the involved enterprises in $d_X$ and 
$n$ is a {\em sequence number} representing the order of transaction $t$ on $d_X$.
In the global part $\gamma$, for {\em every} data collection $d_Y$ where $d_X$ is order-dependent on $d_Y$, $Y{:}m$ is added
to the $\gamma$ of $t$ to capture the state of data collection $d_Y$.
Here, $Y{:}m$ is the local part of the \texttt{ID} of the last transaction that is committed on data collection $d_Y$.
When $t$ is executed, it might read the state of a subset of data collections that are captured in the global part of its \texttt{ID}.

Given two transactions $t$ and $t'$ on data collection $d_X$. Let
$\texttt{ID}(t) = \langle \alpha, \gamma \rangle$ where
$\alpha = [X{:}n]$, $\gamma = [..., Y_p{:}m_p,{\color{purple}Y_q{:}m_q}, ...]$ and
$\texttt{ID}(t') = \langle \alpha', \gamma' \rangle$ where
$\alpha' = [X{:}n']$, and $\gamma = [...,{\color{purple}Y_q{:}m'_q},Y_r{:}m'_r, ...]$.
If $t$ is ordered before $t'$ ($t \rightarrow t'$) then:
\begin{itemize}
    \item $n < n'$ (local consistency), and
    \item $\forall$ data collection $d_{Y_q} \in \gamma \cap \gamma'$, $m_q \leq m'_q$ (global consistency).
\end{itemize}

Figure~\ref{fig:ledger}(a)-(d) shows the blockchain ledger of
four enterprises $A$, $B$, $C$, and $D$
created in the \sys model (following Figure~\ref{fig:data}(a)) for a collaboration workflow.
The collaboration workflow includes two public transactions
on root data collection $d_{ABCD}$,
several internal transactions on local data collections, and
multiple transactions among subsets of enterprises.
In this figure,
$\langle [A{:}1],\varnothing \rangle$
is an internal transaction on data collections $d_A$
(with $\gamma = \varnothing$ because no transactions has been processed yet).
$\langle [ABCD{:}1],\varnothing \rangle$ is a public transaction on $d_{ABCD}$
(with $\gamma = \varnothing$
because the root data collection is not order-dependent on any data collections).
As shown, two transactions $\langle [ABC{:}1],[ABCD{:}1] \rangle$ (on $d_{ABC}$) and
$\langle [BCD{:}1],[ABCD{:}1] \rangle$ (on $d_{BCD}$)
are appended to the ledger of enterprise $B$ (as well as $C$) in parallel
because $d_{ABC}$ and $d_{BCD}$ are not order-dependent.
Both transactions include global part $\gamma = [ABCD{:}1]$ because both $d_{ABC}$ and $d_{BCD}$
are order-dependent on $d_{ABCD}$.
Transactions of Each non-local data collection, e.g., $\langle [ABC{:}1],[ABCD{:}1] \rangle$, which is shared between multiple enterprises,
e.g., $A$, $B$, and $C$, is {\em replicated} on all involved enterprises is the same order.
$\langle [BC{:}1],[ABC{:}1,BCD{:}1] \rangle$ (on $d_{BC}$) has
$\gamma = [ABC{:}1, BCD{:}1]$ as the global part of its \texttt{ID}.
Note that $d_{BC}$ in addition to $d_{ABC}$ and $d_{BCD}$, is order-dependent on $d_{ABCD}$.
However, since the state of $d_{ABCD}$ is captured in the global part of transaction
$\langle [ABC{:}1],[ABCD{:}1] \rangle$ (and $ \langle [BCD{:}1], [ABCD{:}1] \rangle$)
and it has not been changed,
there is no need to add $ABCD{:}1$ to the global part of $\langle [BC{:}1],[ABC{:}1,BCD{:}1] \rangle$.
Dashed lines, e.g., between $\langle [B{:}1], \varnothing \rangle$ and $\langle [ABCD{:}1], \varnothing \rangle$,
are only used to show the order of appending entries to the ledger,
i.e., does not imply any data or ordering dependencies.
Note that while in this example, most transactions are cross-enterprise
(to help illustrate how a blockchain ledger is created),
in practical collaboration workflows, a significant percentage of transactions are internal.

\subsection{\sys Infrastructure}\label{sec:infra}

In \sys, nodes follow either the crash failure model or the Byzantine failure model.
To guarantee {\em fault tolerance}, clusters with crash-only nodes include $2f{+}1$ nodes.
In the presence of Byzantine nodes, $3f{+}1$ nodes are needed to provide fault tolerance \cite{bracha1985asynchronous}.
The malicious nodes, however, can violate data confidentiality by leaking 
requests, replies, and data stored and processed at the nodes.
To prevent confidential data leakage
(known as {\em intrusion tolerance} \cite{khan2021toward,vassantlal2022cobra}),
\sys can use either of the following two techniques:
(1) restricting the data that nodes can access \cite{padilha2011belisarius,vassantlal2022cobra,bessani2008depspace,marsh2004codex,bessani2013depsky}
using secret sharing schemes, or
(2) adding a privacy firewall between the ordering nodes and the execution nodes \cite{yin2003separating,duan2016practical}.

In the secret sharing scheme, clients encode data using an $(f + 1, n)$-threshold secret sharing scheme,
where $f+1$ shares out of $n$ total shares are needed to reconstruct the confidential data \cite{khan2021toward}.
Secret sharing schemes only perform basic store and retrieve operations
(Belisarius \cite{padilha2011belisarius} also supports addition) and 
do not support general transactions that require nodes to manipulate the contents of stored data.
As a result, this technique is not suitable for blockchain systems that are supposed to support complex transactions.

In the privacy firewall mechanism \cite{yin2003separating},
the infrastructure consists of $3f+1$ ordering nodes (where $f$ is the maximum number of malicious ordering nodes)
that run a BFT protocol to order client requests,
$2g+1$ execution nodes (where $g$ is the maximum number of malicious execution nodes) that maintain data and
deterministically execute arbitrary transactions following the order assigned by ordering nodes, and
a privacy firewall consisting of a set of $h+1$ rows of $h+1$ filters 
(where $h$ is the maximum number of malicious filter nodes)
between the ordering nodes and execution nodes.
The privacy firewall architecture assumes a network configuration that physically restricts communication
paths between ordering nodes, filters, and execution nodes, i.e.,
clients can only communicate with ordering (and not execution) nodes and
each filter has a physical network connection only to all (filter) nodes in the rows above and below.
As a result, a malicious node can {\em either} access confidential data (an execution node or a filter)
{\em or} communicate freely with clients (an ordering node) {\em but not both}.
The $h+1$ rows of $h+1$ filters guarantee that
first, there is at least one path between execution nodes and ordering nodes
which only consists of non-faulty filters (liveness) and
second, since there are $h+1$ rows and maximum $h$ malicious filters,
there exist a row consisting of only non-faulty filters
that filters any malicious message (possibly including confidential data).
As a result, the rows below have no access to any confidential data leaked by malicious execution nodes.
Note that if $h \leq 3f$, ordering nodes can be merged with the bottom row of filters
by placing a filter on each ordering node.
In most applications, \req and \reply bodies must also be encrypted, thus, ordering nodes cannot read them (while clients and execution nodes can).

By separating ordering nodes from execution nodes,
a simple majority of non-faulty nodes is sufficient to mask Byzantine failure among execution nodes, i.e.,
$2g+1$ execution nodes can tolerate $g$ Byzantine faults.
This is important because, compared to ordering, executing transactions, and maintaining
the application logic (e.g., smart contracts) and data
require more powerful computation, storage, and I/O resources.
The overhead of the privacy firewall mechanism can be reduced by adding $h$ filters to each row 
while providing a higher degree of confidentiality \cite{duan2016practical}.

Separating ordering nodes from execution nodes and using a privacy firewall comes with extra resource costs
because ordering nodes and execution nodes need to be physically separated.
However, if a cluster is deployed in a cloud platform,
ordering nodes and execution nodes can match the control layer and computing nodes, respectively.
Similarly, tools such as internal authorization
services, node auditors, and load balancers
deployed in existing cloud platforms can be used as privacy firewalls \cite{duan2016practical}.

\sys prevents confidential data leakage despite Byzantine failure using the privacy firewall mechanism presented in \cite{yin2003separating}.

\begin{figure}[t] \centering
\includegraphics[width=0.6\linewidth]{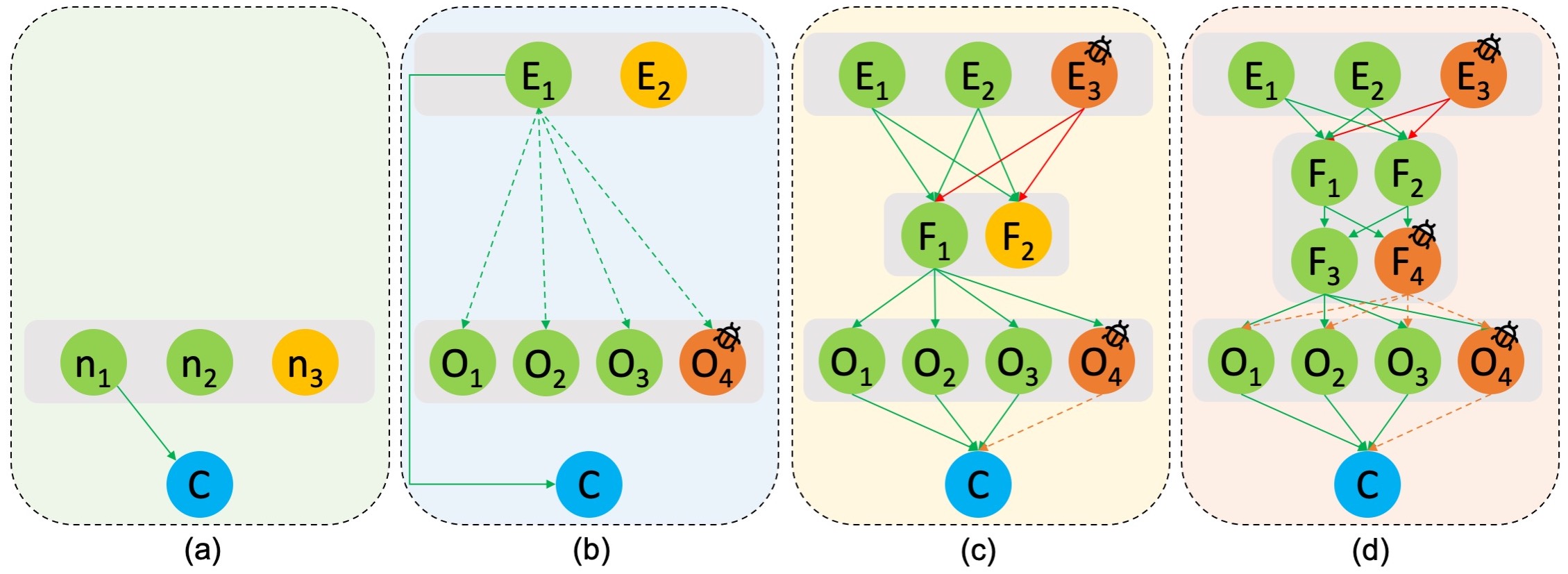}

[\greenp{3pt}: non-faulty,
\orangep{3pt}: malicious,
\yellowp{3pt}: crashed, and
\bluep{3pt}: clients]
\caption{
The flow of \reply messages in an instance of \sys with
(a) $2f+1$ crash-only nodes,
(b) $3f+1$ Byzantine ordering nodes and $g+1$ crash-only execution nodes,
(c) $3f+1$ Byzantine ordering and $2g+1$ Byzantine execution nodes, and a privacy firewall consisting of $h+1$ crash-only filter nodes, and
(d) $3f+1$ Byzantine ordering nodes, $2g+1$ Byzantine execution nodes and $h+1$ rows of $h+1$ Byzantine filter nodes.}
\label{fig:firewall}
\end{figure}

Figure~\ref{fig:firewall} presents a cluster in \sys and shows the flow of \reply messages in different settings
where different types of nodes follow different failure models.
When nodes are crash only, as shown in Figure~\ref{fig:firewall}(a),
a cluster consisting of $2f+1$ nodes can order {\em and} execute transaction while confidential data leakage is prevented.
If ordering nodes follow the Byzantine failure model, the ordering nodes and execution nodes
need to be separated to prevent data leakage.
However, if execution nodes are crash-only, as shown in Figure~\ref{fig:firewall}(b),
there is no need to add a privacy firewall and execution nodes can directly send the \reply to the client and inform
ordering nodes about execution. Since execution nodes are crash-only, $g+1$ nodes are sufficient to execute transactions.
When both ordering and execution nodes might behave maliciously,
a privacy firewall consisting of a set of filters is needed.
The firewall physically restricts communication from execution nodes, hence, 
a malicious execution node cannot send confidential data to
a malicious ordering node that might share the data with clients.
If filters are crash-only (Figure~\ref{fig:firewall}(c)), a row of $h+1$ filters is sufficient
to filter incorrect messages (data leakage) where $h$ is the maximum number of crashed filters.

The general case, as shown in Figure~\ref{fig:firewall}(d), is when
all ordering nodes, execution nodes and filters follow the Byzantine failure model
In this case, a privacy firewall with $h+1$ rows of $h+1$ filters per row is needed.
Communication is restricted so that each filter has a physical network
connection only to all (filter) nodes in the rows above and below.

\subsection{Confidentiality Guarantees}\label{sec:conf}
\sys provides collaboration confidentiality through
its hierarchical data model, which segregates data into ``data collections'' that are only maintained by authorized enterprises.
\sys also prevents confidential data leakage by 
separating ordering nodes from execution nodes and using a privacy firewall in between.
Confidentiality is mainly preserved based on three main rules.

1. Data collections are completely separated.
For example, $d_{AC}$ captures transactions that are executed in a shared collaboration
between $A$ and $C$. This is different from the data records of $d_A$ or $d_C$ (internal transactions of $A$ or $C$).
Moreover, each data collection is maintained {\em only} by its involved enterprises.
Similarly,
each transaction record in the blockchain ledger, as shown in Figure~\ref{fig:ledger},
is only maintained by the involved enterprises.

2. Transactions of a data collection $d_X$ can read transactions of another data collection $d_Y$ {\em if and only if}
$X \subseteq Y$.
For example, transactions of $d_{AB}$ can read records of $d_{ABC}$ because both $A$ and $B$ are involved in $d_{ABC}$.
However, transactions of $d_{ABC}$ can not read records of $d_{AB}$
because enterprise $C$ is not involved in $d_{AB}$.

3. A malicious node can {\em either} access confidential data (an execution node or a filter)
{\em or} communicate freely with clients (an ordering node) {\em but not both}.
This is guaranteed by separating ordering nodes from execution nodes and using a privacy firewall.

Note that a compromised client may leak its own state or updates.
\sys, similar to all other confidential fault-tolerant systems (that we are aware of) does not prevent this.
Moreover, an enterprise might share the data resulting from its cross-enterprise collaborations
with some external entity or even grant its access to the external entity.
\sys cannot restrict such external data leakage. 

\subsection{Multi-Shard Enterprises}\label{sec:multicluster} 

We now show how the data model and the blockchain ledger can be extended to support
multi-shard enterprises.
With single-shard enterprises, every execution node of each enterprise
maintains all data collections that the enterprise is involved in.
Enterprises, however, partition their data into different shards.
Each data shard is then assigned to a cluster of nodes where
ordering nodes of the cluster order the transactions and
execution nodes maintain the data shard and execute transactions on the data shard.
A shard of the enterprise data might not include a shard of every data collection that the enterprise is involved in.
For example, consider a private collaboration between the \textsf{logistic provider}
and the \textsf{pharmaceutical manufacturer} maintained on a shared data collection.
While the \textsf{logistic provider} might partition its data into multiple shards,
it maintains this shared data collection in only one of its data shards on a cluster
that is placed in Michigan (close to the Pfizer manufacturing site), e.g.,
a local collaboration.

We assume that enterprises use the same sharding schema for each shared data collection
to facilitate transaction processing across different enterprises.
The schema is agreed upon by all involved enterprises when a data collection is created, i.e.,
the sharding schema is part of the configuration metadata.
Using the same sharding schema leads to a more efficient ordering phase for cross-enterprise transactions 
because there is no need for all involved enterprises to run a consensus protocol to agree on the order of every transaction.
For each cross-enterprise transaction, one enterprise orders the transaction and other involved enterprises only validate the order
(details in Sections \ref{sec:centralpro} and \ref{sec:decentralpro}).
Furthermore, different data shards of an enterprise are processed by different clusters.
Using the same sharding schema, enterprises can easily communicate with the right cluster that processes the data records of transactions.
If enterprises use different sharding schemas for a shared data collection,
\sys could still process cross-enterprise transactions but with the overhead required to execute a consensus protocol with every transaction on each cluster.

The blockchain ledger of a single-shard enterprise maintains
all transactions that are executed on the enterprise data.
In a multi-shard enterprise, the enterprise data is partitioned into different shards where each shard
is replicated on a cluster of execution nodes.
Since each cluster maintains a separate data shard, it executes a different set of transactions.
As a result, each cluster in a multi-shard enterprise needs to maintain a different ledger.
The ledger of each cluster of a multi-shard enterprise, however, is constructed in the same way as a single-shard enterprise.
Moreover, the notion of global consistency is extended to guarantee that
each cross-shard transaction is ordered across participating shards consistently.
\section{Transaction Processing in \sys}
\label{sec:protocol}

Processing transactions requires establishing consensus on a unique order of requests.
Fault-tolerant protocols use the State Machine Replication (SMR) technique \cite{lamport1978time,schneider1990implementing}
to assign each client request an order in the global service history.
In an SMR fault-tolerant protocol, all non-faulty nodes execute the same requests in the same order ({\em safety}) and
all correct client requests are eventually executed ({\em liveness}).
\sys guarantees safety in an asynchronous network, however,
it considers a synchrony assumption to ensure liveness (due to FLP results \cite{fischer1985impossibility}).

\begin{table}
\caption{Processing transactions in \sys}
\label{tbl:tnx}
\small
\centering
\setlength{\tabcolsep}{0.2em}
\begin{tabular}{|cc|c|cc|}
\hline
\multicolumn{2}{|c|}{Transaction Type}   & Example   & \multicolumn{2}{|c|}{Consensus Protocol}  \\
Shard & Enterprise& Clusters- Shards     & Coordinator-based & Flattened  \\
\hline
intra   & intra  & $A_1$ - $d^1_A$                                   & Pluggable  & Pluggable    \\
intra   & cross     & $C_3$, $D_3$ - $d^3_{CD}$                      & Fig.~\ref{fig:centralized}(a) & Fig.~\ref{fig:decentralized}(a)\\
cross   & intra & $A_2$, $A_3$ - $d^2_A$, $d^3_A$                   & Fig.~\ref{fig:centralized}(b)& Fig.~\ref{fig:decentralized}(b)\\
cross   & cross     & $B_1${,}$C_1${,}$B_2$,$C_2${-} $d^1_{BC}${,}$d^2_{BC}$   & Fig.~\ref{fig:centralized}(c) & Fig.~\ref{fig:decentralized}(c)\\
\hline
\end{tabular}
\end{table}

A transaction might be executed on a single shard or on multiple shards of a data collection
where the data collection belongs to a single enterprise (i.e., a local data collection) or
is shared among multiple enterprises (i.e., a root or an intermediate data collection).
As a result, as shown in Table~\ref{tbl:tnx},
\sys needs to support four different types of transactions.
In Table~\ref{tbl:tnx},  the network consists of four enterprises
where each enterprise $\psi {\in} \{A, B, C,D\}$ owns three clusters of nodes, e.g., $\psi_1$, $\psi_2$ and $\psi_3$
and processes shard $d^i_{\psi}$ by nodes of cluster $\psi_i$.

An {\em intra-shard} transaction is executed on the same shard of a data collection
either {\em intra-enterprise}, e.g.,
on shard $d^1_A$ of a local data collection $d_A$ by cluster $A_1$, or
{\em cross-enterprise}, e.g.,
on shard $d^3_{CD}$ of a shared data collection $d_{CD}$ by clusters $C_3$ and $D_3$.
A {\em cross-shard transaction} is executed on multiple shards of a data collection
either {\em intra-enterprise}, e.g.,
on shards $d^2_A$ and $d^3_A$ of a data collection $d_A$ by clusters $A_2$, $A_3$, or 
{\em cross-enterprise}, e.g.,
on shards $d^1_{BC}$ and $d^2_{BC}$ of a shared data collection $d_{BC}$ by clusters $B_1$, $C_1$, $B_2$, and $C_2$.

A transaction can be executed on multiple shards of the same data collection (i.e., a cross-shard transaction) or
the same data shard that is maintained by multiple enterprises (i.e., a cross-enterprise transaction).
However, as mentioned earlier, a transaction can not be executed or write data records on multiple data collections for two main reasons.
First, the data of each (non-root) data collection is confidentially shared with only its involved enterprises.
For example, transactions of $d_{AB}$ should not write data on $d_{AC}$ as $C$ is not involved in $d_{AB}$.
Note that transactions among enterprises $A$, $B$, and $C$ are executed on $d_{ABC}$.
Moreover, there is no need to write the transaction results of $d_{AB}$ on $d_A$ or $d_B$
as both enterprises $A$ and $B$ store $d_{AB}$ (see Figure~\ref{fig:data}(b))
and can read its data record in their local data collections $d_A$ and $d_B$.
Second, each data collection might have its own business logic to execute transactions.

In \sys, cross-shard intra-enterprise transactions and intra-shard cross-enterprise transactions are handled differently.
In a cross-shard intra-enterprise transaction, different clusters maintain separate data shards.
As a result, each cluster needs to separately reach agreement on the transaction order among transactions of its shard.
However, in an intra-shard cross-enterprise transaction, 
since the involved enterprises use the same sharding schema for each shared data collection,
the cluster that initiates the transaction and
all the involved clusters process the transaction on the same shard of data.
Hence, the order of transactions on different clusters is the same.
As a result, it is sufficient that one (initiator) cluster determines the order and other clusters only validate the suggested order.
Note that while the involved clusters belong to different distrustful enterprises,
all clusters can detect any malicious behavior of the initiator cluster.
We will discuss how \sys addresses all possible malicious behaviors, e.g., assigning invalid \texttt{ID} (order)
or even not sending the transaction to other clusters.
In cross-enterprise collaboration, enterprises cannot trust that the other enterprise nodes might not be compromised.
As a result, independent of the declared failure model of nodes,
\sys uses a BFT protocol to order cross-enterprise transactions.
Furthermore, as it is suggested in Caper \cite{amiri2019caper},
we can assume that less than a {\em one-third} of enterprises might be malicious.
Hence, by a similar argument as in PBFT \cite{castro1999practical},
with agreement from at least {\em two-thirds} of the enterprises,
a cross-enterprise transaction can be committed.
For cross-shard intra-enterprise transactions, on the other hand,
the involved clusters belong to the same enterprise and trust each other.
As a result, if all nodes are crash-only, cross-shard consensus can be achieved using a crash fault-tolerant protocol.

To process transactions across clusters {\em coordinator-based} and {\em flattened}
approaches are used.
In the coordinator-based approach, inspired by existing permissioned blockchains such as
AHL \cite{dang2018towards}, Saguaro \cite{amiri2021saguaro}, and Blockplane \cite{nawab2019blockplane},
a cluster coordinates transaction ordering, whereas,
in the flattened approach, transactions are ordered across the clusters without requiring a coordinator.

Nodes in \sys follow different failure models, i.e., crash or Byzantine.
We mainly focus on two common cases.
First, when nodes follow the crash failure, a cluster contains $2f+1$ nodes that perform both ordering and execution.
Second, when nodes follow the Byzantine failure and each cluster includes $3f+1$ ordering nodes, $2g+1$ execution nodes,
and a privacy firewall consisting of $h+1$ rows of $h+1$ filters.
The number of required matching votes to ensure that
a quorum of ordering nodes within a cluster agrees with the order of a transaction is different in different settings.
We define {\em local-majority} as the number of required matching votes from a cluster.
For crash-only clusters, 
local-majority is $f+1$ (from $2f+1$ nodes), and
for clusters with Byzantine ordering nodes,
local-majority is $2f+1$ (from $3f+1$ ordering nodes).

\subsection{Intra-Cluster Consensus}

In the internal (intra-cluster) consensus protocol,
ordering nodes of a cluster, {\em independently} of other clusters,
agree on the order of a transaction.
The internal consensus protocol is pluggable
and depending on the failure model of nodes of the cluster a crash fault-tolerant (CFT) protocol, e.g., (Multi-)Paxos \cite{lamport2001paxos}
or a Byzantine fault-tolerant (BFT) protocol, e.g., PBFT \cite{castro1999practical}, can be used.
If nodes follow the Byzantine failure model, as discussed before,
\sys separates ordering nodes from execution nodes and
uses a privacy firewall to prevent confidential data leakage.

The protocol
is initiated by a pre-elected ordering node of the cluster, called {\em the primary}.
When the primary node $\pi(P_i)$ of cluster $P_i$ receives a valid signed \req message
$\langle\text{\scriptsize \REQ}, op, t_c, c\rangle_{\sigma_c}$
from an authorized client $c$ (with timestamp $t_c$) to execute (encrypted) operation $op$ on
local data collection $d_{P_i}$,
it initiates the protocol
by multicasting a message, e.g.,
{\sf \small accept} message in Multi-Paxos (assuming the primary is elected) or {\sf \small pre-prepare} message in PBFT,
including the req and its digest to other ordering nodes of the cluster.

To provide a total order among transaction blocks and preserve data consistency,
the primary also assigns an \texttt{ID}, as discussed in Section~\ref{sec:ledger}, to the request.
The \texttt{ID} consists of a local part, e.g., $[P_i{:}n]$, and a global part
including the state of all data collections that $d_{P_i}$ is order-dependent on.
the global part includes the state of {\em all} data collections that $d_{P_i}$ is order-dependent on because
the read-set of the transaction is not known (will be known during the execution time).
The primary includes the current state (i.e., the local sequence number of the last committed transaction)
of such data collections in the global part.
The ordering nodes of the cluster then agree on the order of
the transaction using the utilized protocol.

\subsection{Transaction Execution Routine}\label{sec:execution}

The next step after ordering is to execute a transaction and inform the client.
If nodes follow the Byzantine failure model, ordering and execution are performed by distinct sets of nodes
that are separated by a privacy firewall.
Once a transaction is ordered, the ordering nodes generate a \three certificate 
consisting of signatures from  $2f+1$ different nodes and multicast
both the request and the \three certificate to the bottom row of filters.
The filters check the request and the \three certificate to be valid and multicast them to the next row above.
Filters also track the requests that they have seen in a \three or a \reply certificate and their \texttt{ID}s.

Upon receiving both a valid request and a valid \three certificate, execution
nodes append the transaction and the corresponding \three certificate
to the ledger.
Note that the execution nodes store the last \reply certificate sent to client $c$ to check if the current request
is new (i.e., has a larger timestamp) that needs to be executed and committed to the ledger
or an old request (i.e., has the same or smaller timestamp) where nodes re-send the \reply messages.
If all transactions accessing on $d_{P_i}$ with lower local sequence numbers ($< n$) have been executed,
nodes execute the transaction on data collection $d_{P_i}$ following its application logic.
If, during the execution, the execution nodes need to read values from some other data collection $d_{X}$
where $d_{P_i}$ is order-dependent on $d_X$ (i.e., $P_i \subseteq X$),
the nodes check the public part of the transaction \texttt{ID} to ensure that they read the correct state of $d_{X}$.
Data collections store data in multi-versioned datastores to enable nodes to read the version they need to.
This ensures that all execution nodes execute requests in the same order and on the same state.

The \three certificate are appended to the ledger to guarantee immutability.
Since \three messages include
the digest of transaction blocks, by appending them to the ledger,
any attempt to alter the block data can easily be detected.

Execution nodes then multicast a signed \reply message including the (encrypted) results to the top row of filters.
When a filter node in the top row receives $g+1$ valid matching \reply messages, it generates a \reply certificate
authenticated by $g+1$ signature shares from distinct execution nodes
and multicasts it to the row below.
Each filter then multicasts the \reply certificate to the row of filter nodes or ordering nodes below.
Finally, client $c$ accepts the result once it receives a valid \reply certificate from ordering nodes.

If nodes follow the crash failure model, ordering and execution are performed by the same set of nodes.
In this case, nodes append the transaction to the ledger and execute it as explained before and
then the primary node sends a \reply message to client $c$.

\begin{figure}[t] \center
\includegraphics[width= 0.7\linewidth]{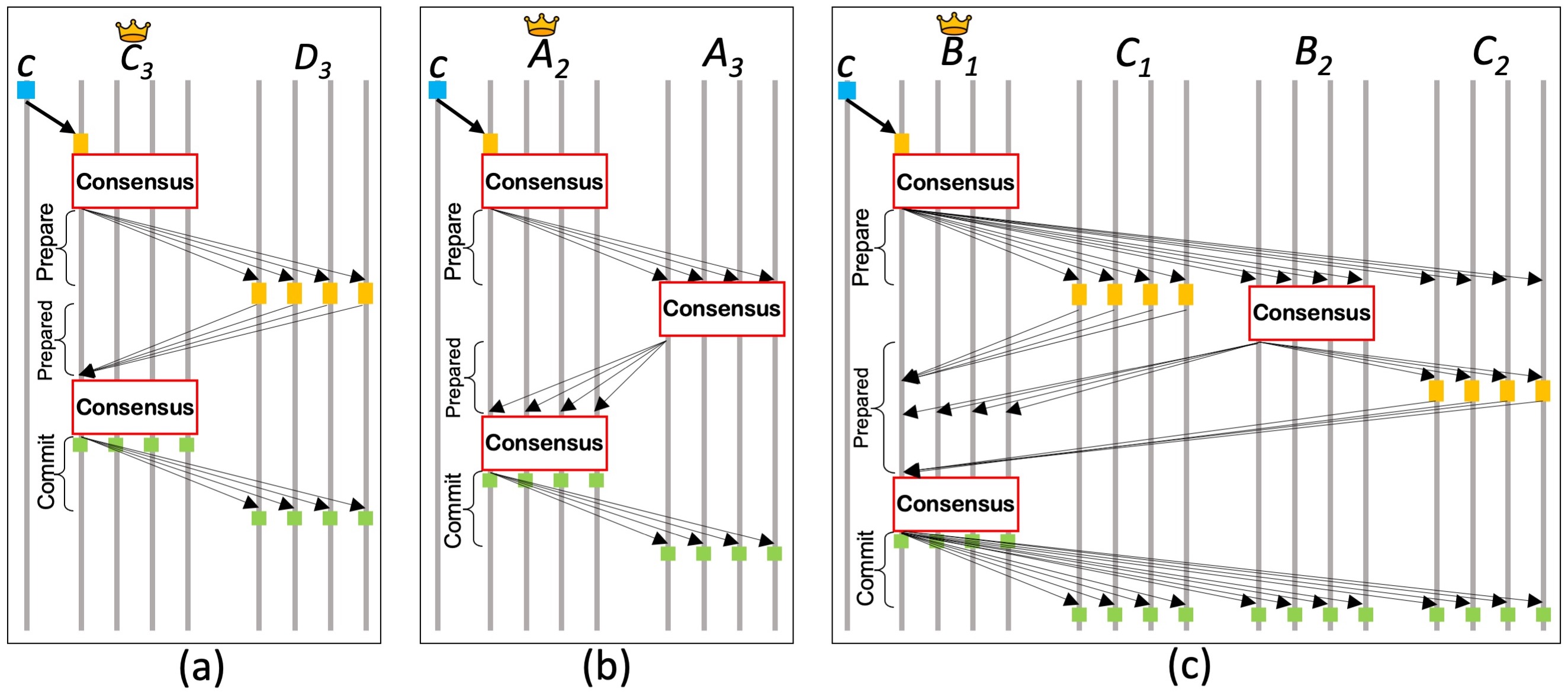}
\caption{(a) intra-shard cross-enterprise, (b) cross-shard intra-enterprise, and (c) cross-shard cross-enterprise
coordinator-based consensus protocols}
\label{fig:centralized}
\end{figure}

\subsection{Coordinator-based Consensus}\label{sec:centralpro}

The coordinator-based approach has been used in distributed databases
to process cross-shard transactions, e.g., two-phase commit.
Coordinator-based protocols have also been used in several permissioned blockchain systems, e.g., 
AHL \cite{dang2018towards}, Blockplane \cite{nawab2019blockplane}, and Saguaro \cite{amiri2021saguaro}.
This section briefly demonstrates how \sys can leverage the coordinator-based approach
to process cross-cluster transactions.
In contrast to the existing coordinator-based databases and blockchains
that address multi-shard single-enterprise contexts, 
\sys deals with multi-shard multi-enterprise environments.
Moreover, since each transaction might read data from multiple data collections,
ordering transactions is \sys is more complex.
All three coordinator-based protocols
(i.e., intra-shard cross-enterprise, cross-shard intra-enterprise, and cross-shard cross-enterprise),
as shown in Figure~\ref{fig:centralized}, consist of \pre, \pred and \three phases.
We briefly explain the normal case operation of the protocols.




\noindent \fcolorbox{black}{boxcolor}{\begin{minipage}{0.97\textwidth}
\noindent \textbf{Prepare:} Upon receiving a transaction, ordering nodes of the {\em coordinator} cluster establish (internal) consensus
on the order of the transaction among the transactions of their shard.
The primary then sends a signed \pre message to the ordering nodes of all involved clusters.
\end{minipage}}

\noindent \fcolorbox{black}{boxcolor}{\begin{minipage}{0.97\textwidth}
\noindent \textbf{Prepared:} This phase is handled differently in different protocols.

\uline{Intra-shard cross-enterprise transactions:}
All involved clusters, e.g., $C_3$ and $D_3$ in Figure~\ref{fig:centralized}(a), maintain the same data shard.
As a result, there is no need to run consensus in non-coordinator clusters, e.g., $D_3$.
Upon receiving a \pre message from the primary of the coordinator cluster, e.g., $C_3$,
each ordering node of a non-coordinator cluster, e.g., $D_3$, validates the order and
sends a \pred message to the primary of the coordinator cluster.

\uline{Cross-shard intra-enterprise transactions:}
The involved clusters, e.g., $A_2$ and $A_3$, maintain different shards of a local data collection, e.g., $d_A$.
Upon receiving a \pre message from the coordinator cluster, e.g., $A_2$,
each non-coordinator cluster separately establishes (internal) consensus on the order of the transaction in its data shard.
The primary of each non-coordinator cluster, e.g., $A_3$, then sends a \pred message to the ordering nodes of the coordinator cluster, e.g., $A_2$.

\uline{Cross-shard cross-enterprise transactions:}
The involved clusters of the initiator enterprise, e.g., $B_1$ and $B_2$ of enterprise $B$,
maintain different shards of a shared data collection, e.g., $d_{BC}$,
while each shard is replicated on
clusters of different enterprises, e.g., $d^1_{BC}$ is maintained by $B_1$ and $C_1$.
Upon receiving a \pre message from the coordinator cluster, e.g., $B_1$,
if the cluster belongs to the initiator enterprise, e.g., $B_2$,
it establishes (internal) consensus on the transaction order.
Otherwise, the cluster waits for a message from the cluster of the initiator enterprise that maintains the same data shard as they do, e.g.,
$C_1$ waits for $B_1$ and $C_2$ waits for $B_2$.
Each node then validates the received message and
sends a \pred message to the primary of the coordinator cluster.
\end{minipage}}

\noindent \fcolorbox{black}{boxcolor}{\begin{minipage}{0.97\textwidth}
\noindent \textbf{Commit:}
When the primary of the coordinator cluster receives valid \pred messages from every involved cluster,
it establishes internal consensus within the coordinator cluster and multicasts a signed \three message
to ordering node of all involved clusters.
\end{minipage}}


\subsubsection{Intra-Shard Cross-Enterprise Consensus}

In the intra-shard cross-enterprise protocol,
multiple clusters (one from each enterprise) process
a transaction on the same shard of a non-local data collection shared between enterprises, e.g.,
clusters $C_3$ and $D_3$ on shard $d^3_{CD}$ of data collection $d_{CD}$ (Figure~\ref{fig:centralized}(a)).

Upon receiving a valid transaction $m$,
the primary of the coordinator cluster $P_c$, i.e., the cluster that receives the transaction,
assigns an \texttt{ID} to the transaction (as discussed in Section~\ref{sec:ledger})
and initiates internal consensus on the order of $m$ in the coordinator cluster.

Once consensus is achieved, the primary
sends a signed \pre message $\langle\text{\scriptsize \PRE}, \textsf{\scriptsize ID}, d, m \rangle_{\sigma_{P_c}}$ 
to the nodes of all involved clusters where $d=D(m)$ is the digest of request $m$.
The \pre messages are signed by local-majority of the clusters (denoted by $\sigma_{P_c}$).
Upon receiving a \pre message with valid \texttt{ID}
(following the local and global consistency rules discussed in Section~\ref{sec:ledger}), digest, and signature,
each node $r$ sends a signed \pred message
$\langle\text{\scriptsize \PRED}, \textsf{\scriptsize ID}, d, r \rangle_{\sigma_r}$
to the primary node of the coordinator cluster.

Upon receiving valid \pred messages from the local-majority of every involved cluster,
the primary establishes internal consensus within the coordinator cluster
and sends a \three message signed by local-majority of the cluster to every node of all involved clusters.
The \three message includes the transaction \texttt{ID}, its digest,
and also \pred messages received from the local-majority of every involved cluster
as evidence of message validity.
When a node receives a valid \three message, the node
appends the transaction to the ledger and executes the transaction.

If a node of an involved cluster waits for a message from another cluster, e.g.,
a node in $D_3$ does not receive a \three message from the coordinator cluster $C_3$,
it sends a query message to all nodes of that cluster resulting in either receiving the message or replacing the faulty primary.
The routine is explained in Section~\ref{sec:cent-fail}.

\subsubsection{Cross-Shard Intra-Enterprise Consensus}\label{sec:cross-intra}

The cross-shard intra-enterprise consensus protocol,
inspired by permissioned blockchains
AHL \cite{dang2018towards} and Saguaro \cite{amiri2021saguaro},
processes transactions on
different shards of a local data collection within an enterprise,
e.g., a transaction on data shards $d^2_A$ and $d^3_A$ of data collection $d_A$ processed by clusters $A_2$ and $A_3$ (Figure~\ref{fig:centralized}(b)).

After receiving transaction $m$,
the primary of the coordinator cluster $P_c$ assigns an \texttt{ID} to the transaction
and establishes internal consensus on the order of $m$
in the coordinator cluster.
$P_c$ then multicasts a signed \pre message
$\langle\text{\scriptsize \PRE}, \textsf{\scriptsize ID}_c, d, m \rangle_{\sigma_{P_c}}$
to the nodes of all involved clusters.
Upon receiving a valid \pre message,
the primary of each involved cluster $P_i$ assigns $\texttt{ID}_i$ to $m$ and establishes
consensus on request $m$ in cluster $P_i$.
$\texttt{ID}_c$ and $\texttt{ID}_i$ represent the order of transaction $m$ in clusters $P_c$ and $P_i$ respectively.
Note that the primary of the coordinator and all involved clusters processes the request only if
there is no concurrent (uncommitted) request $m'$
where $m$ and $m'$ intersect in at least $2$ shards.
This is needed to ensure that requests are committed in different shards in the same order (global consistency).

The primary of each involved cluster $P_i$ then
multicasts a signed \pred message
$\langle\text{\scriptsize \PRED}, \textsf{\scriptsize ID}_c, \textsf{\scriptsize ID}_i, d \rangle_{\sigma_{P_i}}$
to all nodes of the coordinator cluster $P_c$.
Otherwise (if the consensus is not achieved), it multicasts a signed \abort message.
As before, each message is signed by a local-majority of a cluster.
When the primary node of the coordinator cluster receives valid \pred messages from every involved cluster, e.g., $P_i$, $P_j$, ...,
it initiates internal consensus within the coordinator cluster and multicasts a signed \three message
$\langle\text{\scriptsize \THREE}, \textsf{\scriptsize ID}_c, \textsf{\scriptsize ID}_i,
\textsf{\scriptsize ID}_j, ..., d \rangle_{\sigma_{P_c}}$
to every node of all involved clusters.

Otherwise (if some cluster has aborted the transaction), the primary multicasts a signed \abort message.
The \texttt{ID} of the \three messages is a concatenation of the received \texttt{ID}s from
all involved clusters.
Since the transaction takes place within an enterprise,
in contrast to the previous protocol,
the primary of the coordinator cluster does not need to include the received \pred messages in its \three message.
Upon receiving a valid \three message, each node appends the transaction to the ledger and executes it.


\subsubsection{Cross-shard Cross-Enterprise Consensus}\label{sec:cross-cross}

The cross-shard cross-enterprise consensus protocol is needed when
a transaction is executed on different shards of a non-local data collection
shared between different enterprises, e.g.,
a transaction that is executed on data shards $d^1_{BC}$ and $d^2_{BC}$ of shared data collection $d_{BC}$
where $d^1_{BC}$ is maintained by $B_1$ and $C_1$ and $d^2_{BC}$ is maintained by $B_2$, and $C_2$ (Figure~\ref{fig:centralized}(c)).

In this case, different clusters of each enterprise maintain different data shards, e.g., $B_1$ and $B_2$ maintain $d^1_{BC}$ and $d^2_{BC}$, however,
each data shard is replicated on only one cluster of each involved enterprise, e.g., $d^1_{BC}$ is replicated on $B_1$ and $C_1$.
In \sys, in order to improve performance, only the clusters of the initiator enterprise, e.g., $B_1$ and $B_2$,
establish consensus on the transaction order and
clusters of other enterprises, e.g., $C_1$ and $C_2$, validate the order.

Upon receiving a valid transaction $m$,
the primary node of the coordinator cluster $P_c$ establishes internal consensus on the order of $m$
in the coordinator cluster and then
multicasts a signed \pre message
$\langle\text{\scriptsize \PRE}, \textsf{\scriptsize ID}_c, d, m \rangle_{\sigma_{P_c}}$
to the nodes of all involved clusters.
When the primary node of an involved cluster $P_i$ of the {\em initiator enterprise} receives a valid \pre message,
it assigns $\texttt{ID}_i$ to $m$ and initiates internal
consensus on request $m$ in cluster $P_i$.
Same as cross-shard intra-enterprise consensus, the primary of $P_c$ and every $P_i$ processes the request only if
there is no concurrent request $m'$ with shared data shards.
Once consensus is achieved,
the primary of the involved cluster $P_i$
multicasts a signed \pred message
$\langle\text{\scriptsize \PRED}, \textsf{\scriptsize ID}_c, \textsf{\scriptsize ID}_i, d \rangle_{\sigma_{P_i}}$
to every node of $P_c$ and also any other clusters that maintain {\em the same} data shard as cluster $P_i$.
Otherwise, it multicasts an \abort message to those clusters.

A node in a cluster of a non-initiator enterprise either
(1) maintains the same shard as the coordinator cluster and receives
a valid \pre message (with $\texttt{ID}_c$) from primary of coordinator cluster $P_c$ , or
(2) maintains some other shard and receives
a valid \pred message (with $\texttt{ID}_i$) from primary of cluster $P_i$.
In either case, the node sends a signed \pred message including $\texttt{ID}_c$ (and $\texttt{ID}_i$) and message digest
to the primary of the coordinator cluster.

The primary node of the coordinator cluster waits for
(1) signed valid \pred messages from the primary node of all involved clusters of its enterprise, and
(2) signed valid \pred messages from a local-majority of every involved cluster of other enterprises.
The primary then
initiates internal consensus within the coordinator cluster and multicasts a signed \three (or \abort) message
to every node of all involved clusters.

\subsubsection{Primary Failure Handling}\label{sec:cent-fail}

We use the retransmission routine presented in \cite{yin2003separating} to handle unreliable communication
between ordering nodes and execution nodes. In this section, we focus on the failure of the primary ordering node.
If the timer of node $r$ expires before it receives a \three certificate, it suspects that the primary might be faulty.

The timeout mechanism prevents the protocol from
blocking and waiting forever by ensuring that eventually (after GST),
communication between non-faulty ordering nodes is timely and reliable.
Ordering nodes use different timers for intra-cluster and cross-cluster transactions because
processing transactions across clusters usually takes more time.
The timeout value mainly depends on the network latency  (within or across clusters).
It must be long enough to allow nodes to communicate with each other and establish consensus, i.e.,
at least 3 times the WAN round-trip for cross-cluster transactions
to allow a view-change routine to complete without replacing the primary node again.
The timeouts are adjusted to ensure that this period increases exponentially
until some transaction is committed, e.g., in case of consecutive primary failure,
the timeout is doubled as in PBFT \cite{castro1999practical}.

When the primary node of a cluster fails, the primary failure handling routine of
the internal consensus protocol, e.g., view change in PBFT, is triggered by timeouts
to elect a new primary.
For transactions across clusters,
if a node of an involved cluster does not receive
a \three message from the primary of the coordinator cluster for a prepared request and its timer expires,
the node multicasts a signed \cmtq message
to all nodes of the coordinator cluster including the request \texttt{ID} and its digest.
As a result, if the (malicious) primary of the coordinator cluster maliciously has not sent \three messages to other clusters
(while the transaction is committed in the coordinator cluster),
other nodes of the coordinator cluster will be informed (to prevent any inconsistency between clusters).
Similarly, if a node in the coordinator cluster does not receive \pred messages from an involved cluster
(either from the primary node when it requires to run consensus or from a local-majority of nodes),
it multicasts a \predq to all nodes of the involved cluster.
It might also happen between two involved clusters in cross-shard cross-enterprise transactions
where a cluster is waiting for consensus results, e.g.,
$C_2$ and $B_2$ in Figure~\ref{fig:centralized}(c).

In all such cases, if the message has already been processed, the nodes simply re-send the corresponding response.
Nodes also log the query messages to detect denial-of-service attacks initiated by malicious nodes.
If the query message is received from a local-majority of a cluster,
the primary will be suspected to be faulty, triggering the execution of the failure handling routine.
Moreover, since all messages from a primary of a cluster, e.g., \pre, \pred, or \three,
are multicasts to every node of the receiver cluster,
if the primary of the receiver cluster does not initiate consensus on the message among the nodes of its cluster,
it will eventually be suspected to be faulty by the nodes.
Finally, if a client does not receive a \reply soon enough, it multicasts the request
to all nodes of the cluster that it has already sent its request.
If the request has already been processed,
the nodes simply send the execution result back to the client.
Otherwise, if the node is not the primary, it relays the request to the primary.
If the nodes do not receive \pre messages, the primary will be suspected to be faulty.

\subsubsection{Correctness Argument}
We briefly analyze the safety (agreement, validity and consistency) and
liveness properties of the coordinator-based protocols.

\begin{prop} (\textit{Agreement})
If node $r$ commits request $m$ with local \texttt{ID} $\alpha$ in cluster $P$,
no other non-faulty node commits request $m'$ ($m \neq m'$) with the same local \texttt{ID} $\alpha$ in $P$.
\end{prop}

\begin{prf}
We assume that the internal consensus protocols, e.g., Paxos and PBFT,
are correct and ensure safety.
Let $m$ and $m'$ ($m \neq m'$) be two committed requests where
$\texttt{ID}(m) = [\texttt{ID}_i,\texttt{ID}_j,{\color{red}{\texttt{ID}_k}},...]$ and
$\texttt{ID}(m') = [{\color{red}{\texttt{ID}'_k}},\texttt{ID}'_l,\texttt{ID}'_m,..]$ respectively.
Note that in intra-shard transactions, the \texttt{ID} includes a single part.
Given an involved cluster {\color{red}{$P_k$}} in the intersection of $m$ and $m'$ where
$\texttt{ID}_k(m) = \langle \alpha, \gamma \rangle$ and
$\texttt{ID}_k(m') = \langle \alpha', \gamma' \rangle$.
To commit $m$ and $m'$, a local-majority $Q$ agreed with $m$ and local-majority $Q'$ agreed with $m'$
(If nodes of $P_k$ are crash only, local-majority is $f+1$ out of $2f+1$ and
if nodes might be Byzantine, local-majority is $2f+1$ out of $3f+1$).
Since any two local-majority $Q$ and $Q'$ intersect on at least one non-faulty node and
non-faulty nodes do not behave maliciously,
if $m \neq m'$ then $\alpha \neq \alpha'$, hence agreement is guaranteed.
\end{prf}

\begin{prop} (\textit{Validity})
If a non-faulty node $r$ commits $m$, then $m$ must have been proposed by some node $\pi$.
\end{prop}

\begin{prf}
For crash-only clusters validity is guaranteed as crash-only nodes do not send fictitious messages.
Since the adversary cannot subvert standard cryptographic assumptions (as explained in Section~\ref{sec:model}),
validity is guaranteed for Byzantine clusters as well.
All messages are signed and  include
the request or its digest (to prevent changes and alterations to any part of the message).
As a result, if request $m$ is committed by non-faulty node $r$,
the same request must have been proposed earlier by some node $\pi$.
\end{prf}

\begin{prop} (\textit{Consistency})
Let $\mathcal{P}(m)$ denote the set of involved clusters in a request $m$.
For any two committed requests $m$ and $m'$ and any two nodes $r_1$ and $r_2$
such that $r_1 \in P_i$, $r_2 \in P_j$, and $\{P_i,P_j\} \in \mathcal{P}(m) \cap \mathcal{P}(m')$,
if $m$ is committed after $m'$ on $r_1$, then $m$ is committed after $m'$ on $r_2$.
\end{prop}

\begin{prf}
For intra-shard cross-enterprise transactions, consistency is guaranteed as 
all clusters in $\mathcal{P}(m)$ maintain the same data shard.
For cross-shard transactions, as discussed in Sections~\ref{sec:cross-intra} and \ref{sec:cross-cross},
the primary of the coordinator as well as all involved clusters processes the request only if
there is no concurrent (uncommitted) request $m'$ where
$m$ and $m'$ intersect in at least $2$ shards.
Committing request $m$ requires sufficient \pred messages from {\em every} involved cluster, hence,
$m$ cannot be committed in any cluster until $m'$ is committed.
\end{prf}

\begin{prop}(\textit{Liveness})
A request $m$ issued by a correct client eventually completes.
\end{prop}

\begin{prf}
\sys guarantees liveness {\em only} during periods of synchrony (FLP result \cite{fischer1985impossibility}).
If the primary of the coordinator cluster is faulty, e.g.,
does not multicast valid \pre or \three messages, or
the primary of an involved cluster that needs to run consensus is faulty and does not
multicast valid \pred messages, as explained in Section~\ref{sec:cent-fail},
its failure will be detected using timeouts.
As explained, nodes use different timers for intra-cluster and cross-cluster transactions where
the timer for cross-cluster transactions is 
long enough to allow nodes to communicate with each other and establish consensus, i.e.
at least 3 times the WAN round-trip for cross-cluster transactions
to allow a view-change routine to complete without replacing the primary node again.
If a newly elected leader is not able to change the view, the timeout is doubled (as in PBFT \cite{castro1999practical}).
This adjustment is needed to ensure that some transactions will be committed, i.e., nodes are synchronized.
\sys uses the primary failure handling routine of the internal consensus protocol, e.g., view-change in PBFT, to elect a new primary.

\begin{figure}[t] \center
\includegraphics[width= 0.7\linewidth]{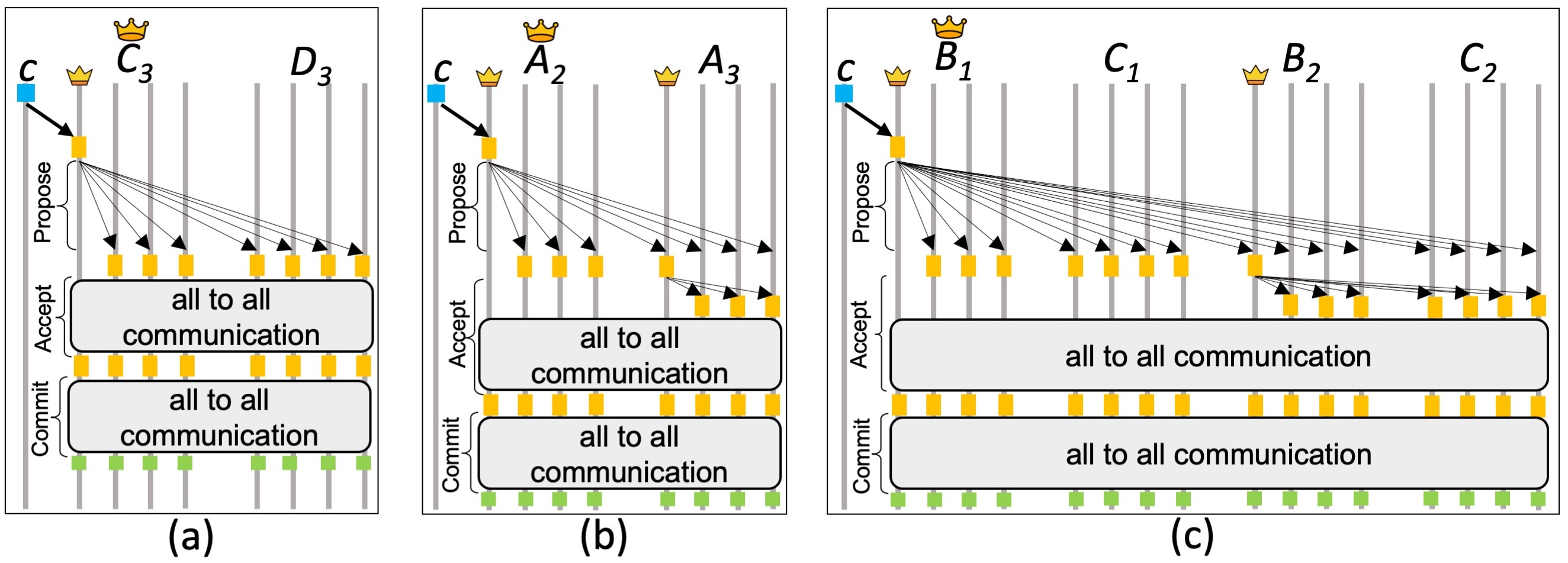}
\caption{(a) intra-shard cross-enterprise,
(b) cross-shard intra-enterprise, and
(c) cross-shard cross-enterprise flattened consensus protocols}
\label{fig:decentralized}
\end{figure}

Two coordinator clusters might initiate concurrent transactions
on the same data shard of the same data collection with inconsistent \texttt{ID}s, e.g., both have the same local \texttt{ID},
and the involved clusters receive these transactions in a different order.
In this case, neither of the concurrent transactions receive \pred messages from all involved clusters
resulting in a deadlock situation.
In such a situation, once the timer of a coordinator cluster for its transaction expires,
the coordinator cluster sends a new \pre message to the involved clusters to resolve the deadlock.
\sys assigns different timers to different clusters to prevent parallel and consecutive deadlock situations.
To prevent deadlock situations, \sys can consider a designated coordinator cluster for each (shard of a) data collection.
An enterprise, however, might not want to consume its resources to
coordinate transactions that are initiated by other enterprises.
\end{prf}

\subsection{Flattened Consensus}\label{sec:decentralpro}

The flattened consensus protocols of \sys,  in contrast 
to the flattened cross-shard consensus protocols, e.g., SharPer \cite{amiri2021sharper},
which are designed for single-enterprise environments,
support multiple multi-shard collaborative enterprises.

\noindent \fcolorbox{black}{boxcolor}{\begin{minipage}{0.97\textwidth}
\noindent \textbf{Propose:} Upon receiving a transaction,
the primary of the coordinator cluster sends a signed \one message to the nodes of all involved clusters.
\end{minipage}}

\noindent \fcolorbox{black}{boxcolor}{\begin{minipage}{0.97\textwidth}
\noindent \textbf{Accept:} This phase is handled differently in different protocols.

\uline{Intra-shard cross-enterprise transactions:}
All involved clusters, e.g., $C_3$ and $D_3$ in Figure~\ref{fig:decentralized}(a), maintain the same data shard.
Upon receiving a \one message from the primary of the coordinator cluster, each node
(in the coordinator, e.g., $C_3$ or another involved cluster, e.g., $D_3$) validates the message and its \texttt{ID} and
multicasts an \two message to all  nodes of every involved cluster.

\uline{Cross-shard intra-enterprise transactions:}
The involved clusters, e.g., $A_2$ and $A_3$, maintain different shards of a local data collection, e.g., $d_A$.
Upon receiving a \one message including $\texttt{ID}_i$ from the coordinator cluster, e.g., $A_2$,
the primary of each non-coordinator cluster, e.g., $A_3$,
assigns an $\texttt{ID}_j$
and sends an \two message (including both \texttt{ID}s) to the nodes of its clusters.
Each node then
multicasts an \two message to all nodes of every involved cluster.

\uline{Cross-shard cross-enterprise transactions:}
While different involved clusters of each enterprise, e.g., $B_1$ and $B_2$ of enterprise $B$,
maintain different shards, e.g., $d^1_{BC}$ and $d^2_{BC}$, of a shared data collection, e.g., $d_{BC}$,
each data shard is replicated on one cluster of each involved enterprise, e.g.,
$d^2_{BC}$ is maintained by $B_2$, and $C_2$.
Upon receiving a valid \one message from the coordinator cluster, e.g., $B_1$,
if the cluster belongs to the initiator enterprise, e.g., $B_2$,
its primary sends an \two message to the nodes of clusters that maintain the same shard, e.g., $B_2$ and $C_2$.
Each node then sends an \two message to all nodes of every involved cluster.
\end{minipage}}

\noindent \fcolorbox{black}{boxcolor}{\begin{minipage}{0.97\textwidth}
\noindent \textbf{Commit:}
Upon receiving valid matching \two messages from a local-majority of every involved cluster, each node
multicasts a \three message to every node of all involved clusters.
Once a node receives valid matching \three messages
from the local-majority of every involved clusters,
the node commits the transaction.
\end{minipage}}

The flattened protocols of \sys enable clusters to commit a transaction in a smaller number of communication phases.
For instance, a cross-shard intra-enterprise transaction can be committed using the flattened protocol in $3$ cross-cluster communication steps. In contrast, using the coordinator-based protocol, the same transaction requires $3$ cross-cluster and $9$ intra-cluster communication steps (3 runs of PBFT).
It also reduces the load on the coordinator cluster.
We now discuss flattened consensus protocols in detail.

\subsubsection{Intra-Shard Cross-Enterprise Consensus}\label{sec:deccrossent}
In the intra-shard cross-enterprise protocol, 
multiple clusters from different enterprises process a transaction on the same data shard of a shared data collection, e.g.,
clusters $C_3$ and $D_3$ on shard $d^3_{CD}$ of shared data collection $d_{CD}$ (Figure~\ref{fig:decentralized}(a)).
Upon receiving a valid request message $m$,
the primary ordering node $\pi(P_i)$ of the initiator cluster $P_i$
assigns transaction \texttt{ID} (as discussed in Section~\ref{sec:ledger}) and
multicasts a signed \one message $\langle\text{\scriptsize \ONE}, \textsf{\scriptsize ID}, d, m \rangle_{\sigma_{\pi(P_i)}}$ 
to the nodes of all involved clusters where $d=D(m)$ is the digest of request $m$.
Upon receiving a \one message, each node $r$ validates the digest, signature and \textsf{\scriptsize ID}, and
multicasts a signed \two message $\langle\text{\scriptsize \TWO}, \textsf{\scriptsize ID}, d \rangle_{\sigma_r}$
to all  nodes of every involved cluster.
Each node $r$ waits for valid matching \two messages from a local-majority of every involved cluster and then
multicasts a \three message including the transaction \texttt{ID} and its digest to every node of all involved clusters.
The \one and \two phases of the protocol, 
similar to {\sf \small pre-prepare} and {\sf \small prepare} phases of PBFT,
guarantee that non-faulty nodes agree on an order for the transactions.
Finally, once an ordering node receives valid \three messages
from the local-majority of all involved clusters that match its \three message,
it generates a \three certificate by merging messages from all involved clusters.
Each cluster then uses the transaction execution routine
to execute the transaction and send \reply back to the client.

\subsubsection{Cross-shard Intra-Enterprise consensus}\label{sec:deccrossshard}

The cross-shard intra-enterprise consensus protocol of \sys is inspired
by the flattened cross-shard consensus protocols of SharPer \cite{amiri2021sharper}.
In the cross-shard intra-enterprise consensus protocol, as shown in Figure~\ref{fig:decentralized}(b),
the involved clusters, e.g., $A_2$ and $A_3$, maintain different shards of a local data collection, e.g., $d_A$.
Upon receiving a valid request message $m$, the primary $\pi(P_i)$ of the initiator cluster $P_i$ 
assigns $\texttt{ID}_i$ to the transaction and multicasts a \one message
$\mu= \langle\text{\scriptsize \ONE}, \textsf{\scriptsize ID}_i, d, m \rangle_{\sigma_{\pi(P_i)}}$
to all nodes of every involved cluster.
Upon receiving a valid \one message, each primary node $\pi(P_j)$ of an involved cluster $P_j$
multicasts
$\langle\text{\scriptsize \TWO}, \textsf{\scriptsize ID}_i, \textsf{\scriptsize ID}_j, d, d_{\mu} \rangle_{\sigma_{\pi(P_j)}}$
to the nodes of its cluster where 
$\texttt{ID}_j$ represents the order of $m$ on data shard of cluster $P_j$.
The digest of the \one message, $d_{\mu} = D(\mu)$, is added to link the \two message with the corresponding \one message
enabling nodes to detect changes and alterations to any part of the message.
The primary of the initiator and all involved clusters process the request only if
there is no (concurrent) uncommitted request $m'$ where
$m$ and $m'$ intersect in at least $2$ shards.
This is needed to ensure that requests are committed in different shards in the same order (global consistency) \cite{amiri2021sharper}.

Each node $r$ (including the primary) of an involved cluster $P_j$ then multicasts a signed \two message
$\langle\text{\scriptsize \TWO}, \textsf{\scriptsize ID}_i, \textsf{\scriptsize ID}_j, d, r \rangle_{\sigma_r}$
to all nodes of every involved cluster.
Upon receiving valid matching \two messages from the local-majority of all involved clusters $P_i$, $P_j$, ..., $P_k$,
each node $r$ multicasts a
$\langle\text{\scriptsize \THREE}, \textsf{\scriptsize ID}_i, \textsf{\scriptsize ID}_j, ..., \textsf{\scriptsize ID}_k, d, r \rangle_{\sigma_r}$
message to every node of all involved clusters.

Upon receiving valid matching \three messages
from the local-majority of every involved cluster,
the node generates a \three certificate.
The transaction execution routine then executes the transaction and appends the transaction to the ledger.

In this protocol, since all clusters belong to the same enterprise,
if nodes of all involved clusters are crash-only,
\sys processes transactions more efficiently.
In that case, and upon receiving
$\langle\text{\scriptsize \ONE}, \textsf{\scriptsize ID}_i, d, m \rangle_{\sigma_{\pi(P_i)}}$
from the initiator primary $\pi(P_i)$,
the primary node of each involved cluster $P_j$ assigns $\texttt{ID}_j$ to the transaction
and multicasts an \two message to the ordering nodes of its cluster.
Nodes of each involved cluster then send an \two message to {\em only} the initiator primary
(in contrast to the previous protocol where nodes multicast \two messages to all nodes of every involved cluster).
The initiator primary waits for $f+1$ valid matching \two messages from different nodes of every involved cluster
and multicasts a \three message to every node of all involved clusters.
Once an ordering node receives a valid \three message from the initiator primary,
considers the transaction as committed.

\subsubsection{Cross-Shard Cross-Enterprise consensus}\label{sec:deccrossboth}

The cross-shard cross-enterprise consensus protocol is needed when
a transaction is executed on different shards of a non-local data collection
shared between different enterprises, e.g.,
a transaction that is executed on data shards $d^1_{BC}$ and $d^2_{BC}$ of shared data collection $d_{BC}$
where $d^1_{BC}$ is maintained by $B_1$ and $C_1$ and $d^2_{BC}$ is maintained by $B_2$, and $C_2$ (Figure~\ref{fig:decentralized}(c)).

In this case, different clusters of each enterprise maintain different data shards, e.g., $B_1$ and $B_2$ maintain $d^1_{BC}$ and $d^2_{BC}$, however,
each data shard is replicated on only one cluster of each involved enterprise, e.g., $d^1_{BC}$ is replicated on $B_1$ and $C_1$.
In \sys, in order to improve performance, only the clusters of the initiator enterprise, e.g., $B_1$ and $B_2$,
establish consensus on the transaction order and
clusters of other enterprises, e.g., $C_1$ and $C_2$, validate the order.

The primary node $\pi(P_i)$ of the initiator cluster $P_i$ multicasts a signed \one message
$\mu=\langle\text{\scriptsize \ONE}, \textsf{\scriptsize ID}_i, d, m \rangle_{\sigma_{\pi(P_i)}}$
to the nodes of all involved clusters.
Upon receiving a valid \one message, each primary node $\pi(P_j)$ of an involved cluster $P_j$
belonging to the initiator enterprise
assigns an $\texttt{ID}_j$ to the transaction and
multicasts a signed $\langle\text{\scriptsize \TWO}, \textsf{\scriptsize ID}_i, \textsf{\scriptsize ID}_j, d, d_{\mu} \rangle_{\sigma_{\pi(P_j)}}$ message
to all the nodes of its cluster {\it and} all other clusters that maintain the same data shard.
As before, primary nodes ensure consistency (i.e., no concurrent request with more than one shared shard) before processing a request.
The \two and \three phases process the transaction in the same way as cross-shard intra-enterprise protocol.

\subsubsection{Primary Failure Handling}\label{sec:dec-fail}

We use the retransmission routine presented in \cite{yin2003separating} to handle unreliable communication
between ordering nodes and execution nodes. In this section, we focus on the failure of the primary ordering node.
If the timer of node $r$ expires before it receives a \three certificate, it suspects that the primary might be faulty.
When the primary node of a cluster fails, the primary failure handling routine of
the internal consensus protocol, e.g., view-change in PBFT, is triggered by timeouts
to elect a new primary.

While in the intra-shard cross-enterprise consensus, only the failure of the initiator primary node needs to be handled,
in the second and third protocols, the primary node of any of the involved clusters that assign an \texttt{ID}, i.e.,
belongs to the initiator enterprise, might fail.
In the intra-shard cross-enterprise consensus, if the timer of node $r$ expires and $r$ belongs to the initiator cluster,
it initiates the failure handling routine of the internal consensus protocol, e.g., view-change in PBFT.
Otherwise, if the node is in some other involved cluster, it multicasts a \cmtq 
message including the request \texttt{ID} and its digest to the nodes of the initiator cluster.
If nodes of the initiator cluster receive \cmtq from the local-majority of an involved cluster
for a request, they suspect their primary is faulty and did not send consistent \one messages.
Nodes also log the query messages to detect denial-of-service attacks initiated by malicious nodes.

In the second and third protocols, each transaction needs to be ordered by the initiator primary as well as
the primary nodes of all clusters belonging to the initiator enterprise (i.e., which maintain different data shards).
Suppose a malicious primary node (from the initiator or an involved cluster) sends a request with inconsistent \texttt{ID} to different nodes. In that case, nodes detect the failure in all to all communication phases of the protocol triggering the primary failure handling routine within the faulty primary cluster.
Note that since \one and \two messages need to be multicast to all nodes,
and all nodes of all involved clusters multicast \two and \three messages to each other,
even if a malicious primary decides not to multicast a request to a group of nodes, 
it will be easily detected.

Finally, if a client does not receive a \reply soon enough, it multicasts its \req message
to all ordering nodes of the cluster that it has already sent its request.
If an ordering node has both the \three certificate (i.e., the request has already been ordered) and
the \reply certificate (i.e., the request has already been executed),
the node simply sends the \reply certificate to the client.
If the node has not received the \reply certificate, it re-sends the \three certificate to the filter nodes.
Otherwise, if the node is not the primary, it relays the request to the primary.
If the nodes do not receive \one messages, the primary will be suspected to be faulty.

\subsubsection{Correctness Argument}

We briefly analyze the agreement and liveness properties of the flattened protocols.
Validity, and consistency are proven in the same way as coordinator-based protocols.

\begin{lmm} (\textit{Agreement})
If node $r$ commits request $m$ with local \texttt{ID} $\alpha$ in cluster $P$,
no other non-faulty node commits request $m'$ ($m \neq m'$) with the same local \texttt{ID} $\alpha$ in $P$.
\end{lmm}

\begin{prf}
The \one and \two phases of each presented consensus protocol
guarantee that non-faulty nodes agree on the order of a transaction.
To commit a request, matching messages from the local-majority of every involved cluster are needed.
Given two requests $m$ with local \texttt{ID} $\alpha$ and $m'$ with local \texttt{ID} $\alpha'$ where
local-majority $Q$ agreed with $m$ and local-majority $Q'$ agreed with $m'$.
Let $m$ and $m'$ ($m \neq m'$) be two committed requests where
$\texttt{ID}(m) = [\texttt{ID}_i,\texttt{ID}_j,{\color{violet}{\texttt{ID}_k}},...]$ and
$\texttt{ID}(m') = [{\color{violet}{\texttt{ID}'_k}},\texttt{ID}'_l,\texttt{ID}'_m,..]$ respectively.
Note that in intra-shard transactions, the \texttt{ID} includes a single part.
Given an involved cluster {\color{violet}{$P_k$}} in the intersection of $m$ and $m'$ where
$\texttt{ID}_k(m) = \langle \alpha, \gamma \rangle$ and
$\texttt{ID}_k(m') = \langle \alpha', \gamma' \rangle$.
To commit $m$ and $m'$, a local-majority $Q$ agreed with $m$ and local-majority $Q'$ agreed with $m'$.
Since $Q$ and $Q'$ intersect on at least one non-faulty node
and non-faulty nodes do not behave maliciously,
if $m \neq m'$ then $\alpha \neq \alpha'$, hence agreement is guaranteed.
\end{prf}

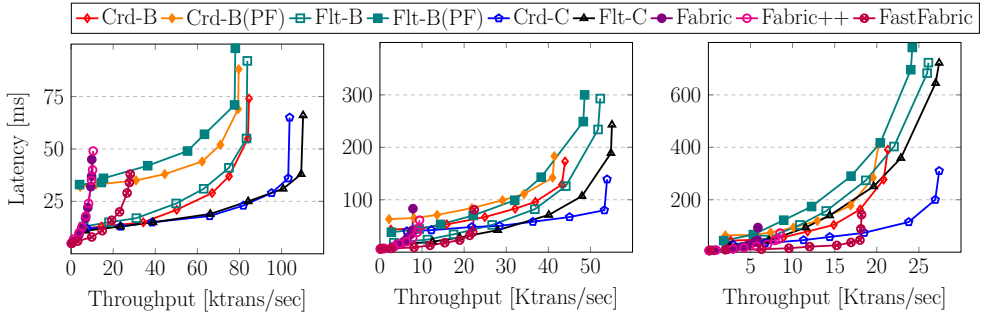
\begin{figure}[t!]
\centering
\LARGE
\begin{minipage}{.25\textwidth}\centering
\begin{tikzpicture}[scale=0.49]
\begin{axis}[
    xlabel={Throughput [ktrans/sec]},
    ylabel={Latency [ms]},
    xmin=0, xmax=120,
    ymin=0, ymax=100,
    xtick={0,20,40,60,80,100},
    ytick={25,50,75},
    legend columns=9,
    legend style={at={(0,1.05)},anchor=south west},
    ymajorgrids=true,
    grid style=dashed,
]

\addplot[
    color=red,
    mark=diamond,
    mark size=3pt,
    line width=0.5mm,
    ]
    coordinates {
    (5.743,12)(14.405,13)(34.302,15)(50.20,21)(67.023,29)(75.04,37)(84.037,55)(84.532,74)};

\addplot[
    color=orange,
    mark=diamond*,
    mark size=3pt,
    line width=0.5mm,
    ]
    coordinates {
    (4.331,32)(11.813,33)(30.872,35)(44.473,38)(62.249,44)(70.93,52)(79.351,69)(79.643,88)};
    
\addplot[
    color=teal,
    mark=square,
    mark size=3pt,
    line width=0.5mm,
    ]
    coordinates {
    (5.97,13)(18.01,15)(31.01,17)(49.913,24)(63.011,31)(74.983,41)(83.321,55)(83.724,92)};
   
\addplot[
    color=teal,
    mark=square*,
    mark size=3pt,
    line width=0.5mm,
    ]
    coordinates {
    (4.03,33)(14.392,34)(15.384,36)(36.387,42)(55.324,49)(63.387,57)(77.682,71)(78.021,98)};

\addplot[
    color=blue,
    mark=pentagon,
    mark size=3pt,
    line width=0.5mm,
    ]
    coordinates {
    (7.541,12)(23.594,13)(38.487,15)(65.845,18)(81.834,23)(95.234,29)(103.143,36)(103.821,65)};

\addplot[
    color=black,
    mark=triangle,
    mark size=3pt,
    line width=0.5mm,
    ]
    coordinates {
    (6.832,11)(23.397,13)(38.902,15)(66.121,19)(84.109,25)(100.743,31)(109.321,38)(110.221,66)};
    
\addplot[
    color=violet,
    mark=*,
    mark size=3pt,
    line width=0.5mm,
    ]
    coordinates {
   (0.134,5)(1.159,6)(3.704,9)(5.048,12)(6.876,17)(7.932,22)(9.381,32)(9.743,37)(9.833,45)};
   
\addplot[
    color=magenta,
    mark=o,
    mark size=3pt,
    line width=0.5mm,
    ]
    coordinates {
   (0.143,5)(1.233,7)(3.801,10)(5.282,13)(7.023,18)(8.311,24)(9.719,35)(10.212,40)(10.511,49)};
   
\addplot[
    color=purple,
    mark=otimes,
    mark size=3pt,
    line width=0.5mm,
    ]
    coordinates {
   (0.409,5)(3.119,6)(9.811,8)(14.803,11)(19.302,16)(23.034,20)(26.614,29)(27.702,34)(28.101,38)};

\addlegendentry{Crd-B}
\addlegendentry{Crd-B(PF)}
\addlegendentry{Flt-B}
\addlegendentry{Flt-B(PF)}
\addlegendentry{Crd-C}
\addlegendentry{Flt-C}
\addlegendentry{Fabric}
\addlegendentry{Fabric++}
\addlegendentry{FastFabric}

\end{axis}
\end{tikzpicture}
\end{minipage}\hspace{0.1em}
\begin{minipage}{.25\textwidth}\centering
\vspace{0.9em}
\begin{tikzpicture}[scale=0.49]
\begin{axis}[
    xlabel={Throughput [Ktrans/sec]},
    xmin=0, xmax=60,
    ymin=0, ymax=400,
    xtick={0,10,20,30,40,50},
    ytick={100,200,300},
    ymajorgrids=true,
    grid style=dashed,
]

\addplot[
    color=red,
    mark=diamond,
    mark size=3pt,
    line width=0.5mm,
    ]
    coordinates {
    (2.605,43)(9.818,46)(15.902,53)(24.911,67)(32.111,83)(36.910,96)(43.219,129)(43.972,173)};

\addplot[
    color=orange,
    mark=diamond*,
    mark size=3pt,
    line width=0.5mm,
    ]
    coordinates {
    (2.163,63)(8.073,65)(13.584,71)(21.852,84)(29.119,99)(34.223,111)(41.012,142)(41.432,183)};

\addplot[
    color=teal,
    mark=square,
    mark size=3pt,
    line width=0.5mm,
    ]
    coordinates {
    (3.303,18)(11.302,24)(17.498,34)(26.653,52)(36.793,82)(44.143,126)(51.809,234)(52.332,293)
    };
   
\addplot[
    color=teal,
    mark=square*,
    mark size=3pt,
    line width=0.5mm,
    ]
    coordinates {
    (2.720,38)(9.323,44)(14.408,53)(22.143,70)(32.02,99)(38.303,143)(48.302,249)(48.622,300)};

\addplot[
    color=blue,
    mark=pentagon,
    mark size=3pt,
    line width=0.5mm,
    ]
    coordinates {
    (6.406,40)(12.232,42)(21.802,48)(28.408,50)(36.330,58)(45.033,67)(53.309,80)(53.954,139)};

\addplot[
    color=black,
    mark=triangle,
    mark size=3pt,
    line width=0.5mm,
    ]
    coordinates {
    (3.673,14)(12.505,20)(18.913,28)(27.993,42)(40.104,71)(47.990,107)(54.873,189)(55.103,243)
    };
    
\addplot[
    color=violet,
    mark=*,
    mark size=3pt,
    line width=0.5mm,
    ]
    coordinates {
   (0.105,6)(0.901,7)(3.043,10)(4.011,14)(5.442,19)(6.325,25)(7.440,36)(7.769,42)(7.843,83)};
   
\addplot[
    color=magenta,
    mark=o,
    mark size=3pt,
    line width=0.5mm,
    ]
    coordinates {
   (0.121,6)(1.064,7)(3.301,10)(4.604,16)(6.301,20)(7.783,27)(8.881,38)(9.172,45)(9.401,61)};
   
\addplot[
    color=purple,
    mark=otimes,
    mark size=3pt,
    line width=0.5mm,
    ]
    coordinates {
   (0.359,7)(2.603,7)(8.101,9)(11.903,13)(15.412,18)(18.954,23)(21.309,32)(22.013,38)(22.404,81)};

\end{axis}
\end{tikzpicture}
\end{minipage}\hspace{0.1em}
\begin{minipage}{.25\textwidth}\centering
\vspace{0.9em}
\begin{tikzpicture}[scale=0.49]
\begin{axis}[
    xlabel={Throughput [Ktrans/sec]},
    xmin=0, xmax=30,
    ymin=0, ymax=800,
    xtick={5,10,15,20,25},
    ytick={200,400,600},
    ymajorgrids=true,
    grid style=dashed,
]

\addplot[
    color=red,
    mark=diamond,
    mark size=3pt,
    line width=0.5mm,
    ]
    coordinates {
   (2.632,44)(5.403,47)(8.942,57)(11.781,79)(14.892,104)(18.018,163)(20.820,277)(21.313,391)
    };
    
\addplot[
    color=orange,
    mark=diamond*,
    mark size=3pt,
    line width=0.5mm,
    ]
    coordinates {
   (2.012,64)(5.391,66)(7.402,75)(10.093,94)(12.934,119)(16.894,178)(19.545,284)(20.211,411)};
    
\addplot[
    color=teal,
    mark=square,
    mark size=3pt,
    line width=0.5mm,
    ]
    coordinates {
   (2.14,25)(6.593,48)(10.873,104)(13.993,157)(18.702,274)(22.043,403)(25.981,683)(26.133,723)
   };

\addplot[
    color=teal,
    mark=square*,
    mark size=3pt,
    line width=0.5mm,
    ]
    coordinates {
   (1.82,44)(5.335,67)(8.943,122)(12.254,174)(16.943,290)(20.420,417)(24.041,696)(24.219,781)
   };

\addplot[
    color=blue,
    mark=pentagon,
    mark size=3pt,
    line width=0.5mm,
    ]
    coordinates {
   (2.93,28)(6.743,34)(11.302,46)(14.411,58)(18.512,74)(23.791,115)(26.941,200)(27.413,310)};
   
\addplot[
    color=black,
    mark=triangle,
    mark size=3pt,
    line width=0.5mm,
    ]
    coordinates {
   (2.531,22)(7.201,44)(11.541,94)(14.402,140)(19.632,251)(22.862,359)(26.983,645)(27.404,722)};

\addplot[
    color=violet,
    mark=*,
    mark size=3pt,
    line width=0.5mm,
    ]
    coordinates {
   (0.088,6)(0.667,7)(2.230,11)(3.092,15)(4.081,20)(4.720,27)(5.581,41)(5.822,45)(5.883,94)};
   
\addplot[
    color=magenta,
    mark=o,
    mark size=3pt,
    line width=0.5mm,
    ]
    coordinates {
   (0.105,6)(0.913,8)(2.911,12)(4.211,16)(5.703,23)(7.014,29)(7.913,45)(8.302,49)(8.511,73)};
   
\addplot[
    color=purple,
    mark=otimes,
    mark size=3pt,
    line width=0.5mm,
    ]
    coordinates {
   (0.289,7)(2.011,9)(6.302,11)(9.562,15)(12.023,21)(15.243,26)(16.943,37)(18.002,46)(18.110,89)(18.193,143)};

\end{axis}
\end{tikzpicture}
\end{minipage}
\caption{Workloads with (a) $10\%$, (b) $50\%$, (c) $90\%$ intra-shard cross-enterprise transactions}
\label{fig:cross-E}
\end{figure}

\begin{prop}(\textit{Liveness})
A request $m$ issued by a correct client eventually completes.
\end{prop}

\begin{prf}
\sys guarantees liveness {\em only} during periods of synchrony.
If the primary of a cluster that needs to order the request and assign \texttt{ID} is faulty,
as explained in Section~\ref{sec:dec-fail},
its failure will be detected using timers and \sys uses
the primary failure handling routine of the internal consensus protocol to elect a new primary.
As explained for the coordinator-based protocols,
nodes use different timers for intra-cluster and cross-cluster transactions where
the timer for cross-cluster transactions is 
long enough to allow nodes to communicate with each other and establish consensus, i.e.
at least 3 times the WAN round-trip for cross-cluster transactions
to allow a view-change routine to complete without replacing the primary node again.
If a newly elected leader is not able to change the view, the timeout is doubled (as in PBFT \cite{castro1999practical}).
This adjustment is needed to ensure that some transactions will be committed, i.e., nodes are synchronized.
The deadlock situation is addressed in the same way as the coordinator-based protocols.
\end{prf}
\section{Experimental Evaluation}
\label{sec:eval}

 Our evaluation has three goals:
 (1) comparing the performance of the
coordinator-based and flattened
consensus protocols in different workloads
with different types of transactions;
(2) demonstrating the overhead of using the privacy firewall mechanism \cite{yin2003separating}
to provide confidentiality despite Byzantine faults; 
(3) comparing the performance of \sys with
Hyperledger Fabric \cite{androulaki2018hyperledger}
and two of its variants, FastFabric \cite{fastfabric2019} and Fabric++ \cite{sharma2019blurring},
to analyze the impact of sharding and the performance trade-off between
a higher number of shards versus a higher percentage of cross-shard transactions.

We analyze the impact of
(1) the failure model of nodes,
(2) the percentage of cross-enterprise transactions, 
(3) the percentage of cross-shard transactions,
(4) the geo-distribution of clusters,
(5) the number of involved enterprises,
(6) the failure of nodes, and
(7) the degrees of contention
 on the performance of these protocols and systems.

We have implemented a prototype of \sys using Java and
deployed a simple asset management collaboration workflow.
We use {\sf \small SmallBank} benchmarks
(modified it to be able to control cross-shard and cross-enterprise transactions).
The workloads are write-heavy ({\sf \small sendPayment} transactions are used to perform
read and write in a single or multiple shards of  data collection).
The key selection distribution in each data collection is uniform with {\sf \small s-value} = $0$.

Other than Section \ref{sec:enteprise} where we vary the number of enterprises,
we consider an infrastructure with $4$ enterprises.
Each enterprise partitions
its data into $4$ shards.
Each crash-only clusters includes $2f+1$ nodes that order and execute transactions
while each Byzantine cluster includes $3f+1$ ordering nodes,
$2g+1$ execution nodes and $h+1$ rows of $h+1$ filter nodes.
To demonstrate the overhead of the privacy firewall mechanism, we also measure the performance of
Byzantine clusters with only $3f+1$ nodes that order and execute transactions.
In all experiments $f=g=h=1$.
We use Paxos and PBFT as the internal consensus protocols.

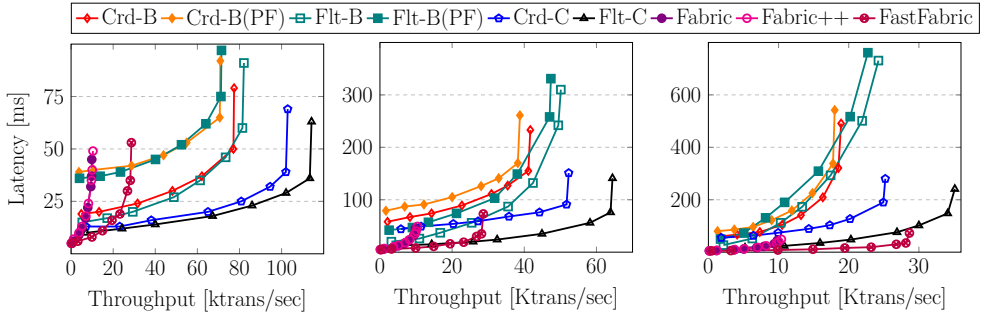
\begin{figure}[t!]
\centering
\LARGE
\begin{minipage}{.25\textwidth}\centering
\begin{tikzpicture}[scale=0.49]
\begin{axis}[
    xlabel={Throughput [ktrans/sec]},
    ylabel={Latency [ms]},
    xmin=0, xmax=120,
    ymin=0, ymax=100,
    xtick={0,20,40,60,80,100},
    ytick={25,50,75},
    legend columns=9,
    legend style={at={(0,1.05)},anchor=south west},
    ymajorgrids=true,
    grid style=dashed,
]

\addplot[
    color=red,
    mark=diamond,
    mark size=3pt,
    line width=0.5mm,
    ]
    coordinates {
    (4.843,19)(13.305,20)(31.60,24)(48.123,30)(62.14,37)(77.137,50)(77.432,79)};
   
\addplot[
    color=orange,
    mark=diamond*,
    mark size=3pt,
    line width=0.5mm,
    ]
    coordinates {
    (3.719,39)(10.792,40)(28.731,42)(43.923,47)(54.932,53)(70.643,65)(70.913,92)};

\addplot[
    color=teal,
    mark=square,
    mark size=3pt,
    line width=0.5mm,
    ]
    coordinates {
    (5.32,15)(17.16,17)(29.302,20)(48.82,27)(61.281,35)(73.421,46)(81.392,60)(82.102,91)};

\addplot[
    color=teal,
    mark=square*,
    mark size=3pt,
    line width=0.5mm,
    ]
    coordinates {
    (4.014,36)(13.802,37)(23.459,39)(40.106,45)(52.451,52)(64.031,62)(71.192,75)(71.451,97)};

\addplot[
    color=blue,
    mark=pentagon,
    mark size=3pt,
    line width=0.5mm,
    ]
    coordinates {
    (7.03,13)(23.091,13)(38.087,16)(65.045,20)(80.834,25)(94.53,32)(102.04,39)(102.911,69)};
  
\addplot[
    color=black,
    mark=triangle,
    mark size=3pt,
    line width=0.5mm,
    ]
    coordinates {
    (7.1,10)(24.217,12)(40.019,14)(67.23,18)(85.982,23)(102.111,29)(113.409,36)(114.221,63)};
    
\addplot[
    color=violet,
    mark=*,
    mark size=3pt,
    line width=0.5mm,
    ]
    coordinates {
   (0.134,5)(1.159,6)(3.704,9)(5.048,12)(6.876,17)(7.932,22)(9.381,32)(9.743,37)(9.833,45)};

\addplot[
    color=magenta,
    mark=o,
    mark size=3pt,
    line width=0.5mm,
    ]
    coordinates {
   (0.141,5)(1.193,7)(3.742,10)(5.119,13)(7.023,18)(8.243,24)(9.684,35)(10.034,40)(10.402,49)};
   
\addplot[
    color=purple,
    mark=otimes,
    mark size=3pt,
    line width=0.5mm,
    ]
    coordinates {
   (0.453,5)(3.203,6)(9.899,8)(14.902,11)(19.441,16)(23.203,19)(26.784,30)(28.142,35)(28.654,53)};
   
\addlegendentry{Crd-B}
\addlegendentry{Crd-B(PF)}
\addlegendentry{Flt-B}
\addlegendentry{Flt-B(PF)}
\addlegendentry{Crd-C}
\addlegendentry{Flt-C}
\addlegendentry{Fabric}
\addlegendentry{Fabric++}
\addlegendentry{FastFabric}

\end{axis}
\end{tikzpicture}
\end{minipage}\hspace{0.1em}
\begin{minipage}{.25\textwidth}\centering
\vspace{0.9em}
\begin{tikzpicture}[scale=0.49]
\begin{axis}[
    xlabel={Throughput [Ktrans/sec]},
    xmin=0, xmax=70,
    ymin=0, ymax=400,
    xtick={0,20,40,60},
    ytick={100,200,300},
    ymajorgrids=true,
    grid style=dashed,
]

\addplot[
    color=red,
    mark=diamond,
    mark size=3pt,
    line width=0.5mm,
    ]
    coordinates {
    (2.105,58)(8.318,67)(14.302,74)(22.711,89)(30.911,111)(35.410,127)(41.119,155)(41.672,233)
    };
    
\addplot[
    color=orange,
    mark=diamond*,
    mark size=3pt,
    line width=0.5mm,
    ]
    coordinates {
    (1.651,79)(6.863,87)(12.198,91)(20.014,105)(28.034,126)(32.913,141)(38.212,170)(38.792,261)};
    
\addplot[
    color=teal,
    mark=square,
    mark size=3pt,
    line width=0.5mm,
    ]
    coordinates {
    (3.170,20)(10.93,26)(16.711,37)(25.411,56)(35.510,87)(42.410,132)(49.431,242)(50.132,310)};

\addplot[
    color=teal,
    mark=square*,
    mark size=3pt,
    line width=0.5mm,
    ]
    coordinates {
    (2.549,42)(9.002,47)(13.438,57)(21.219,74)(31.792,103)(38.072,149)(47.021,258)(47.371,331)};

\addplot[
    color=blue,
    mark=pentagon,
    mark size=3pt,
    line width=0.5mm,
    ]
    coordinates {
    (5.809,44)(11.401,49)(20.331,54)(27.188,59)(35.751,68)(44.211,76)(51.719,91)(52.309,151)};

\addplot[
    color=black,
    mark=triangle,
    mark size=3pt,
    line width=0.5mm,
    ]
    coordinates {
    (6.14,13)(14.343,15)(25.821,20)(32.482,24)(44.902,35)(58.213,56)(63.909,76)(64.442,141)};
    
\addplot[
    color=violet,
    mark=*,
    mark size=3pt,
    line width=0.5mm,
    ]
    coordinates {
   (0.134,5)(1.159,6)(3.704,9)(5.048,12)(6.876,17)(7.932,22)(9.381,32)(9.743,37)(9.833,45)};

\addplot[
    color=magenta,
    mark=o,
    mark size=3pt,
    line width=0.5mm,
    ]
    coordinates {
   (0.141,5)(1.193,7)(3.742,10)(5.119,13)(7.023,18)(8.243,24)(9.684,35)(10.034,40)(10.402,49)};
   
\addplot[
    color=purple,
    mark=otimes,
    mark size=3pt,
    line width=0.5mm,
    ]
    coordinates {
   (0.453,5)(3.203,6)(9.899,8)(14.902,11)(19.441,16)(23.203,19)(26.784,30)(28.142,35)(28.654,73)};

\end{axis}
\end{tikzpicture}
\end{minipage}\hspace{0.1em}
\begin{minipage}{.25\textwidth}\centering
\vspace{0.9em}
\begin{tikzpicture}[scale=0.49]
\begin{axis}[
    xlabel={Throughput [Ktrans/sec]},
    xmin=0, xmax=36,
    ymin=0, ymax=800,
    xtick={0,10,20,30},
    ytick={200,400,600},
    ymajorgrids=true,
    grid style=dashed,
]

\addplot[
    color=red,
    mark=diamond,
    mark size=3pt,
    line width=0.5mm,
    ]
    coordinates {
   (1.62,60)(4.099,67)(7.327,77)(10.514,104)(13.201,141)(16.311,209)(18.531,321)(18.913,491)
    };
    
\addplot[
    color=orange,
    mark=diamond*,
    mark size=3pt,
    line width=0.5mm,
    ]
    coordinates {
      (1.301,80)(3.692,86)(6.381,96)(9.013,122)(11.892,159)(14.823,225)(17.761,338)(18.001,542)};
    
\addplot[
    color=teal,
    mark=square,
    mark size=3pt,
    line width=0.5mm,
    ]
    coordinates {
   (2.14,27)(6.254,52)(10.294,112)(13.363,173)(17.418,293)(21.910,501)(24.235,731)};

\addplot[
    color=teal,
    mark=square*,
    mark size=3pt,
    line width=0.5mm,
    ]
    coordinates {
   (1.73,50)(5.031,73)(8.152,131)(10.872,190)(15.681,309)(20.204,517)(22.731,761)};

\addplot[
    color=blue,
    mark=pentagon,
    mark size=3pt,
    line width=0.5mm,
    ]
    coordinates {
   (1.83,55)(6.409,63)(9.94,74)(14.314,89)(17.301,103)(20.191,127)(24.931,190)(25.213,280)
    };
   
\addplot[
    color=black,
    mark=triangle,
    mark size=3pt,
    line width=0.5mm,
    ]
    coordinates {
    (5.221,17)(10.802,24)(15.911,35)(20.311,48)(26.881,76)(30.001,102)(34.104,148)(35.119,241)
    };

\addplot[
    color=violet,
    mark=*,
    mark size=3pt,
    line width=0.5mm,
    ]
    coordinates {
   (0.134,5)(1.159,6)(3.704,9)(5.048,12)(6.876,17)(7.932,22)(9.381,32)(9.743,37)(9.833,45)};
   
\addplot[
    color=magenta,
    mark=o,
    mark size=3pt,
    line width=0.5mm,
    ]
    coordinates {
   (0.141,5)(1.193,7)(3.742,10)(5.119,13)(7.023,18)(8.243,24)(9.684,35)(10.034,40)(10.402,49)};
   
\addplot[
    color=purple,
    mark=otimes,
    mark size=3pt,
    line width=0.5mm,
    ]
    coordinates {
   (0.453,5)(3.203,6)(9.899,8)(14.902,11)(19.441,16)(23.203,19)(26.784,30)(28.142,35)(28.654,73)};

\end{axis}
\end{tikzpicture}
\end{minipage}
\caption{Workloads with (a) $10\%$, (b) $50\%$, (c) $90\%$ cross-shard intra-enterprise transactions}
\label{fig:cross-shard}
\end{figure}

We consider a single-channel Fabric deployment ($v2.2$) where Raft \cite{ongaro2014search} is its consensus protocol.
We deploy four enterprises on the channel where enterprises can form public or private collaboration.
Sharding was not possible in a single-channel Fabric deployment, however, we defined
four endorsers to execute transactions of enterprises in parallel.
We use a similar setting for FastFabric and Fabric++.
FastFabric re-architects Fabric and provides efficient optimizations such as
separating endorsers from storage nodes and sending transaction hashes to orderers.
Fabric++, on the other hand, presents transaction reordering and early abort mechanisms to
improve the performance of Fabric, especially in contentious workloads.
We do not compare \sys with Caper \cite{amiri2019caper} because Caper supports
neither cross-enterprise transactions among a subset of enterprises nor sharding.
Similarly, sharded permissioned blockchains like AHL \cite{dang2018towards} and SharPer \cite{amiri2021sharper}
can only be compared to cross-shard intra-enterprise transactions as they do not support multi-enterprise environments.

The experiments were conducted on
the Amazon EC2 platform on multiple VM instances.
Other than the fourth set of experiments where clusters are distributed over $4$ different AWS regions,
in all experiments,
clusters are placed in the same data center (California) with $<1$ ms ping latency.
Each VM is a c4.2xlarge instance with 8 vCPUs and 15GB RAM,
Intel Xeon E5-2666 v3 processor clocked at 3.50 GHz.
The results reflect end-to-end measurements from the clients.
Each experiment is run for 90 seconds (30s warmup/cool-down).
The reported results are the average of five runs.
we use an increasing number of requests,
until the end-to-end throughput is saturated,
and state the throughput and latency just below saturation.

\subsection{Intra-Shard Cross-Enterprise Transactions}\label{sec:ishardcenterprise}

In the first set of experiments, we measure the performance of all protocols
with different percentages, i.e., $10\%$, $50\%$, and $90\%$,
of intra-shard cross-enterprise transactions.
Each transaction is randomly initiated on a single data shard of a data collection shared among multiple enterprises.
The number of involved enterprises depends on the data collection.
Figure~\ref{fig:cross-E}(a) demonstrate the results for the workload with
$10\%$ intra-shard cross-enterprise (and $90\%$ internal) transactions.
In this scenario and with crash-only nodes,
\sys processes more than
$110$ ktps with $38$ ms latency using the flattened protocol ({\sf\small Flt-C})
and $103$ ktps with $36$ ms latency using the coordinator-based protocol ({\sf\small Crd-C})
before the end-to-end throughput is saturated.

Fabric processes only $9.7$ ktps ($8\%$ of the throughput of {\sf\small Flt-C}) with $37$ ms latency.
While different endorsers of different enterprises execute their transactions in parallel,
ordering the transactions of all enterprises by a single set of orderers becomes a bottleneck.
This clearly demonstrates the impact of parallel ordering (due to sharding) in \sys, where different clusters
order and execute their transactions independently.
Fabric++ demonstrates only $3\%$ higher throughput compared to Fabric
with the same latency as it is able to reorder and early abort
invalidated transactions.
FastFabric, however, demonstrates $189\%$ throughput improvement compared Fabric due to its optimized architecture.
However, its throughput is still $26\%$ of the throughput of {\sf\small Flt-C} with the same latency.

With Byzantine nodes, the performance of the flattened protocol ({\sf\small Flt-B})
and the coordinator-based protocol ({\sf\small Crd-B}) is reduced,
which is expected due to the higher complexity of BFT protocols.
Using the privacy firewall mechanism results in $8\%$ and $6\%$ lower throughput
in the coordinator-based ({\sf\small Crd-B(PF)}) and flattened ({\sf\small Flt-B(PF)}) protocols respectively
before the end-to-end throughput is saturated.
This slight throughput reduction is the result of two infrastructural changes.
On one hand, using the privacy firewall mechanism,  \req and \reply messages go through filters resulting in lower performance.
On the other hand, separating ordering from execution reduces the performance overhead by decreasing the load on ordering nodes.
The privacy firewall mechanism also increases the latency of transaction processing in different workloads by a constant coefficient.
This is expected because the increased latency comes from sending messages through the filters,
which is separated from the consensus routine; the bottleneck in heavy workloads.
While this latency is considerable in light workloads, it becomes much lower in heavy workloads, e.g.,
$166\%$ vs. $25\%$ higher latency in {\sf\small Crd-B(PF)} protocol.

\begin{figure}[t!]
\centering
\LARGE
\begin{minipage}{.25\textwidth}\centering
\begin{tikzpicture}[scale=0.49]
\begin{axis}[
    xlabel={Throughput [ktrans/sec]},
    ylabel={Latency [ms]},
    xmin=0, xmax=120,
    ymin=0, ymax=100,
    xtick={0,20,40,60,80,100},
    ytick={30,60,90},
    legend columns=9,
    legend style={at={(0,1.05)},anchor=south west},
    ymajorgrids=true,
    grid style=dashed,
]

\addplot[
    color=red,
    mark=diamond,
    mark size=3pt,
    line width=0.5mm,
    ]
    coordinates {
    (4.673,20)(13.033,21)(31.228,25)(46.612,31)(61.563,38)(77.374,53)(77.432,81)};
    
\addplot[
    color=orange,
    mark=diamond*,
    mark size=3pt,
    line width=0.5mm,
    ]
    coordinates {
    (3.581,42)(11.491,43)(25.821,45)(39.301,50)(51.023,56)(68.092,71)(68.302,92)};
    
\addplot[
    color=teal,
    mark=square,
    mark size=3pt,
    line width=0.5mm,
    ]
    coordinates {
    (4.983,18)(16.305,22)(27.594,24)(46.923,31)(58.643,37)(71.382,49)(75.211,66)(75.381,72)
    };
   
\addplot[
    color=teal,
    mark=square*,
    mark size=3pt,
    line width=0.5mm,
    ]
    coordinates {
    (3.761,40)(13.153,41)(22.891,43)(39.393,49)(51.331,54)(64.822,64)(69.018,81)(69.811,88)};

\addplot[
    color=blue,
    mark=pentagon,
    mark size=3pt,
    line width=0.5mm,
    ]
    coordinates {
   (7.399,12)(23.301,14)(38.109,16)(65.394,20)(81.594,25)(95.002,31)(102.731,38)(103.140,68)
    };

\addplot[
    color=black,
    mark=triangle,
    mark size=3pt,
    line width=0.5mm,
    ]
    coordinates {
   (6.011,11)(21.291,14)(36.563,17)(61.349,23)(80.011,28)(95.421,34)(102.011,42)(103.203,70)
    };
    
\addplot[
    color=violet,
    mark=*,
    mark size=3pt,
    line width=0.5mm,
    ]
    coordinates {
   (0.134,5)(1.159,6)(3.704,9)(5.048,12)(6.876,17)(7.932,22)(9.381,32)(9.743,37)(9.833,45)};

\addplot[
    color=magenta,
    mark=o,
    mark size=3pt,
    line width=0.5mm,
    ]
    coordinates {
   (0.143,5)(1.233,7)(3.801,10)(5.282,13)(7.023,18)(8.311,24)(9.719,35)(10.212,40)(10.511,49)};
   
\addplot[
    color=purple,
    mark=otimes,
    mark size=3pt,
    line width=0.5mm,
    ]
    coordinates {
   (0.348,5)(3.021,6)(9.692,9)(14.632,12)(19.119,17)(22.652,21)(26.391,31)(27.287,36)(27.442,67)};

\addlegendentry{Crd-B}
\addlegendentry{Crd-B(PF)}
\addlegendentry{Flt-B}
\addlegendentry{Flt-B(PF)}
\addlegendentry{Crd-C}
\addlegendentry{Flt-C}
\addlegendentry{Fabric}
\addlegendentry{Fabric++}
\addlegendentry{FastFabric}

\end{axis}
\end{tikzpicture}
\end{minipage}\hspace{0.1em}
\begin{minipage}{.25\textwidth}\centering
\vspace{0.9em}
\begin{tikzpicture}[scale=0.49]
\begin{axis}[
    xlabel={Throughput [Ktrans/sec]},
    xmin=0, xmax=55,
    ymin=0, ymax=500,
    xtick={0,10,20,30,40,50},
    ytick={150,300,450},
    ymajorgrids=true,
    grid style=dashed,
]

\addplot[
    color=red,
    mark=diamond,
    mark size=3pt,
    line width=0.5mm,
    ]
    coordinates {
   (2.105,58)(6.204,63)(12.109,74)(19.034,89)(23.794,131)(27.091,177)(31.082,235)(31.981,341)
   };
   
\addplot[
    color=orange,
    mark=diamond*,
    mark size=3pt,
    line width=0.5mm,
    ]
    coordinates {
   (1.762,78)(4.913,72)(10.049,92)(16.211,106)(20.281,147)(24.001,192)(27.832,250)(28.103,371)};
    
\addplot[
    color=teal,
    mark=square,
    mark size=3pt,
    line width=0.5mm,
    ]
    coordinates {
   (2.643,34)(7.739,52)(12.409,71)(18.380,103)(22.204,139)(27.209,212)(31.431,280)(32.132,403)
    };
   
\addplot[
    color=teal,
    mark=square*,
    mark size=3pt,
    line width=0.5mm,
    ]
    coordinates {
      (2.001,54)(6.201,72)(10.219,90)(15.210,121)(19.192,157)(24.311,228)(29.092,296)(29.302,400)};

\addplot[
    color=blue,
    mark=pentagon,
    mark size=3pt,
    line width=0.5mm,
    ]
    coordinates {
    (5.572,44)(11.091,53)(19.587,57)(26.403,63)(34.414,73)(43.173,82)(49.903,101)(50.411,169)
    };

\addplot[
    color=black,
    mark=triangle,
    mark size=3pt,
    line width=0.5mm,
    ]
    coordinates {
    (6.394,23)(15.343,27)(21.821,36)(29.482,53)(38.902,78)(45.872,112)(52.114,174)(52.842,235)};
    
\addplot[
    color=violet,
    mark=*,
    mark size=3pt,
    line width=0.5mm,
    ]
    coordinates {
   (0.105,6)(0.901,7)(3.043,10)(4.011,14)(5.442,19)(6.325,25)(7.440,36)(7.769,42)(7.843,83)};
   
\addplot[
    color=magenta,
    mark=o,
    mark size=3pt,
    line width=0.5mm,
    ]
    coordinates {
   (0.121,6)(1.064,7)(3.301,10)(4.604,14)(6.301,19)(7.783,25)(8.881,37)(9.172,43)(9.401,61)};
   
\addplot[
    color=purple,
    mark=otimes,
    mark size=3pt,
    line width=0.5mm,
    ]
    coordinates {
   (0.328,5)(2.890,6)(9.023,9)(13.781,12)(17.642,17)(19.325,21)(20.383,31)(21.391,36)(21.572,67)};

\end{axis}
\end{tikzpicture}
\end{minipage}\hspace{0.05em}
\begin{minipage}{.25\textwidth}\centering
\vspace{0.9em}
\begin{tikzpicture}[scale=0.49]
\begin{axis}[
    xlabel={Throughput [Ktrans/sec]},
    xmin=0, xmax=22,
    ymin=0, ymax=1400,
    xtick={5,10,15,20},
    ytick={400,800,1200},
    ymajorgrids=true,
    grid style=dashed,
]

\addplot[
    color=red,
    mark=diamond,
    mark size=3pt,
    line width=0.5mm,
    ]
    coordinates {
   (1.523,80)(3.657,87)(6.213,94)(9.118,134)(11.653,184)(14.671,289)(16.743,537)(17.119,784)
    };
    
\addplot[
    color=orange,
    mark=diamond*,
    mark size=3pt,
    line width=0.5mm,
    ]
    coordinates {
   (1.141,98)(2.903,106)(5.011,112)(7.623,151)(10.133,198)(13.219,299)(15.843,547)(15.981,734)};
    
\addplot[
    color=teal,
    mark=square,
    mark size=3pt,
    line width=0.5mm,
    ]
    coordinates {
   (1.430,54)(2.98,73)(5.034,142)(7.504,237)(9.522,433)(11.732,782)(12.821,1041)(13.032,1303)
   };
   
\addplot[
    color=teal,
    mark=square*,
    mark size=3pt,
    line width=0.5mm,
    ]
    coordinates {
   (1.102,74)(2.372,92)(4.211,160)(6.044,255)(8.202,445)(10.611,792)(11.719,1046)(12.011,1298)};

\addplot[
    color=blue,
    mark=pentagon,
    mark size=3pt,
    line width=0.5mm,
    ]
    coordinates {
   (1.847,67)(4.430,76)(7.409,85)(11.013,119)(13.871,159)(17.419,248)(19.932,467)(20.341,634)
   };
   
\addplot[
    color=black,
    mark=triangle,
    mark size=3pt,
    line width=0.5mm,
    ]
    coordinates {
   (1.593,58)(3.330,77)(5.409,109)(7.894,192)(9.998,377)(12.305,701)(13.443,921)(13.694,1102)
   };
   
\addplot[
    color=violet,
    mark=*,
    mark size=3pt,
    line width=0.5mm,
    ]
    coordinates {
   (0.088,6)(0.667,7)(2.230,10)(3.092,13)(4.081,19)(4.720,24)(5.581,35)(5.822,40)(5.883,94)};
   
\addplot[
    color=magenta,
    mark=o,
    mark size=3pt,
    line width=0.5mm,
    ]
    coordinates {
   (0.105,6)(0.913,7)(2.911,10)(4.211,14)(5.703,20)(7.014,26)(7.913,38)(8.302,44)(8.511,73)};
   
\addplot[
    color=purple,
    mark=otimes,
    mark size=3pt,
    line width=0.5mm,
    ]
    coordinates {
   (0.293,5)(2.304,7)(7.341,10)(11.193,14)(14.293,20)(16.024,24)(16.840,36)(17.103,42)(17.205,92)(17.403,154)};

\end{axis}
\end{tikzpicture}
\end{minipage}
\caption{Workloads with (a) $10\%$, (b) $50\%$, (c) $90\%$ cross-shard cross-enterprise transactions}
\label{fig:cross-SHE}
\end{figure}
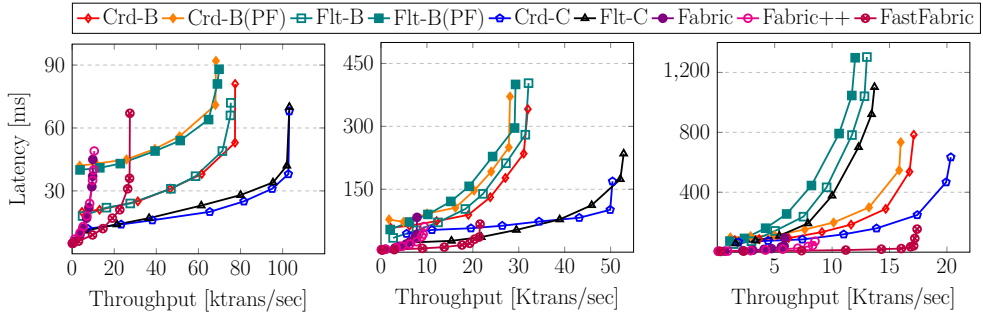

Increasing the percentage of cross-enterprise transactions to $50\%$, as shown in Figure~\ref{fig:cross-E}(b),
reduces the throughput of all protocols.
This is expected because a higher percentage of transactions requires cross-enterprise consensus.
In this experiment, {\sf\small Flt-B} processes $52$ ktps with $230$ ms latency
while {\sf\small Crd-B} processes $43$ ktps ($18\%$ lower) with $130$ ms latency ($44\%$ lower).
This shows a trade-off between the number of communication phases and the quorum size.
While the coordinator-based approach requires more phases of message passing (resulting in lower throughput),
the quorum size of the flattened approach is larger, i.e., all nodes communicate with each other (resulting in higher latency).
Using the privacy firewall mechanism, 
the throughput of {\sf\small Crd-B(PF)} and {\sf\small Flt-B(PF)} is increased by $5\%$ and $7\%$ respectively
and their latency is increased by only $9\%$ and $6\%$ 
(compared to {\sf\small Crd-B} and {\sf\small Flt-B})
before the end-to-end throughput is saturated.
These results demonstrate that as the ordering routine becomes heavily loaded,
the overhead of using the privacy firewall mechanism is alleviated.

The performance of {\sf\small Crd-C} is significantly better than {\sf\small Crd-B}
(i.e., $23\%$ higher throughput, $39\%$ lower latency) in this scenario.
This difference, however, is not remarkable in the flattened protocols.
The reason is that in the coordinator-based protocol, consensus takes place
within each cluster using the internal consensus protocols (Paxos in this case).
However, in the flattened protocol, as shown in Figure~\ref{fig:decentralized}(a),
there is no internal consensus within clusters and
a BFT protocol across enterprises establishes agreement.

The performance of Fabric is also affected by increasing the percentage of cross-enterprise transactions due to
(1) the overhead of hashing techniques and (2) conflicting transactions
\cite{amiri2019parblockchain,gorenflo2019fastfabric,gorenflo2020xox,sharma2019blurring}.
Interestingly, the throughput gap between Fabric and Fabric++ is increased to $18\%$ (from $3\%$) in this scenario due
to early abort and reordering mechanisms presented in Fabric++.

With $90\%$ cross-enterprise transactions, as shown in Figure~\ref{fig:cross-E}(c),
the latency of {\sf\small Flt-B} becomes very high ($680$ ms) due to its $O(n^2)$ 
message communication.
The performance of {\sf\small Flt-C} and {\sf\small Flt-B} becomes close to each other since
in both cases, a BFT protocol is used for cross-enterprise consensus.
In this experiment, FastFabric demonstrates the lowest latency ($35\%$ lower than {\sf\small Crd-C}
with $18$ ktps) as it does not suffer from the overhead of message communication across clusters.

\subsection{Cross-Shard Intra-Enterprise Transactions}\label{sec:cshardienterprise}

In the second set of experiments, we measure the performance of
coordinator-based and flattened
cross-shard intra-enterprise protocols.
Compared to the first set, these two protocols, in general, demonstrate lower throughput and higher latency due to
the need for establishing consensus in {\em multiple} data shards.
With Byzantine nodes and with $10\%$ cross-shard transactions, as shown in Figure~\ref{fig:cross-shard}(a),
the performance of {\sf\small Crd-B} is still close to {\sf\small Flt-B}.
However, by increasing the percentage of cross-shard transactions to $50\%$,
{\sf\small Flt-B} shows $20\%$ higher throughput.
In this set of experiments, since all shards belong to a single enterprise, as explained in Section~\ref{sec:deccrossshard},
{\sf\small Flt-C} is implemented as a CFT protocol, and,
as shown in Figure~\ref{fig:cross-shard}(a)-(c), has the best performance in all three workloads.
Similar to the previous section, the overhead of using the privacy firewall mechanism is alleviated when
the ordering nodes become highly loaded, e.g., the gap between the latency of {\sf\small Flt-B(PF)} and {\sf\small Flt-B}
is reduced from $25\%$ to $4\%$ by increasing the percentage of cross-shard transactions from $10\%$ to $90\%$.
Since enterprises maintain their data on a single data shard,
Fabric demonstrates the same performance in all three workloads, which is significantly worse than \sys.
However, with $90\%$ cross-shard transactions, similar to cross-enterprise transactions,
FastFabric shows the lowest latency because only one cluster orders all transactions.

\subsection{Cross-Shard Cross-Enterprise Transactions}\label{sec:cshardcenterprise}

\begin{figure}[t!]
\centering
\LARGE
\begin{minipage}{.25\textwidth}\centering
\begin{tikzpicture}[scale=0.49]
\begin{axis}[
    xlabel={Throughput [ktrans/sec]},
    ylabel={Latency [ms]},
    xmin=0, xmax=75,
    ymin=0, ymax=120,
    xtick={0,20,40,60},
    ytick={30,60,90},
    legend columns=9,
    legend style={at={(0.65,1.05)},anchor=south west},
    ymajorgrids=true,
    grid style=dashed,
]

\addplot[
    color=red,
    mark=diamond,
    mark size=3pt,
    line width=0.5mm,
    ]
    coordinates {
    (4.403,37)(15.332,39)(24.398,41)(36.302,46)(47.352,53)(57.902,63)(64.862,70)(65.762,91)};

\addplot[
    color=orange,
    mark=diamond*,
    mark size=3pt,
    line width=0.5mm,
    ]
    coordinates {
    (4.034,47)(13.803,49)(22.182,50)(34.051,55)(45.292,62)(55.203,71)(63.082,78)(63.803,97)};
   
\addplot[
    color=teal,
    mark=square,
    mark size=3pt,
    line width=0.5mm,
    ]
    coordinates {
    (3.243,26)(12.101,30)(20.981,34)(32.084,40)(41.548,50)(50.761,60)(56.287,80)(57.201,103)};

\addplot[
    color=teal,
    mark=square*,
    mark size=3pt,
    line width=0.5mm,
    ]
    coordinates {
    (2.637,46)(9.903,50)(17.382,52)(27.834,58)(35.911,67)(44.014,76)(51.109,95)(51.334,105)};
   
\addplot[
    color=blue,
    mark=pentagon,
    mark size=3pt,
    line width=0.5mm,
    ]
    coordinates {
   (5.643,29)(15.693,32)(27.903,36)(41.634,40)(50.904,44)(59.594,51)(71.193,62)(71.945,85)};

\addplot[
    color=black,
    mark=triangle,
    mark size=3pt,
    line width=0.5mm,
    ]
    coordinates {
    (4.132,20)(15.302,23)(24.309,26)(37.873,33)(48.772,39)(59.092,49)(67.872,69)(68.772,89)};

\addlegendentry{Crd-B}
\addlegendentry{Crd-B(PF)}
\addlegendentry{Flt-B}
\addlegendentry{Flt-B(PF)}
\addlegendentry{Crd-C}
\addlegendentry{Flt-C}

\end{axis}
\end{tikzpicture}
\end{minipage}\hspace{0.1em}
\begin{minipage}{.25\textwidth}\centering
\vspace{0.9em}
\begin{tikzpicture}[scale=0.49]
\begin{axis}[
    xlabel={Throughput [ktrans/sec]},
    xmin=0, xmax=95,
    ymin=0, ymax=120,
    xtick={0,20,40,60,80},
    ytick={30,60,90},
    ymajorgrids=true,
    grid style=dashed,
]

\addplot[
    color=red,
    mark=diamond,
    mark size=3pt,
    line width=0.5mm,
    ]
    coordinates {
    (3.94,38)(14.542,41)(23.878,45)(35.982,49)(46.352,56)(56.902,67)(62.439,74)(63.321,96)};

\addplot[
    color=orange,
    mark=diamond*,
    mark size=3pt,
    line width=0.5mm,
    ]
    coordinates {
    (3.182,58)(11.792,61)(18.283,64)(30.192,67)(39.452,73)(49.332,83)(56.013,90)(56.291,101)};
    
\addplot[
    color=teal,
    mark=square,
    mark size=3pt,
    line width=0.5mm,
    ]
    coordinates {
    (3.137,28)(12.091,32)(20.461,36)(31.732,43)(41.211,53)(50.321,63)(56.033,84)(57.032,106)};

\addplot[
    color=teal,
    mark=square*,
    mark size=3pt,
    line width=0.5mm,
    ]
    coordinates {
    (2.671,49)(9.802,51)(16.801,54)(25.913,60)(36.011,69)(44.241,78)(51.329,98)(51.421,110)};
   
\addplot[
    color=blue,
    mark=pentagon,
    mark size=3pt,
    line width=0.5mm,
    ]
    coordinates {
   (5.643,29)(14.794,33)(27.903,37)(40.872,41)(49.893,46)(58.883,53)(69.542,66)(71.302,92)};
  
\addplot[
    color=black,
    mark=triangle,
    mark size=3pt,
    line width=0.5mm,
    ]
    coordinates {
    (5.132,15)(17.302,17)(32.309,20)(63.873,25)(73.772,30)(84.092,40)(90.872,51)(91.772,79)};
 
\end{axis}
\end{tikzpicture}
\end{minipage}\hspace{0.1em}
\begin{minipage}{.25\textwidth}\centering
\vspace{0.9em}
\begin{tikzpicture}[scale=0.49]
\begin{axis}[
    xlabel={Throughput [ktrans/sec]},
    xmin=0, xmax=72,
    ymin=0, ymax=200,
    xtick={0,20,40,60},
    ytick={50,100,150},
    ymajorgrids=true,
    grid style=dashed,
]

\addplot[
    color=red,
    mark=diamond,
    mark size=3pt,
    line width=0.5mm,
    ]
    coordinates {
    (3.043,40)(13.111,45)(21.902,49)(33.892,53)(43.943,60)(53.877,72)(57.544,80)(58.322,112)};
  
\addplot[
    color=orange,
    mark=diamond*,
    mark size=3pt,
    line width=0.5mm,
    ]
    coordinates {
    (2.481,62)(10.891,65)(18.034,66)(28.903,70)(38.023,77)(46.922,88)(52.001,96)(52.402,113)};

\addplot[
    color=teal,
    mark=square,
    mark size=3pt,
    line width=0.5mm,
    ]
    coordinates {
    (2.137,43)(9.091,47)(14.461,52)(19.732,57)(24.211,61)(35.321,87)(42.033,142)(43.032,179)};

\addplot[
    color=teal,
    mark=square*,
    mark size=3pt,
    line width=0.5mm,
    ]
    coordinates {
    (1.816,63)(7.235,66)(12.032,70)(17.201,74)(21.902,77)(31.832,103)(38.651,152)(38.913,169)};
   
\addplot[
    color=blue,
    mark=pentagon,
    mark size=3pt,
    line width=0.5mm,
    ]
    coordinates {
   (5.399,30)(14.301,34)(27.109,38)(40.394,43)(49.594,48)(58.002,56)(68.841,69)(70.109,95)
    };

\addplot[
    color=black,
    mark=triangle,
    mark size=3pt,
    line width=0.5mm,
    ]
    coordinates {
    (4.132,33)(14.302,37)(19.309,43)(24.873,49)(32.772,58)(40.092,73)(48.872,119)(50.772,148)};
 
\end{axis}
\end{tikzpicture}
\end{minipage}
\caption{Scalability over spatial domains with
(a) $10\%$ intra-shard cross-enterprise,
(b) $10\%$ cross-shard intra-enterprise,
(c) $10\%$ cross-shard cross-enterprise transactions}
\label{fig:geoscalability}
\end{figure}
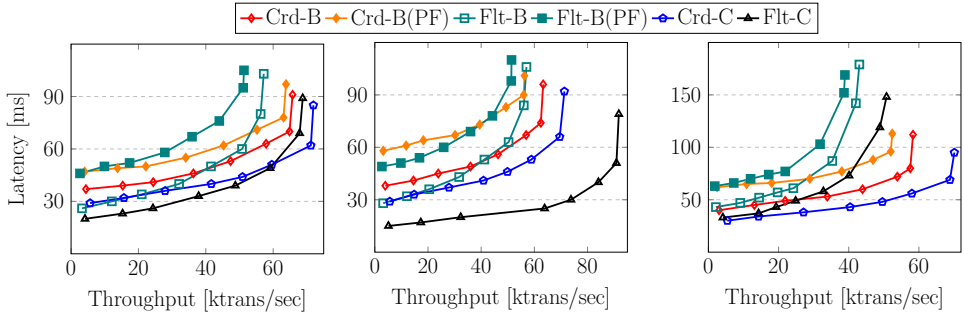

In workloads consisting of cross-shard cross-enterprise transactions,
as shown in Figure~\ref{fig:cross-SHE}(a)-(c),
the coordinator-based protocol shows better performance, especially with a higher percentage of cross-cluster transactions.
In particular, with $90\%$ cross-cluster transactions,
{\sf\small Flt-B}
demonstrates $24\%$ lower throughput and $93\%$ higher latency compared to {\sf\small Crd-B}.
This is expected because cross-shard cross-enterprise transactions,
as shown in Figures~\ref{fig:centralized}(c) and \ref{fig:decentralized}(c),
require consensus among
multiple clusters of multiple enterprises. As a result, the all to all communication phases of
the flattened protocol result in high latency.
Note that since clusters belong to different enterprises, even if nodes follow the crash failure model,
a BFT across enterprises needs to establish agreement, hence,
the performance is not significantly improved, i.e., 
{\sf\small Flt-C} processes only $13.4$ k transaction with $920$ ms latency.
Since cross-shard transactions do not affect the performance of Fabric,
it demonstrates similar performance as intra-shard cross-enterprise transactions.

\subsection{Scalability Across Spatial Domains}\label{sec:geoscalability}

In the next set of experiments, we measure the impact of 
network distance on the performance of the protocols.
Clusters of different enterprises are  distributed over four different AWS regions, i.e.,
Tokyo ({\em TY}), Seoul ({\em SU}), Virginia ({\em VA}), and California ({\em CA}) with
the average Round-Trip Time (RTT):
{\em TY} $\rightleftharpoons$ {\em SU}: $33$ ms,
{\em TY} $\rightleftharpoons$ {\em VA}: $148$ ms,
{\em TY} $\rightleftharpoons$ {\em CA}: $107$ ms,
{\em SU} $\rightleftharpoons$ {\em VA}: $175$ ms,
{\em SU} $\rightleftharpoons$ {\em CA}: $135$ ms, and
{\em VA} $\rightleftharpoons$ {\em CA}: $62$ ms.
We consider workloads with {\em $90\%$ internal} transactions (the typical setting in partitioned databases \cite{taft2014store})
and repeat the previous experiments, i.e.,
intra-shard cross-enterprise transactions (Figure~\ref{fig:cross-E}(a)),
cross-shard intra-enterprise transactions (Figure~\ref{fig:cross-shard}(a)), and
cross-shard cross-enterprise transactions (Figure~\ref{fig:cross-SHE}(a)).
Since in Fabric, Fabric++, and FastFabric, all transactions are ordered by the same set of orderers, placing
endorser nodes that execute transactions of different enterprises and orderer nodes in different locations far from each other
is not plausible. Hence, we do not perform this set of experiments for Fabric and its variants.

With $10\%$ intra-shard cross-enterprise transactions (Figure~\ref{fig:geoscalability}(a)),
{\sf\small Flt-B}
demonstrates higher latency
due to the message complexity of the protocol that requires all nodes to multicast messages to each other
over a wide area network.
With $10\%$ cross-shard intra-enterprise transactions (Figure~\ref{fig:geoscalability}(b)),
{\sf\small Flt-C} demonstrates the best performance
because clusters belong to the same enterprise and \sys processes transactions using a CFT protocol.
Finally, With $10\%$ cross-shard cross-enterprise transactions (Figure~\ref{fig:geoscalability}(c)),
the coordinator-based protocols show better performance because of the
two all to all communication phases of the flattened protocols that take place among distant clusters.
Compared to the single datacenter setting, the overhead of using the privacy firewall mechanism is also reduced.
With $10\%$ intra-shard cross-enterprise transactions,
{\sf\small Crd-B(PF)} shows $3\%$ lower throughput and $10\%$ higher latency Compared to {\sf\small Crd-B}
while with the same workload and in the single datacenter setting,
{\sf\small Crd-B(PF)} demonstrates $6\%$ lower throughput and $20\%$ higher latency.

\subsection{Varying the number of Enterprises}\label{sec:enteprise}

\begin{table}[t]
\centering
\caption{Performance with different number of Enterprises}
\label{tbl:ent}
\begin{tabular}{c|cc|cc|cc|cc}
         & \multicolumn{2}{|c|}{2 Enterprises} & \multicolumn{2}{|c|}{4 Enterprises} & \multicolumn{2}{|c|}{6 Enterprises} & \multicolumn{2}{c}{8 Enterprises}  \\
Protocols  & T[tps] & L[ms]  & T[tps] & L[ms]  & T[tps] & L[ms]  & T[tps] & L[ms]    \\ \hline
Crd-B      &  42023 &   45     & 79213  &   52   & 108672 &   53   & 141502 & 56 \\
Crd-B(PF)  &    34011     &    65      &  72629 &  68     &    99810    &    70    &   129402      & 75\\
Flt-B      &  44209 &  58      & 81111  &   60   & 116559 &   61   & 154951 &  64 \\
Flt-B(PF)  &   36091      &   70     &    73530     &    74    &      105023   &   77     &    143509     & 79 \\
Crd-C      &  53581 &  33      & 104511  &   39   & 144311 &  41    & 189311 & 41 \\
Flt-C      & 59302  &   31     & 108192 &   36   & 161864 &   37   & 214729 & 36 \\ \hline
\end{tabular}
\end{table}

We next measure the performance of different protocols by varying the number of enterprises
from $2$ to $8$.
The workloads include
$90\%$ internal and $10\%$ cross-cluster transactions (typical workload).
We do not report the results for Fabric and its variants because the ordering routine in those systems is the bottleneck,
and increasing the number of enterprises does not significantly affect their performance.
As shown in Table~\ref{tbl:ent},
by increasing the number of enterprises (clusters)
the throughput of all protocols is increased almost linearly.
This is expected because $90\%$ of transactions are internal where for such
transactions, the throughput of the entire system will increase linearly with the increasing number of clusters.
In addition, since cross-cluster transactions access two clusters, by increasing the number of clusters,
the chance of parallel processing of such transactions is increased.
With $8$ enterprises, {\sf\small Flt-C} processes $214$ ktps with $36$ ms latency.

\begin{table}[]
\centering
\caption{Performance with faulty nodes}
\label{tbl:faulty}
\begin{tabular}{c|c|c|c|c|c|c|c|c|c|c}
     \multicolumn{1}{c}{}     &  & Crd-B & \pbox{1cm}{Crd-B \\ (PF)} & Flt-B & \pbox{1cm}{Flt-B \\ (PF)} & Crd-C & Flt-C & Fabric & \pbox{0.8cm}{Fabric \\ ++} & \pbox{1cm}{Fast \\ Fabric}  \\ \hline
\parbox[t]{5mm}{\multirow{2}{*}{\rotatebox[origin=c]{90}{\pbox{1cm}{{\tiny Throu.} \\ (tps)}}}} & no fail    &  79213 & 72629  & 81111 & 73530 & 104511 & 108192 & 9743 & 10212 & 28142   \\
& $1$ fail      & 69541 & 64032 & 67023 & 61672 & 98032 & 97034 & 9211 &  9661 & 26392  \\ \hline
\parbox[t]{5mm}{\multirow{2}{*}{\rotatebox[origin=c]{90}{\pbox{1cm}{{\tiny Laten.} \\ (ms)} }}} & no fail     &  52 & 68  & 60 & 74 & 39 & 36 & 37 & 40 &  35  \\
& $1$ fail      & 56 &  74 & 68 & 79 & 40 & 38 & 38 & 42 & 38   \\ \hline

\end{tabular}
\end{table}

\subsection{Performance with Faulty Nodes}\label{sec:faulty}
In this set of experiments, we force a non-primary ordering node to fail  ($f=1$) and
repeat the first set of experiments.
In case of {\sf\small Crd-B(PF)} and {\sf\small Flt-B(PF)},
one ordering node, one execution node and one filter are failed.
As can be seen in Table~\ref{tbl:faulty}, since all protocols are pessimistic and quorums can be constructed
even if $f$ nodes (in our experiments, $1$) fail,
the failure of a single node does not reduce the performance of different systems significantly.
For example, {\sf\small Crd-B(PF)} incurs $12\%$ throughput reduction and $9\%$ higher latency.

\subsection{Varying the Degree of Contention}\label{sec:contention}

\begin{figure}[t]
\centering
\Large
\begin{minipage}{.33\textwidth} \centering
\begin{tikzpicture}[scale=0.57]
    \begin{axis}[
       ybar=2*\pgflinewidth,
        bar width=0.17cm,
        ymajorgrids = true,
        grid style=dashed,
        xlabel={Zipfian skewness},
        ylabel={Throughput [ktrans/sec]},
        symbolic x coords={0,1,2},
        xtick = data,
        scaled y ticks = false,
        enlarge x limits=0.25,
        ymin=0,
        legend columns=9,
        legend style={at={(-0.2,1.05)},anchor=south west},
    ]
    
    \addplot[style={black,pattern color=red,pattern = crosshatch}]
    coordinates {(0,79.213)(1,68.309)(2,56.110)};
 
    \addplot[style={orange,fill=orange,mark=none}]
    coordinates {(0,72.629)(1,62.012)(2,52.591)};

    \addplot[style={black,pattern color=teal,pattern = crosshatch}]
    coordinates {(0,81.111)(1,64.112)(2,52.402)};
    
    \addplot[style={teal,fill=teal,mark=none}]
    coordinates {(0,73.530)(1,57.203)(2,46.432)};
    
    \addplot[style={blue,fill=blue,mark=none}]
    coordinates {(0,104.511)(1,90.13)(2,81.023)};
   
    \addplot[style={black,pattern color=black,pattern = horizontal lines}]
    coordinates {(0,108.199)(1,89.502)(2,80.33)};

    \addplot[style={violet,fill=violet,mark=none}]
    coordinates {(0,9.743)(1,3.711)(2,0.872)};
    
    \addplot[style={black,pattern color=magenta,pattern = crosshatch}]
    coordinates {(0,10.212)(1,6.503)(2,4.342)};
    
    \addplot[style={purple,fill=purple,mark=none}]
    coordinates {(0,28.142)(1,7.239)(2,2.172)};

\addlegendentry{Crd-B}
\addlegendentry{Crd-B(PF)}
\addlegendentry{Flt-B}
\addlegendentry{Flt-B(PF)}
\addlegendentry{Crd-C}
\addlegendentry{Flt-C}
\addlegendentry{Fabric}
\addlegendentry{Fabric++}
\addlegendentry{FastFabric}

    \end{axis}
\end{tikzpicture}
\vspace{0.1em}
\end{minipage}\hspace{0.3em}
\begin{minipage}{.33\textwidth} \centering
\begin{tikzpicture}[scale=0.57]
    \begin{axis}[
       ybar=2*\pgflinewidth,
        bar width=0.17cm,
        ymajorgrids = true,
        grid style=dashed,
        xlabel={Zipfian skewness},
        ylabel={Latency [ms]},
        symbolic x coords={0,1,2},
        xtick = data,
        scaled y ticks = false,
        enlarge x limits=0.25,
        ymin=0, 
    ]
    
    \addplot[style={black,pattern color=red,pattern = crosshatch}]
    coordinates {(0,52)(1,57)(2,63)};
 
    \addplot[style={orange,fill=orange,mark=none}]
    coordinates {(0,68)(1,75)(2,78)};

    \addplot[style={black,pattern color=teal,pattern = crosshatch}]
    coordinates {(0,60)(1,70)(2,77)};
    
    \addplot[style={teal,fill=teal,mark=none}]
    coordinates {(0,74)(1,86)(2,93)};
    
    \addplot[style={blue,fill=blue,mark=none}]
    coordinates {(0,39)(1,47)(2,52)};
   
    \addplot[style={black,pattern color=black,pattern = horizontal lines}]
    coordinates {(0,36)(1,44)(2,50)};

    \addplot[style={violet,fill=violet,mark=none}]
    coordinates {(0,37)(1,38)(2,38)};
    
    \addplot[style={black,pattern color=magenta,pattern = crosshatch}]
    coordinates {(0,40)(1,45)(2,51)};
    
    \addplot[style={purple,fill=purple,mark=none}]
    coordinates {(0,35)(1,36)(2,37)};

    \end{axis}
\end{tikzpicture}
\end{minipage}
\caption{Performance with different Zipfian skewness}
  \label{fig:batching}
\end{figure}
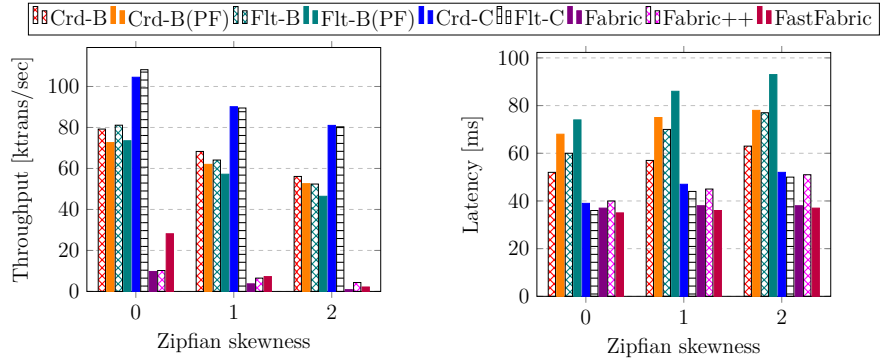

We next measure the impact of conflicting transactions on the performance of different systems.
We use three workloads with Zipfian skewness equal to $0$ (uniform distribution) $1$ and $2$ (high contention).
The workloads include {\em $90\%$ internal} and $10\%$ cross-cluster transactions.
Increasing the degree of contention slightly affects the performance of different protocols in \sys.
This is because transactions are first ordered and then executed and the execution is performed sequentially.
The performance of Fabric, however, reduces significantly (i.e., $91\%$ throughput reduction)
by increasing the workload contention. This is because in Fabric, transactions are simulated on the same state and
if two transactions write the same record, one of them needs to be invalidated during the validation phase.
FastFabric incurs a similar throughput reduction ($92\%$) as it follows the workflow of Fabric.
Fabric++, however, reorders transactions to resolve w-r conflicts and early aborts transactions with w-w conflicts.
As a result, Fabric++ incurs $58\%$ throughput reduction by increasing the skewness from $0$ to $2$.

\section{Related Work}
\label{sec:related}

A {\em permissioned} blockchain system, e.g., \cite{kwon2014tendermint,morgan2016quorum,androulaki2018hyperledger,nathan2019blockchain,gorenflo2019fastfabric,gorenflo2020xox,sharma2019blurring,istvan2018streamchain,ruan2020transactional,agarwal2020consentio,qi2021bidl}
consists of a set of known, identified but possibly untrusted participants
(permissioned blockchains are analyzed in several surveys and empirical studies  
\cite{dinh2018untangling,cachin2017blockchain,chacko2021my,ruan2021blockchains,dinh2017blockbench,sit2021experimental,sit2021experimental,amiri2021permissioned}).
Several blockchains support confidential transactions in both the permissioned \cite{cecchetti2017solidus,narula2018zkledger}
and permissionless \cite{hopwood2016zcash,bunz2020zether} settings.
These systems, however, are not scalable and
only support simple, financial transfers, not more complex database updates.
Hyperledger Fabric~\cite{androulaki2018hyperledger} stores
confidential data in private data collections \cite{fabric2018collections}
replicated on authorized enterprises.
The hash of all private transactions, however, is appended to the single global ledger of every node
resulting in low performance.
Several variants of Fabric, e.g.,
Fast Fabric \cite{gorenflo2019fastfabric}, 
XOX Fabric  \cite{gorenflo2020xox},
Fabric++ \cite{sharma2019blurring}, and
FabricSharp \cite{ruan2020transactional},
have been presented to improve its performance (as demonstrated in Section~\ref{sec:eval}).
Such systems, however, do not address the confidential data leakage and consistency challenges of Fabric.
Fabric also can be combined with secure multiparty computation \cite{benhamouda2019supporting}
but this does not address performance issues.
Caper~\cite{amiri2019caper} supports private internal and public global transactions of collaborative enterprises.
Caper, however, does not address transactions among a subset of enterprises,
consistency across collaboration workflows, confidential data leakage prevention,
and multi-shard enterprises.

The zero-knowledge proof techniques used in Quorum \cite{morgan2016quorum} to support
private transactions have also considerable overhead \cite{androulaki2018hyperledger} especially in
an environment where most transactions might be local.
Enterprises can also maintain their independent blockchains and
use cross-chain transactions \cite{herlihy2018atomic,zakhary2020atomic}
for confidential cross-enterprise collaboration.
Such techniques, however, are often costly, complex, and mainly designed for permissionless blockchains.
Hawk~\cite{kosba2016hawk}, Arbitrum~\cite{kalodner2018arbitrum}, Ekiden~\cite{cheng2019ekiden},
Zkay \cite{steffen2019zkay} and zexe \cite{bowe2020zexe} are other blockchains that
support confidential transactions at the application layer.
These systems do not address the consensus and data layers that are the focus of \sys.
To support confidential transactions, e.g. Hawk~\cite{kosba2016hawk} (which requires a semi-trusted manager),
Arbitrum~\cite{kalodner2018arbitrum} (which allows users to delegate managers, who can execute contracts in an off-chain Virtual Machine), Ekiden~\cite{cheng2019ekiden} (which relies on  trusted hardware) or
Zkay \cite{steffen2019zkay} and zexe \cite{bowe2020zexe} (which use Zero-Knowledge proofs) have been proposed.
These systems focus on the application layer and do not attempt to address the consensus and data layers that are the focus of \sys.
Hence, these systems are complementary to \sys, and some of these tools could be adapted to provide private contracting capabilities on top of the \sys ledger.
Related to blockchain confidentiality, tracking handover of calls \cite{ma2021fraud}, secure data provenance \cite{ruan2019fine}, verifiable data sharing \cite{el2019blockchaindb}, tamper evidence storage \cite{wang2018forkbase},
global asset management \cite{zakhary2019towards},
verifiable query processing \cite{xu2019vchain,peng2020falcondb,zhang2021authenticated},
on-chain secret management \cite{kokoris2020calypso} and
private membership attestation \cite{xu2021div}
have been studied.

Data sharding techniques are used in distributed databases
\cite{kallman2008h,thomson2012calvin,corbett2013spanner,glendenning2011scalable,baker2011megastore,decandia2007dynamo,bronson2013tao,taft2014store}
with crash-only nodes and
in permissioned blockchain systems, e.g.,
AHL \cite{dang2018towards}, Blockplane \cite{nawab2019blockplane}, chainspace \cite{al2017chainspace},
SharPer \cite{amiri2021sharper}, and Saguaro \cite{amiri2021saguaro}
with Byzantine nodes.

AHL \cite{dang2018towards} uses a similar technique as
permissionless blockchains Elastico \cite{luu2016secure},
OmniLedger \cite{kokoris2018omniledger}, and Rapidchain \cite{zamani2018rapidchain},
provides probabilistic safety guarantees, and
processes cross-shard transactions in a centralized manner.
SharPer \cite{amiri2021sharper} 
in contrast to AHL, guarantees deterministic safety and
processes transactions in a flattened manner.
Scalability over spatial domains has also been studied in
ResilientDB \cite{gupta2020resilientdb,gupta2020permissioned},
and Saguaro \cite{amiri2021saguaro}.
\sys, in contrast to all these systems, supports multi-enterprise environments and guarantees
confidentiality.
\section{Conclusion}
\label{sec:conc}

In this paper, we proposed \sys, a permissioned blockchain system to support the
scalability and confidentiality requirements of multi-enterprise applications.
To guarantee collaboration confidentiality,
\sys constructs a hierarchical data model consisting of a set of data collections for each collaboration workflow.
Every subset of enterprises is able to
form a confidential collaboration private from other enterprises and execute
transactions on a private data collection shared between only the involved enterprises.
To prevent confidential data leakage, \sys utilizes the privacy firewall mechanism \cite{yin2003separating}.
To support scalability, each enterprise partitions its data into different data shards. 
\sys further presents a transaction ordering scheme that enforces only the  necessary and sufficient constraints
to guarantee data consistency.
Finally, a suite of consensus protocols is presented to
process different types of intra-shard and cross-shard transactions within and across enterprises.
Our experimental results clearly demonstrate the scalability of \sys in comparison to Fabric and its variants.
Moreover, while coordinator-based protocols demonstrate better performance in cross-enterprise transactions,
flattened protocols show higher performance in cross-shard transactions.

\section*{Acknowledgement}
This work is funded by NSF grants CNS-1703560, CNS-1815733, and CNS-2104882 and by ONR grant 130.1303.4.573135.3000.2000.0292.

\bibliographystyle{abbrv}
\bibliography{_blockchain,_misc,_privacy,_system}

\begin{thebibliography}{100}

\bibitem{aborode2022fake}
A.~T. Aborode, W.~A. Awuah, S.~Talukder, A.~A. Oyeyemi, E.~P. Nansubuga,
  P.~Machai, H.~Tillewein, and C.~I. Oko.
\newblock Fake covid-19 vaccinations in africa.
\newblock {\em Postgraduate medical journal}, 98(1159):317--318, 2022.

\bibitem{agarwal2020consentio}
R.~R. Agarwal, D.~Kumar, L.~Golab, and S.~Keshav.
\newblock Consentio: Managing consent to data access using permissioned
  blockchains.
\newblock In {\em Int. Conf. on Blockchain and Cryptocurrency (ICBC)}, pages
  1--9. IEEE, 2020.

\bibitem{al2017chainspace}
M.~Al-Bassam, A.~Sonnino, S.~Bano, D.~Hrycyszyn, and G.~Danezis.
\newblock Chainspace: A sharded smart contracts platform.
\newblock In {\em Network and Distributed System Security Symposium (NDSS)},
  2018.

\bibitem{alam2021challenges}
S.~T. Alam, S.~Ahmed, S.~M. Ali, S.~Sarker, G.~Kabir, et~al.
\newblock Challenges to covid-19 vaccine supply chain: Implications for
  sustainable development goals.
\newblock {\em International Journal of Production Economics}, 239:108193,
  2021.

\bibitem{amankwah2022covid}
J.~Amankwah-Amoah.
\newblock Covid-19 and counterfeit vaccines: Global implications, new
  challenges and opportunities.
\newblock {\em Health Policy and Technology}, page 100630, 2022.

\bibitem{amiri2019caper}
M.~J. Amiri, D.~Agrawal, and A.~El~Abbadi.
\newblock Caper: a cross-application permissioned blockchain.
\newblock {\em Proc. of the VLDB Endowment}, 12(11):1385--1398, 2019.

\bibitem{amiri2019parblockchain}
M.~J. Amiri, D.~Agrawal, and A.~El~Abbadi.
\newblock Parblockchain: Leveraging transaction parallelism in permissioned
  blockchain systems.
\newblock In {\em Int. Conf. on Distributed Computing Systems (ICDCS)}, pages
  1337--1347. IEEE, 2019.

\bibitem{amiri2021permissioned}
M.~J. Amiri, D.~Agrawal, and A.~El~Abbadi.
\newblock Permissioned blockchains: Properties, techniques and applications.
\newblock In {\em SIGMOD Int. Conf. on Management of Data}, pages 2813--2820,
  2021.

\bibitem{amiri2021sharper}
M.~J. Amiri, D.~Agrawal, and A.~El~Abbadi.
\newblock Sharper: Sharding permissioned blockchains over network clusters.
\newblock In {\em SIGMOD Int. Conf. on Management of Data}, pages 76--88. ACM,
  2021.

\bibitem{amiri2021separ}
M.~J. Amiri, J.~Duguépéroux, T.~Allard, D.~Agrawal, and A.~El~Abbadi.
\newblock Separ: Separ: Towards regulating future of work multi-platform
  crowdworking environments with privacy guarantees.
\newblock In {\em Proceedings of The Web Conf. (WWW)}, pages 1891--1903, 2021.

\bibitem{amiri2021saguaro}
M.~J. Amiri, Z.~Lai, L.~Patel, B.~Thau~Loo, E.~Loo, and W.~Zhou.
\newblock Saguaro: Efficient processing of transactions in wide area networks
  using a hierarchical permissioned blockchain.
\newblock {\em arXiv preprint arXiv:2101.08819}, 2021.

\bibitem{androulaki2018hyperledger}
E.~Androulaki, A.~Barger, V.~Bortnikov, C.~Cachin, et~al.
\newblock Hyperledger fabric: a distributed operating system for permissioned
  blockchains.
\newblock In {\em European Conf. on Computer Systems (EuroSys)}, page~30. ACM,
  2018.

\bibitem{androulaki2018channels}
E.~Androulaki, C.~Cachin, A.~De~Caro, and E.~Kokoris-Kogias.
\newblock Channels: Horizontal scaling and confidentiality on permissioned
  blockchains.
\newblock In {\em European Symposium on Research in Computer Security
  (ESORICS)}, pages 111--131. Springer, 2018.

\bibitem{archer2018keys}
D.~W. Archer, D.~Bogdanov, Y.~Lindell, L.~Kamm, K.~Nielsen, J.~I. Pagter, N.~P.
  Smart, and R.~N. Wright.
\newblock From keys to databases—real-world applications of secure
  multi-party computation.
\newblock {\em The Computer Journal}, 61(12):1749--1771, 2018.

\bibitem{azaria2016medrec}
A.~Azaria, A.~Ekblaw, T.~Vieira, and A.~Lippman.
\newblock Medrec: Using blockchain for medical data access and permission
  management.
\newblock In {\em Int. Conf. on Open and Big Data (OBD)}, pages 25--30. IEEE,
  2016.

\bibitem{baker2011megastore}
J.~Baker, C.~Bond, J.~C. Corbett, J.~Furman, A.~Khorlin, J.~Larson, J.-M. Leon,
  Y.~Li, A.~Lloyd, and V.~Yushprakh.
\newblock Megastore: Providing scalable, highly available storage for
  interactive services.
\newblock In {\em Conf. on Innovative Data Systems Research (CIDR)}, 2011.

\bibitem{VaccineCounterfeit}
{BBC News}.
\newblock Coronavirus: Pfizer confirms fake versions of vaccine in poland and
  mexico.
\newblock \url{https://www.bbc.com/news/world-56844149}, April 2021.

\bibitem{benhamouda2019supporting}
F.~Benhamouda, S.~Halevi, and T.~Halevi.
\newblock Supporting private data on hyperledger fabric with secure multiparty
  computation.
\newblock {\em IBM Journal of Research and Development}, 63(2/3):3--1, 2019.

\bibitem{global2022liberty}
G.~Benigno, J.~di~Giovanni, J.~J. Groen, and A.~Noble.
\newblock Global supply chain pressure index: May 2022 update.
\newblock
  \url{https://libertystreeteconomics.newyorkfed.org/2022/05/global-supply-chain-pressure-index-may-2022-update/},
  2022.

\bibitem{bessani2013depsky}
A.~Bessani, M.~Correia, B.~Quaresma, F.~Andr{\'e}, and P.~Sousa.
\newblock Depsky: dependable and secure storage in a cloud-of-clouds.
\newblock {\em Transactions on Storage (TOS)}, 9(4):12, 2013.

\bibitem{bessani2008depspace}
A.~N. Bessani, E.~P. Alchieri, M.~Correia, and J.~S. Fraga.
\newblock Depspace: a byzantine fault-tolerant coordination service.
\newblock In {\em ACM SIGOPS/EuroSys European Conference on Computer Systems},
  pages 163--176, 2008.

\bibitem{bowe2020zexe}
S.~Bowe, A.~Chiesa, M.~Green, I.~Miers, P.~Mishra, and H.~Wu.
\newblock Zexe: Enabling decentralized private computation.
\newblock In {\em Symposium on Security and Privacy (SP)}, pages 947--964.
  IEEE, 2020.

\bibitem{bracha1985asynchronous}
G.~Bracha and S.~Toueg.
\newblock Asynchronous consensus and broadcast protocols.
\newblock {\em Journal of the ACM (JACM)}, 32(4):824--840, 1985.

\bibitem{bronson2013tao}
N.~Bronson, Z.~Amsden, G.~Cabrera, P.~Chakka, P.~Dimov, H.~Ding, J.~Ferris,
  A.~Giardullo, S.~Kulkarni, H.~Li, et~al.
\newblock Tao: Facebook's distributed data store for the social graph.
\newblock In {\em Annual Technical Conf. (ATC)}, pages 49--60. USENIX
  Association, 2013.

\bibitem{emergent2022bruggeman}
L.~Bruggeman and S.~Pezenik.
\newblock Emergent biosolutions discarded ingredients for 400 million covid-19
  vaccines, probe finds.
\newblock
  \url{https://abcnews.go.com/US/emergent-biosolutions-discarded-ingredients-400-million-covid-19/story?id=84604285},
  2022.

\bibitem{bunz2020zether}
B.~B{\"u}nz, S.~Agrawal, M.~Zamani, and D.~Boneh.
\newblock Zether: Towards privacy in a smart contract world.
\newblock In {\em Int. Conf. on Financial Cryptography and Data Security},
  pages 423--443. Springer, 2020.

\bibitem{cachin2017blockchain}
C.~Cachin and M.~Vukoli{\'c}.
\newblock Blockchain consensus protocols in the wild.
\newblock In {\em Int. Symposium on Distributed Computing (DISC)}, pages 1--16,
  2017.

\bibitem{castro1999practical}
M.~Castro, B.~Liskov, et~al.
\newblock Practical byzantine fault tolerance.
\newblock In {\em Symposium on Operating systems design and implementation
  (OSDI)}, volume~99, pages 173--186. USENIX Association, 1999.

\bibitem{cecchetti2017solidus}
E.~Cecchetti, F.~Zhang, Y.~Ji, A.~Kosba, A.~Juels, and E.~Shi.
\newblock Solidus: Confidential distributed ledger transactions via pvorm.
\newblock In {\em SIGSAC Conf. on Computer and Communications Security}, pages
  701--717. ACM, 2017.

\bibitem{chacko2021my}
J.~A. Chacko, R.~Mayer, and H.-A. Jacobsen.
\newblock Why do my blockchain transactions fail? a study of hyperledger
  fabric.
\newblock In {\em SIGMOD Int. Conf. on Management of Data}, pages 221--234.
  ACM, 2021.

\bibitem{morgan2016quorum}
J.~M. Chase.
\newblock Quorum white paper, 2016.

\bibitem{chen2020blockchain}
J.~Chen, T.~Cai, W.~He, L.~Chen, G.~Zhao, W.~Zou, and L.~Guo.
\newblock A blockchain-driven supply chain finance application for auto retail
  industry.
\newblock {\em Entropy}, 22(1):95, 2020.

\bibitem{cheng2019ekiden}
R.~Cheng, F.~Zhang, J.~Kos, W.~He, N.~Hynes, N.~Johnson, A.~Juels, A.~Miller,
  and D.~Song.
\newblock Ekiden: A platform for confidentiality-preserving, trustworthy, and
  performant smart contracts.
\newblock In {\em European Symposium on Security and Privacy (EuroS\&P)}, pages
  185--200. IEEE, 2019.

\bibitem{choudhary2021fake}
O.~P. Choudhary, Priyanka, I.~Singh, T.~A. Mohammed, and A.~J.
  Rodriguez-Morales.
\newblock Fake covid-19 vaccines: scams hampering the vaccination drive in
  india and possibly other countries.
\newblock {\em Human Vaccines \& Immunotherapeutics}, pages 1--2, 2021.

\bibitem{corbett2013spanner}
J.~C. Corbett, J.~Dean, M.~Epstein, A.~Fikes, et~al.
\newblock Spanner: Google's globally distributed database.
\newblock {\em Transactions on Computer Systems (TOCS)}, 31(3):8, 2013.

\bibitem{dang2018towards}
H.~Dang, T.~T.~A. Dinh, D.~Loghin, E.-C. Chang, Q.~Lin, and B.~C. Ooi.
\newblock Towards scaling blockchain systems via sharding.
\newblock In {\em SIGMOD Int. Conf. on Management of Data}. ACM, 2019.

\bibitem{decandia2007dynamo}
G.~DeCandia, D.~Hastorun, M.~Jampani, G.~Kakulapati, A.~Lakshman, A.~Pilchin,
  S.~Sivasubramanian, P.~Vosshall, and W.~Vogels.
\newblock Dynamo: amazon's highly available key-value store.
\newblock In {\em Operating Systems Review (OSR)}, volume~41, pages 205--220.
  ACM SIGOPS, 2007.

\bibitem{dinh2018untangling}
T.~T.~A. Dinh, R.~Liu, M.~Zhang, G.~Chen, B.~C. Ooi, and J.~Wang.
\newblock Untangling blockchain: A data processing view of blockchain systems.
\newblock {\em IEEE transactions on knowledge and data engineering},
  30(7):1366--1385, 2018.

\bibitem{dinh2017blockbench}
T.~T.~A. Dinh, J.~Wang, G.~Chen, R.~Liu, B.~C. Ooi, and K.-L. Tan.
\newblock Blockbench: A framework for analyzing private blockchains.
\newblock In {\em SIGMOD Int. Conf. on Management of Data}, pages 1085--1100.
  ACM, 2017.

\bibitem{duan2016practical}
S.~Duan and H.~Zhang.
\newblock Practical state machine replication with confidentiality.
\newblock In {\em Symposium on Reliable Distributed Systems (SRDS)}, pages
  187--196. IEEE, 2016.

\bibitem{el2019blockchaindb}
M.~El-Hindi, C.~Binnig, A.~Arasu, D.~Kossmann, and R.~Ramamurthy.
\newblock Blockchaindb: A shared database on blockchains.
\newblock {\em Proceedings of the VLDB Endowment}, 12(11):1597--1609, 2019.

\bibitem{evans2017pragmatic}
D.~Evans, V.~Kolesnikov, and M.~Rosulek.
\newblock A pragmatic introduction to secure multi-party computation.
\newblock {\em Foundations and Trends{\textregistered} in Privacy and
  Security}, 2(2-3), 2017.

\bibitem{fischer1985impossibility}
M.~J. Fischer, N.~A. Lynch, and M.~S. Paterson.
\newblock Impossibility of distributed consensus with one faulty process.
\newblock {\em Journal of the ACM (JACM)}, 32(2):374--382, 1985.

\bibitem{gabizon2019plonk}
A.~Gabizon and Z.~J. Williamson.
\newblock Plonk: Permutations over lagrange-bases for oecumenical
  noninteractive arguments of knowledge.
\newblock 2019.

\bibitem{glendenning2011scalable}
L.~Glendenning, I.~Beschastnikh, A.~Krishnamurthy, and T.~Anderson.
\newblock Scalable consistency in scatter.
\newblock In {\em Symposium on Operating Systems Principles (SOSP)}, pages
  15--28. ACM, 2011.

\bibitem{gorenflo2020xox}
C.~Gorenflo, L.~Golab, and S.~Keshav.
\newblock Xox fabric: A hybrid approach to transaction execution.
\newblock In {\em Int. Conf. on Blockchain and Cryptocurrency (ICBC)}, pages
  1--9. IEEE, 2020.

\bibitem{gorenflo2019fastfabric}
C.~Gorenflo, S.~Lee, L.~Golab, and S.~Keshav.
\newblock Fastfabric: Scaling hyperledger fabric to 20,000 transactions per
  second.
\newblock In {\em Int. Conf. on Blockchain and Cryptocurrency (ICBC)}, pages
  455--463. IEEE, 2019.

\bibitem{fastfabric2019}
C.~Gorenflo, S.~Lee, L.~Golab, and S.~Keshav.
\newblock Fastfabric: Scaling hyperledger fabric to 20,000 transactions per
  second.
\newblock {\em arXiv preprint arXiv:1901.00910}, 2019.

\bibitem{gupta2020resilientdb}
S.~Gupta, S.~Rahnama, J.~Hellings, and M.~Sadoghi.
\newblock Resilientdb: Global scale resilient blockchain fabric.
\newblock {\em Proceedings of the VLDB Endowment}, 13(6):868--883, 2020.

\bibitem{gupta2020permissioned}
S.~Gupta, S.~Rahnama, and M.~Sadoghi.
\newblock Permissioned blockchain through the looking glass: Architectural and
  implementation lessons learned.
\newblock In {\em Int. Conf. on Distributed Computing Systems (ICDCS)}, pages
  754--764. IEEE, 2020.

\bibitem{han2019fluid}
S.~Han, Z.~Xu, Y.~Zeng, and L.~Chen.
\newblock Fluid: A blockchain based framework for crowdsourcing.
\newblock In {\em SIGMOD Int. Conf. on Management of Data}, pages 1921--1924.
  ACM, 2019.

\bibitem{herlihy2018atomic}
M.~Herlihy.
\newblock Atomic cross-chain swaps.
\newblock In {\em Symposium on Principles of Distributed Computing (PODC)},
  pages 245--254. ACM, 2018.

\bibitem{hopwood2016zcash}
D.~Hopwood, S.~Bowe, T.~Hornby, and N.~Wilcox.
\newblock Zcash protocol specification.
\newblock {\em GitHub: San Francisco, CA, USA}, 2016.

\bibitem{fabric2018collections}
Hyperledger.
\newblock Private data collections: A high-level overview.
\newblock
  {https://www.hyperledger.org/blog/2018/10/23/private-data-collections-a-high-level-overview}.

\bibitem{istvan2018streamchain}
Z.~Istv{\'a}n, A.~Sorniotti, and M.~Vukoli{\'c}.
\newblock Streamchain: Do blockchains need blocks?
\newblock In {\em Workshop on Scalable and Resilient Infrastructures for
  Distributed Ledgers (SERIAL)}, pages 1--6. ACM, 2018.

\bibitem{kallman2008h}
R.~Kallman, H.~Kimura, J.~Natkins, A.~Pavlo, A.~Rasin, S.~Zdonik, E.~P. Jones,
  S.~Madden, M.~Stonebraker, Y.~Zhang, et~al.
\newblock H-store: a high-performance, distributed main memory transaction
  processing system.
\newblock {\em Proc. of the VLDB Endowment}, 1(2):1496--1499, 2008.

\bibitem{kalodner2018arbitrum}
H.~Kalodner, S.~Goldfeder, X.~Chen, S.~M. Weinberg, and E.~W. Felten.
\newblock Arbitrum: Scalable, private smart contracts.
\newblock In {\em USENIX Security Symposium)}, pages 1353--1370, 2018.

\bibitem{khan2021toward}
M.~Khan and A.~Babay.
\newblock Toward intrusion tolerance as a service: Confidentiality in partially
  cloud-based bft systems.
\newblock In {\em Int. Conf. on Dependable Systems and Networks (DSN)}, pages
  14--25. IEEE, 2021.

\bibitem{kokoris2020calypso}
E.~Kokoris-Kogias, E.~C. Alp, L.~Gasser, P.~Jovanovic, E.~Syta, and B.~Ford.
\newblock Calypso: Private data management for decentralized ledgers.
\newblock {\em Proceedings of the VLDB Endowment}, 14(4):586--599, 2020.

\bibitem{kokoris2018omniledger}
E.~Kokoris-Kogias, P.~Jovanovic, L.~Gasser, N.~Gailly, E.~Syta, and B.~Ford.
\newblock Omniledger: A secure, scale-out, decentralized ledger via sharding.
\newblock In {\em Symposium on Security and Privacy (SP)}, pages 583--598.
  IEEE, 2018.

\bibitem{korpela2017digital}
K.~Korpela, J.~Hallikas, and T.~Dahlberg.
\newblock Digital supply chain transformation toward blockchain integration.
\newblock In {\em Hawaii Int. Conf. on system sciences (HICSS)}, 2017.

\bibitem{kosba2016hawk}
A.~Kosba, A.~Miller, E.~Shi, Z.~Wen, and C.~Papamanthou.
\newblock Hawk: The blockchain model of cryptography and privacy-preserving
  smart contracts.
\newblock In {\em Symposium on security and privacy (SP)}, pages 839--858.
  IEEE, 2016.

\bibitem{kosba2018xjsnark}
A.~Kosba, C.~Papamanthou, and E.~Shi.
\newblock xjsnark: A framework for efficient verifiable computation.
\newblock In {\em 2018 IEEE Symposium on Security and Privacy (SP)}, pages
  944--961. IEEE, 2018.

\bibitem{kwon2014tendermint}
J.~Kwon.
\newblock Tendermint: Consensus without mining.
\newblock 2014.

\bibitem{lamport1978time}
L.~Lamport.
\newblock Time, clocks, and the ordering of events in a distributed system.
\newblock {\em Communications of the ACM}, 21(7):558--565, 1978.

\bibitem{lamport2001paxos}
L.~Lamport.
\newblock Paxos made simple.
\newblock {\em ACM Sigact News}, 32(4):18--25, 2001.

\bibitem{luu2016secure}
L.~Luu, V.~Narayanan, C.~Zheng, K.~Baweja, S.~Gilbert, and P.~Saxena.
\newblock A secure sharding protocol for open blockchains.
\newblock In {\em SIGSAC Conf. on Computer and Communications Security (CCS)},
  pages 17--30. ACM, 2016.

\bibitem{ma2021fraud}
S.~Ma, T.~Dasu, Y.~Kanza, D.~Srivastava, and L.~Xiong.
\newblock Fraud buster: Tracking irsf using blockchain while protecting
  business confidentiality.
\newblock In {\em CIDR}, 2021.

\bibitem{maller2019sonic}
M.~Maller, S.~Bowe, M.~Kohlweiss, and S.~Meiklejohn.
\newblock Sonic: Zero-knowledge snarks from linear-size universal and updatable
  structured reference strings.
\newblock In {\em Proceedings of the 2019 ACM SIGSAC Conference on Computer and
  Communications Security}, pages 2111--2128, 2019.

\bibitem{marsh2004codex}
M.~A. Marsh and F.~B. Schneider.
\newblock Codex: A robust and secure secret distribution system.
\newblock {\em Transactions on Dependable and secure Computing}, 1(1):34--47,
  2004.

\bibitem{narula2018zkledger}
N.~Narula, W.~Vasquez, and M.~Virza.
\newblock zkledger: Privacy-preserving auditing for distributed ledgers.
\newblock In {\em 15th $\{$USENIX$\}$ Symposium on Networked Systems Design and
  Implementation ($\{$NSDI$\}$ 18)}, pages 65--80, 2018.

\bibitem{nathan2019blockchain}
S.~Nathan, C.~Govindarajan, A.~Saraf, M.~Sethi, and P.~Jayachandran.
\newblock Blockchain meets database: design and implementation of a blockchain
  relational database.
\newblock {\em Proceedings of the VLDB Endowment}, 12(11):1539--1552, 2019.

\bibitem{nawab2019blockplane}
F.~Nawab and M.~Sadoghi.
\newblock Blockplane: A global-scale byzantizing middleware.
\newblock In {\em 2019 IEEE 35th Int. Conf. on Data Engineering (ICDE)}, pages
  124--135. IEEE, 2019.

\bibitem{ongaro2014search}
D.~Ongaro and J.~K. Ousterhout.
\newblock In search of an understandable consensus algorithm.
\newblock In {\em Annual Technical Conf. (ATC)}, pages 305--319. USENIX
  Association, 2014.

\bibitem{medical2021WHO}
W.~H. Organization.
\newblock Medical product alert n°5/2021: Falsified covishield vaccine.
\newblock
  \url{https://www.who.int/news/item/31-08-2021-medical-product-alert-n-5-2021-falsified-covishield-vaccine},
  2021.

\bibitem{world2017study}
W.~H. Organization et~al.
\newblock A study on the public health and socioeconomic impact of substandard
  and falsified medical products.
\newblock 2017.

\bibitem{VaccineTheft}
P.~Owen.
\newblock Vaccine supply under threat from theft and counterfeits.
\newblock
  \url{https://www.ttclub.com/news-and-resources/news/press-releases/2021/vaccine-supply-under-threat-from-theft-and-counterfeits/},
  March 2021.

\bibitem{padilha2011belisarius}
R.~Padilha and F.~Pedone.
\newblock Belisarius: Bft storage with confidentiality.
\newblock In {\em 2011 IEEE 10th International Symposium on Network Computing
  and Applications}, pages 9--16. IEEE, 2011.

\bibitem{covid21}
D.~Patterson.
\newblock Hackers are attacking the covid-19 vaccine supply chain.
\newblock
  \url{https://www.cbsnews.com/news/covid-19-vaccine-hackers-supply-chain/},
  2021.

\bibitem{peng2020falcondb}
Y.~Peng, M.~Du, F.~Li, R.~Cheng, and D.~Song.
\newblock Falcondb: Blockchain-based collaborative database.
\newblock In {\em SIGMOD Int. Conf. on Management of Data}, pages 637--652,
  2020.

\bibitem{peng2021p2b}
Z.~Peng, C.~Xu, H.~Wang, J.~Huang, J.~Xu, and X.~Chu.
\newblock P2b-trace: Privacy-preserving blockchain-based contact tracing to
  combat pandemics.
\newblock In {\em SIGMOD Int. Conf. on Management of Data}, pages 2389--2393,
  2021.

\bibitem{peng2021vfchain}
Z.~Peng, J.~Xu, X.~Chu, S.~Gao, Y.~Yao, R.~Gu, and Y.~Tang.
\newblock Vfchain: Enabling verifiable and auditable federated learning via
  blockchain systems.
\newblock {\em IEEE Transactions on Network Science and Engineering}, 2021.

\bibitem{qi2021bidl}
J.~Qi, X.~Chen, Y.~Jiang, J.~Jiang, T.~Shen, S.~Zhao, S.~Wang, G.~Zhang,
  L.~Chen, M.~H. Au, et~al.
\newblock Bidl: A high-throughput, low-latency permissioned blockchain
  framework for datacenter networks.
\newblock In {\em Symposium on Operating Systems Principles (SOSP)}, pages
  18--34. ACM SIGOPS, 2021.

\bibitem{fake2020reilly}
S.~Reilly, J.~Paladino, J.~Lambert, and M.~Stiles.
\newblock Fake vaccine cards are everywhere. it’s a public health nightmare.
\newblock
  \url{https://www.grid.news/story/science/2022/01/25/fake-vaccine-cards-are-everywhere-its-a-public-health-nightmare/},
  2022.

\bibitem{ruan2019fine}
P.~Ruan, G.~Chen, T.~T.~A. Dinh, Q.~Lin, B.~C. Ooi, and M.~Zhang.
\newblock Fine-grained, secure and efficient data provenance on blockchain
  systems.
\newblock {\em Proceedings of the VLDB Endowment}, 12(9):975--988, 2019.

\bibitem{ruan2021blockchains}
P.~Ruan, T.~T.~A. Dinh, D.~Loghin, M.~Zhang, G.~Chen, Q.~Lin, and B.~C. Ooi.
\newblock Blockchains vs. distributed databases: Dichotomy and fusion.
\newblock In {\em SIGMOD Int. Conf. on Management of Data}, pages 1504--1517,
  2021.

\bibitem{ruan2020transactional}
P.~Ruan, D.~Loghin, Q.-T. Ta, M.~Zhang, G.~Chen, and B.~C. Ooi.
\newblock A transactional perspective on execute-order-validate blockchains.
\newblock In {\em SIGMOD Int. Conf. on Management of Data}, pages 543--557.
  ACM, 2020.

\bibitem{sanofi2019}
Sanofi.
\newblock Journey of vaccine: a complex manufacturing process.
\newblock {https://www.sanofi.com/en/your-health/vaccines/production}, 2019.

\bibitem{Shanghai2022Schiffling}
S.~Schiffling and N.~Valantasis~Kanellos.
\newblock Shanghai: world's biggest port is returning to normal, but supply
  chains will get worse before they get better.
\newblock
  \url{https://theconversation.com/shanghai-worlds-biggest-port-is-returning-to-normal-but-supply-chains-will-get-worse/before-they-get-better-182720},
  2022.

\bibitem{schneider1990implementing}
F.~B. Schneider.
\newblock Implementing fault-tolerant services using the state machine
  approach: A tutorial.
\newblock {\em Computing Surveys (CSUR)}, 22(4):299--319, 1990.

\bibitem{sharma2019blurring}
A.~Sharma, F.~M. Schuhknecht, D.~Agrawal, and J.~Dittrich.
\newblock Blurring the lines between blockchains and database systems: the case
  of hyperledger fabric.
\newblock In {\em SIGMOD Int. Conf. on Management of Data}, pages 105--122.
  ACM, 2019.

\bibitem{sit2021experimental}
M.-K. Sit, M.~Bravo, and Z.~Istv{\'a}n.
\newblock An experimental framework for improving the performance of bft
  consensus for future permissioned blockchains.
\newblock In {\em Proceedings of the 15th ACM Int. Conf. on Distributed and
  Event-based Systems}, pages 55--65, 2021.

\bibitem{steffen2019zkay}
S.~Steffen, B.~Bichsel, M.~Gersbach, N.~Melchior, P.~Tsankov, and M.~Vechev.
\newblock zkay: Specifying and enforcing data privacy in smart contracts.
\newblock In {\em ACM SIGSAC Conf. on Computer and Communications Security
  (CCS)}, pages 1759--1776, 2019.

\bibitem{counterfeit2021stone}
J.~Stone.
\newblock How counterfeit covid-19 vaccines and vaccination cards endanger us
  all.
\newblock
  \url{https://www.forbes.com/sites/judystone/2021/03/31/how-counterfeit-covid-19-vaccines-and-vaccination-cards-endanger-us-all/?sh=eaddb0e36495},
  2021.

\bibitem{taft2014store}
R.~Taft, E.~Mansour, M.~Serafini, J.~Duggan, A.~J. Elmore, A.~Aboulnaga,
  A.~Pavlo, and M.~Stonebraker.
\newblock E-store: Fine-grained elastic partitioning for distributed
  transaction processing systems.
\newblock {\em Proc. of the VLDB Endowment}, 8(3):245--256, 2014.

\bibitem{thomson2012calvin}
A.~Thomson, T.~Diamond, S.-C. Weng, K.~Ren, P.~Shao, and D.~J. Abadi.
\newblock Calvin: fast distributed transactions for partitioned database
  systems.
\newblock In {\em SIGMOD Int. Conf. on Management of Data}, pages 1--12. ACM,
  2012.

\bibitem{tian2017supply}
F.~Tian.
\newblock A supply chain traceability system for food safety based on haccp,
  blockchain \& internet of things.
\newblock In {\em Int. Conf. on service systems and service management
  (ICSSSM)}, pages 1--6. IEEE, 2017.

\bibitem{vassantlal2022cobra}
R.~Vassantlal, E.~Alchieri, B.~Ferreira, and A.~Bessani.
\newblock Cobra: Dynamic proactive secret sharing for confidential bft
  services.
\newblock In {\em Symposium on Security and Privacy (SP)}, pages 1528--1528.
  IEEE, 2022.

\bibitem{vaccine2016}
V.~E. (VE).
\newblock Vaccines manufacturing. how are vaccines produced?
\newblock
  {https://www.vaccineseurope.eu/about-vaccines/how-are-vaccines-produced},
  2016.

\bibitem{wang2018forkbase}
S.~Wang, T.~T.~A. Dinh, Q.~Lin, Z.~Xie, M.~Zhang, Q.~Cai, G.~Chen, B.~C. Ooi,
  and P.~Ruan.
\newblock Forkbase: An efficient storage engine for blockchain and forkable
  applications.
\newblock {\em Proceedings of the VLDB Endowment}, 11(10):1137--1150, 2018.

\bibitem{xu2019vchain}
C.~Xu, C.~Zhang, and J.~Xu.
\newblock vchain: Enabling verifiable boolean range queries over blockchain
  databases.
\newblock In {\em SIGMOD Int. Conf. on Management of Data}, pages 141--158,
  2019.

\bibitem{xu2021div}
Z.~Xu and L.~Chen.
\newblock Div: Resolving the dynamic issues of zero-knowledge set membership
  proof in the blockchain.
\newblock In {\em SIGMOD Int. Conf. on Management of Data}, pages 2036--2048.
  ACM, 2021.

\bibitem{yin2003separating}
J.~Yin, J.-P. Martin, A.~Venkataramani, L.~Alvisi, and M.~Dahlin.
\newblock Separating agreement from execution for byzantine fault tolerant
  services.
\newblock {\em Operating Systems Review (OSR)}, 37(5):253--267, 2003.

\bibitem{zakhary2020atomic}
V.~Zakhary, D.~Agrawal, and A.~El~Abbadi.
\newblock Atomic commitment across blockchains.
\newblock {\em Proc. of the VLDB Endowment}, 13:1319--1331, 2020.

\bibitem{zakhary2019towards}
V.~Zakhary, M.~J. Amiri, S.~Maiyya, D.~Agrawal, and A.~El~Abbadi.
\newblock Towards global asset management in blockchain systems.
\newblock {\em arXiv preprint arXiv:1905.09359}, 2019.

\bibitem{zamani2018rapidchain}
M.~Zamani, M.~Movahedi, and M.~Raykova.
\newblock Rapidchain: Scaling blockchain via full sharding.
\newblock In {\em SIGSAC Conf. on Computer and Communications Security}, pages
  931--948. ACM, 2018.

\bibitem{zhang2021authenticated}
C.~Zhang, C.~Xu, H.~Wang, J.~Xu, and B.~Choi.
\newblock Authenticated keyword search in scalable hybrid-storage blockchains.
\newblock In {\em Int. Conf. on Data Engineering (ICDE)}, pages 996--1007.
  IEEE, 2021.

\end{thebibliography}

\end{document}